\RequirePackage{fix-cm}

\documentclass[pdftex,twocolumn,epjc3]{svjour3}          

\usepackage[margin=1.5cm]{geometry} 

\RequirePackage[T1]{fontenc}
\smartqed  

\RequirePackage{mathptmx}      
\RequirePackage[numbers,sort&compress]{natbib}
\RequirePackage[colorlinks,citecolor=blue,urlcolor=blue,linkcolor=blue]{hyperref}

\journalname{Eur.~Phys.~J.~C}
\usepackage{subfig}
\usepackage{xspace}
\usepackage{verbatim}
\usepackage{color}\definecolor{darkblue}{rgb}{0,0,0.5}
\usepackage{upgreek}
\usepackage{booktabs}
\usepackage{hyperref}
\hypersetup{
    bookmarks=true,         
    unicode=false,          
    pdftoolbar=true,        
    pdfmenubar=true,        
    pdffitwindow=true,      
    pdftitle={My title},    
    pdfauthor={Author},     
    pdfsubject={Subject},   
    pdfnewwindow=true,      
    pdfkeywords={keywords}, 
    colorlinks=true,       
    linkcolor=darkblue,          
    citecolor=darkblue,        
    filecolor=magenta,      
    urlcolor=darkblue,         
}

\newcommand{\pth}{$\{ p, \theta \} \,$}

\newcommand{\Venus}{{\scshape Venus}\xspace}
\newcommand{\VenusLong}{{\scshape Venus4.12}\xspace}
\newcommand{\GiBUU}{{\scshape GiBUU}\xspace}
\newcommand{\GiBUULong}{{\scshape GiBUU1.6}\xspace}
\newcommand{\FlukaNew}{{\scshape Fluka2011}\xspace}

\newcommand{\Epos}{{\scshape Epos}\xspace}
\newcommand{\EposLong}{{\scshape Epos1.99}\xspace}
\newcommand{\NASixtyOne}{NA61\slash SHINE\xspace}
\newcommand{\avg}[1]{\langle{#1}\rangle}
\newcommand{\dd}{\mathrm{d}}

\usepackage[latin2]{inputenc}
\usepackage{indentfirst}
\usepackage{enumerate}

\usepackage{amsmath}
\usepackage{bm}
\usepackage{amssymb}
\usepackage[english]{babel}
\usepackage{url}
\usepackage{natbib}

\usepackage{multirow}
\usepackage[abs]{overpic}

\newif\ifpdf

\ifx\pdfoutput\undefined

\pdffalse

\else

\pdfoutput=1

\pdftrue

\fi

\ifpdf

\else

\usepackage{graphicx}

\fi

\usepackage[switch]{lineno}

\journalname{Eur. Phys. J. C}

\begin{document}
\newcommand{\ra}[1]{\renewcommand{\arraystretch}{#1}}
\DeclareGraphicsExtensions{.pdf,.png,.eps,.jpg,.ps}
\cleardoublepage

\pagenumbering{roman}

\title{\large \bf
  Measurements
  of $\pi^\pm$, $K^\pm$, $K^0_S$, $\Lambda$
  and proton production
  in proton--carbon interactions at 31\,GeV/$c$
  with the NA61/SHINE spectrometer at the CERN SPS
}

\clearpage

\institute{
{National Nuclear Research Center, Baku, Azerbaijan}\label{inst0}
\and{Faculty of Physics, University of Sofia, Sofia, Bulgaria}\label{inst1}
\and{Ru{\dj}er Bo\v{s}kovi\'c Institute, Zagreb, Croatia}\label{inst2}
\and{LPNHE, University of Paris VI and VII, Paris, France}\label{inst3}
\and{Karlsruhe Institute of Technology, Karlsruhe, Germany}\label{inst4}
\and{Fachhochschule Frankfurt, Frankfurt, Germany}\label{inst5}
\and{University of Frankfurt, Frankfurt, Germany}\label{inst6}
\and{University of Athens, Athens, Greece}\label{inst7}
\and{Wigner Research Centre for Physics of the Hungarian Academy of Sciences, Budapest, Hungary}\label{inst8}
\and{Institute for Particle and Nuclear Studies, KEK, Tsukuba, Japan}\label{inst9}
\and{University of Bergen, Bergen, Norway}\label{inst10}
\and{Institute for Nuclear Research, Moscow, Russia}\label{inst11}
\and{Joint Institute for Nuclear Research, Dubna, Russia}\label{inst12}
\and{St. Petersburg State University, St. Petersburg, Russia}\label{inst13}
\and{University of Belgrade, Belgrade, Serbia}\label{inst14}
\and{ETH Z\"urich, Z\"urich, Switzerland}\label{inst15}
\and{University of Bern, Bern, Switzerland}\label{inst16}
\and{University of Geneva, Geneva, Switzerland}\label{inst17}
\and{Jan Kochanowski University in Kielce, Poland}\label{inst18}
\and{National Center for Nuclear Research, Warsaw, Poland}\label{inst19}
\and{Jagiellonian University, Cracow, Poland}\label{inst20}
\and{University of Silesia, Katowice, Poland}\label{inst21}
\and{Faculty of Physics, University of Warsaw, Warsaw, Poland}\label{inst22}
\and{University of Wroc{\l}aw,  Wroc{\l}aw, Poland}\label{inst23}
\and{Warsaw University of Technology, Warsaw, Poland}\label{inst24}
\and{National Research Nuclear University ``MEPhI'' (Moscow Engineering Physics Institute), Moscow, Russia}\label{inst25}
\and{Los Alamos National Laboratory, Los Alamos, USA}\label{inst27}
\and{University of Colorado, Boulder, USA}\label{inst28}
\and{University of Pittsburgh, Pittsburgh, USA}\label{inst29}
\\
$^*${present address: Department of Physics, COMSATS Institute of Information Technology, Islamabad 44000 Pakistan}
}
\author{
{N.~Abgrall}\thanksref{inst17}
\and{A.~Aduszkiewicz}\thanksref{inst22}
\and{Y.~Ali}\thanksref{inst20}$^{,*}$
\and{E.~Andronov}\thanksref{inst13}
\and{T.~Anti\'ci\'c}\thanksref{inst2}
\and{N.~Antoniou}\thanksref{inst7}
\and{B.~Baatar}\thanksref{inst12}
\and{F.~Bay}\thanksref{inst15}
\and{A.~Blondel}\thanksref{inst17}
\and{J.~Bl\"umer}\thanksref{inst4}
% ...
\\(\NASixtyOne Collaboration) 
}

 \date{\today}

 \maketitle
 \clearpage

\begin{abstract}
  Measurements
  of hadron production 
  in p+C interactions at 31\,GeV/$c$ are
  performed using the \NASixtyOne spectrometer at the CERN SPS. 
  The analysis is based on the full set of data collected in 2009 
  using
  a graphite target with a thickness of 4\% 
  of a nuclear interaction length. 
  Inelastic and production cross sections as well as spectra of 
  $\pi^\pm$, $K^\pm$, p, $K^0_S$ and $\Lambda$
  are measured with high precision. 
  These measurements are essential for improved calculations 
  of the initial neutrino fluxes in
  the T2K long-baseline neutrino oscillation experiment in Japan.
  A comparison of the \NASixtyOne measurements with predictions of
  several hadropro\-duction models is presented. 
\end{abstract}


 \tableofcontents

 \pagenumbering{arabic}

\section{Introduction}

 The \NASixtyOne (SPS Heavy Ion and Neutrino Experiment) at CERN pursues
 a rich physics programme in various fields of 
 physics~\cite{proposala,add1,proposalb,Status_Report_2008}.
 Hadron production
 measurements in p+C~\cite{pion_paper,kaon_paper} and $\pi$+C interactions 
 are performed which are required to improve calculations of neutrino fluxes 
 for the T2K/J-PARC~\cite{T2K} and Fermilab neutrino experiments~\cite{USNA61}  
 as well as for simulations of cosmic-ray air showers in
 the Pierre Auger and KASCADE experiments~\cite{Auger,KASCADE}. The
 programme on strong interactions investigates p+p~\cite{NA61_2009_pp_EPJC}, 
 p+Pb and nucleus-nucleus collisions
 at SPS energies, to study the onset of deconfinement and to search for
 the critical point of strongly interacting matter~\cite{Status_Report_2014,Status_Report_2015}. 

\begin{figure*}
\centering
\subfloat[$\pi^+$]{
\includegraphics[width=0.32\textwidth]{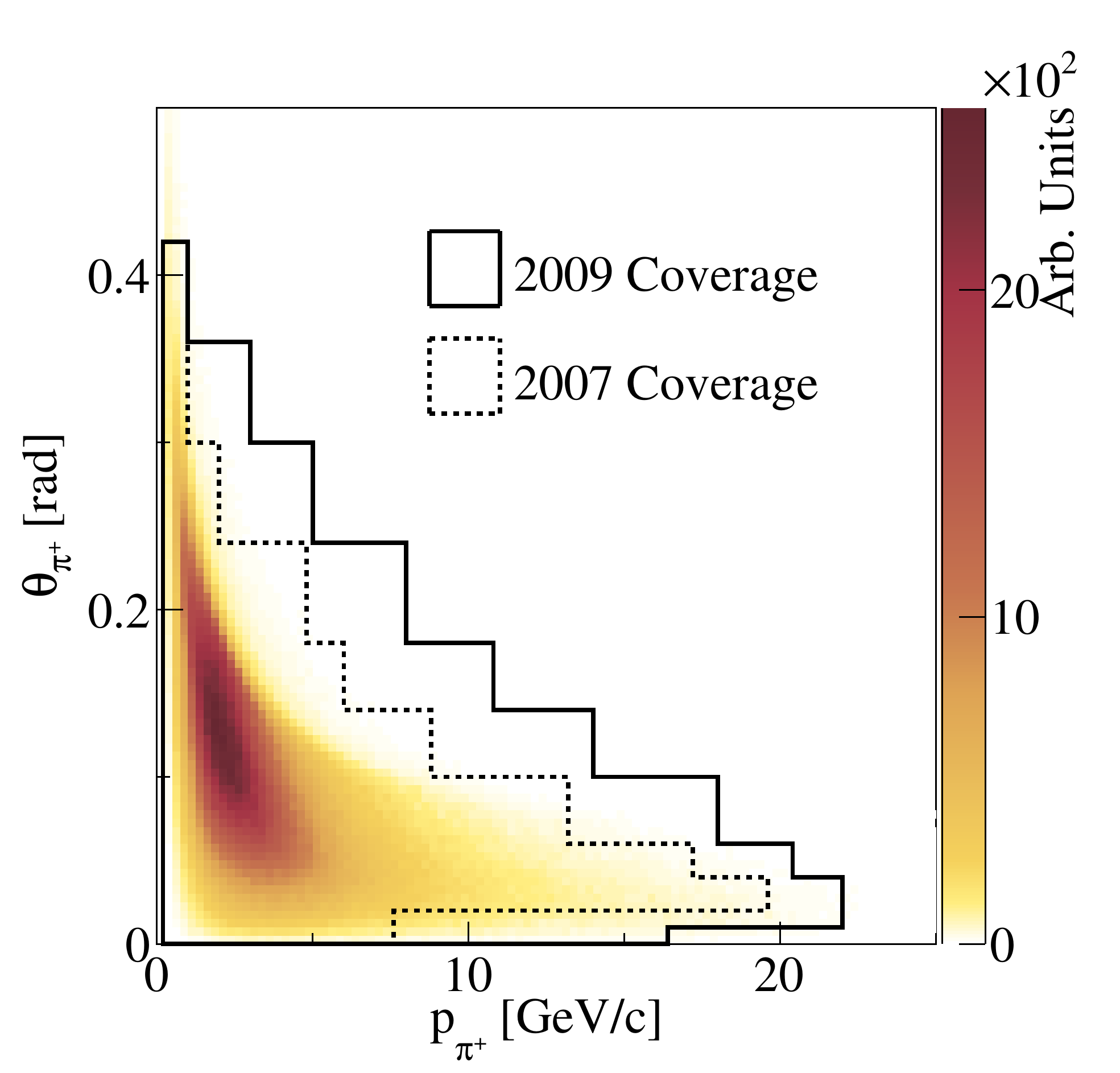}}
\subfloat[$\pi^-$]{
\includegraphics[width=0.32\textwidth]{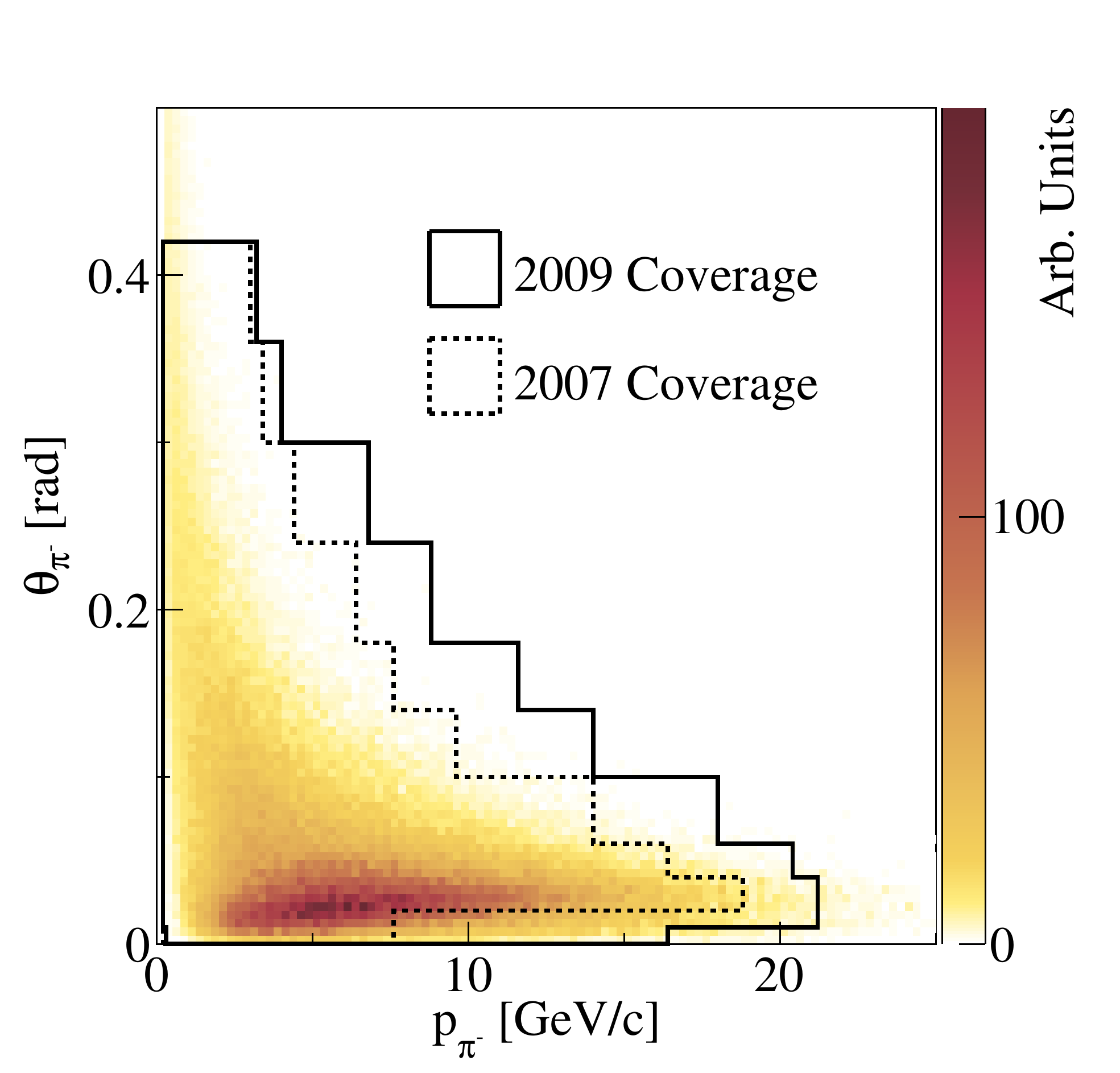}} 
\subfloat[$K^+$]{
\includegraphics[width=0.32\textwidth]{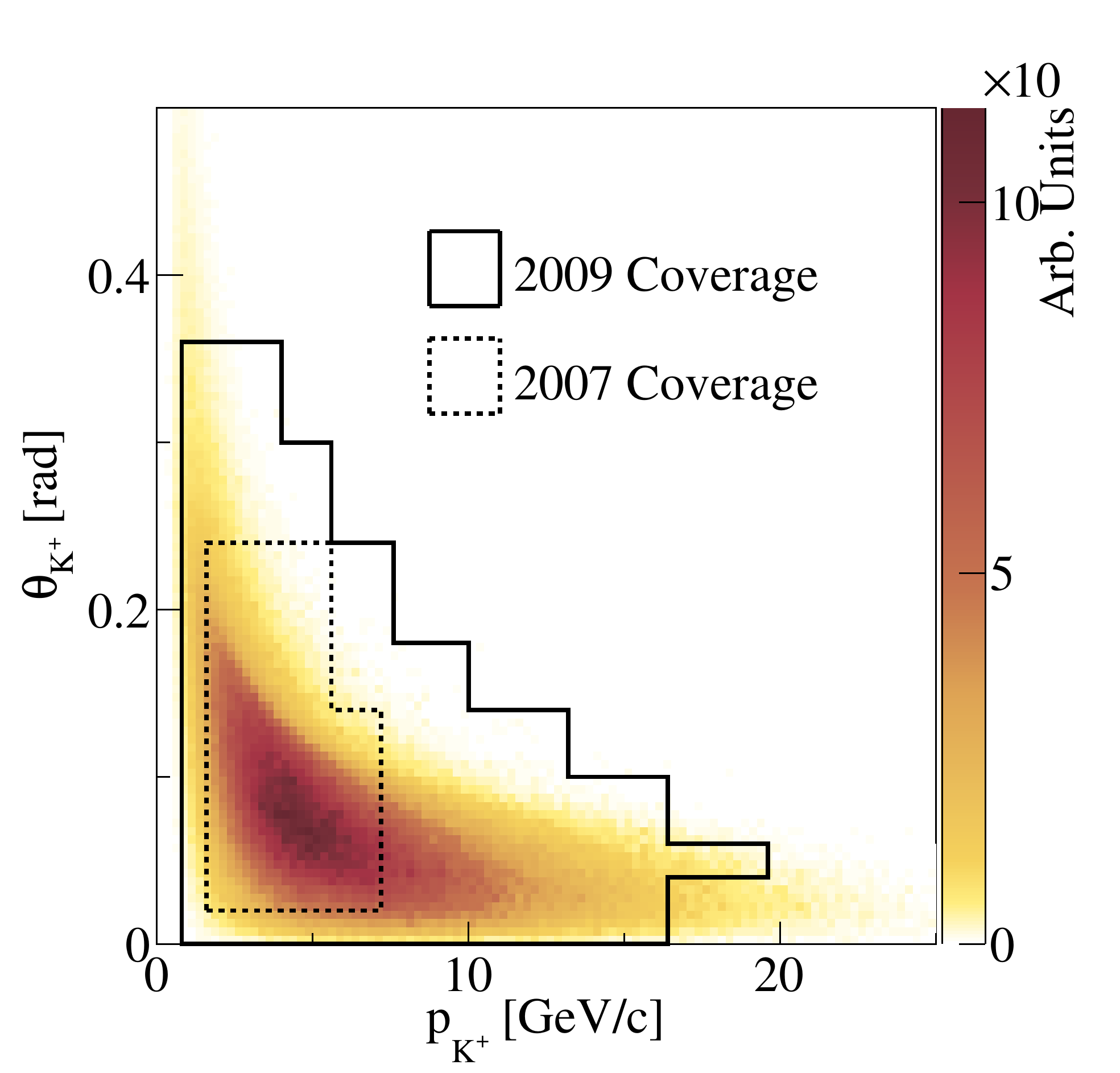}} \\
\subfloat[$K^-$]{
\includegraphics[width=0.32\textwidth]{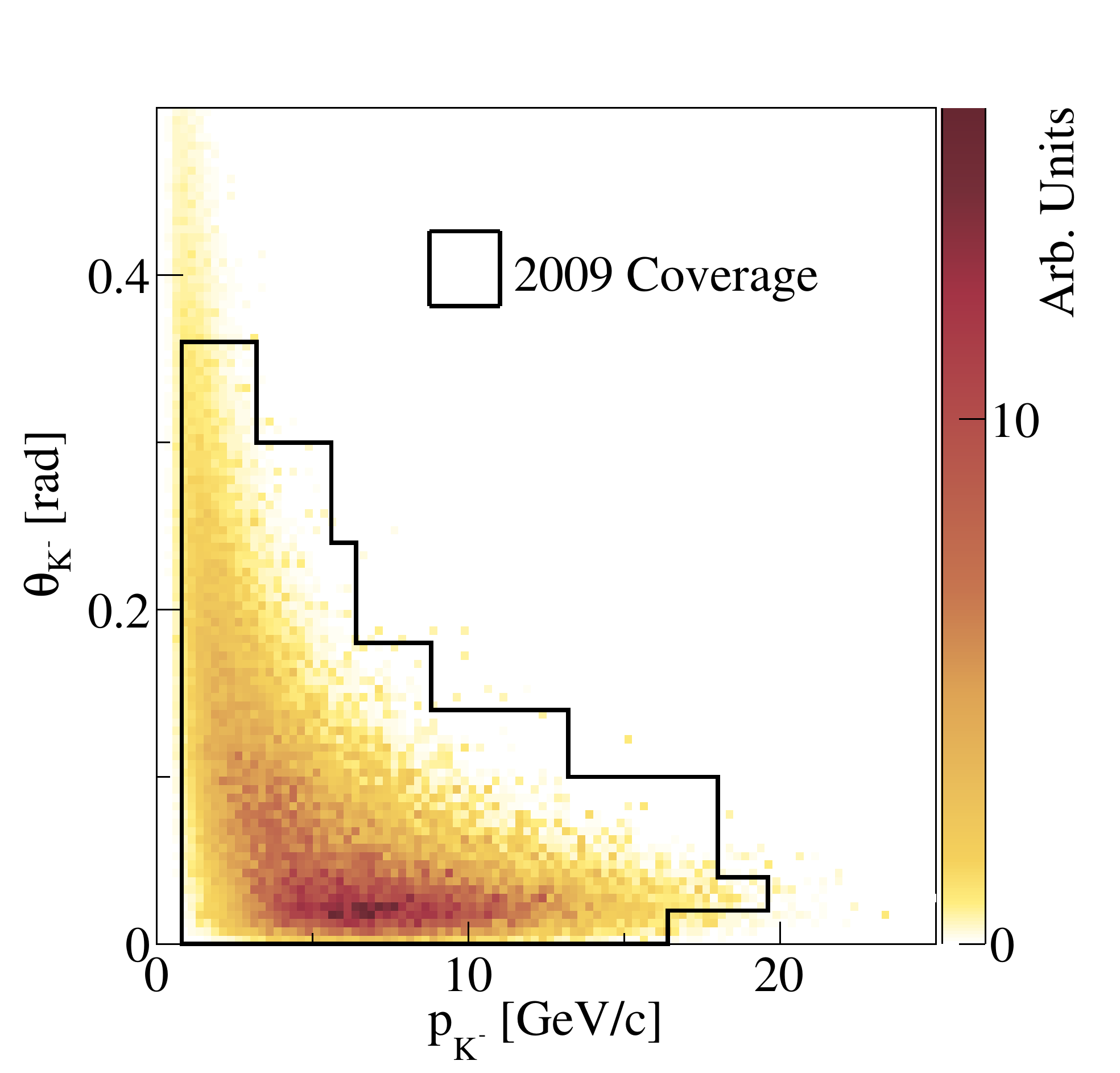}}
\subfloat[$K^0_S$]{
\includegraphics[width=0.32\textwidth]{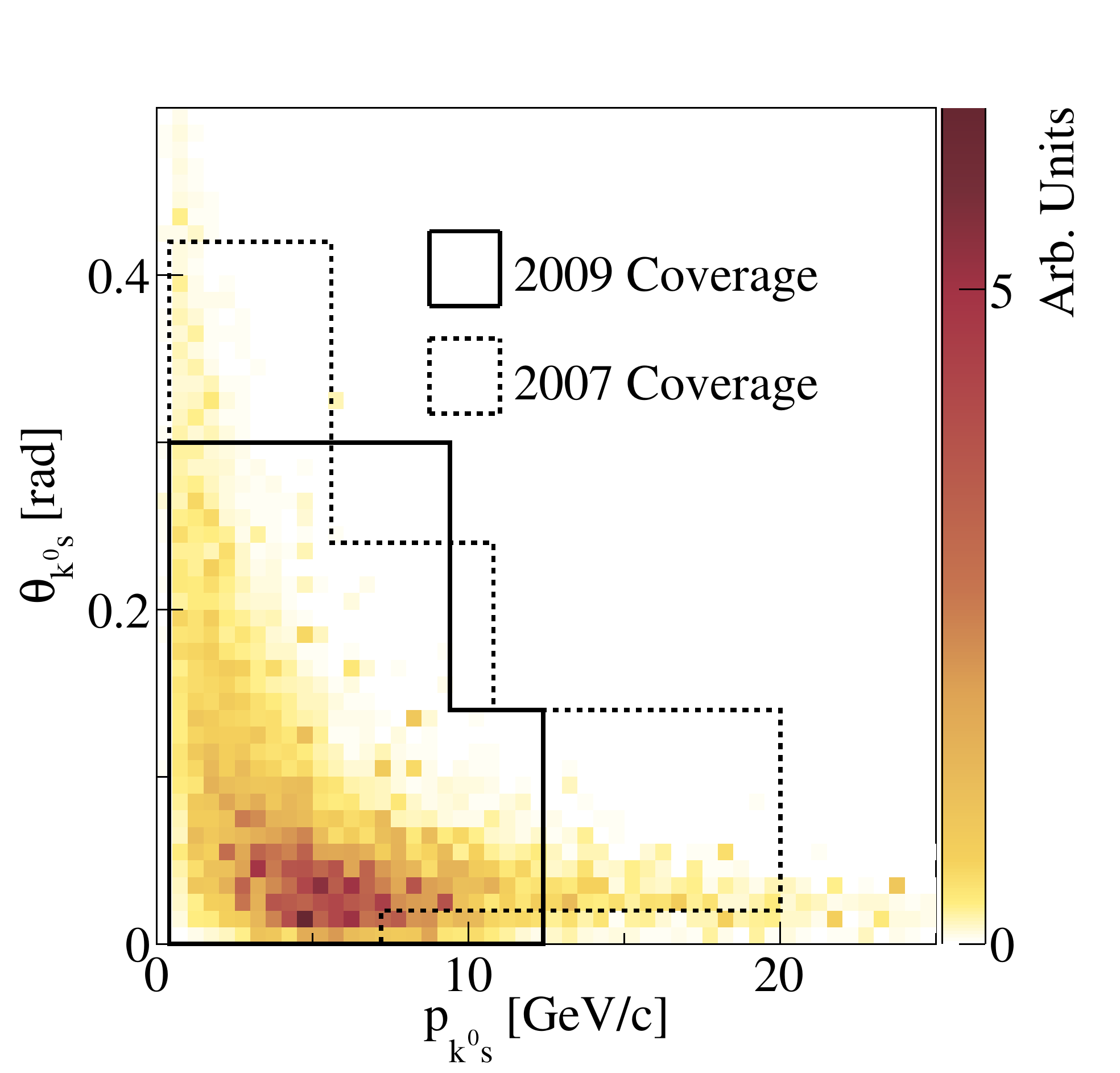}}
\subfloat[proton]{
\includegraphics[width=0.32\textwidth]{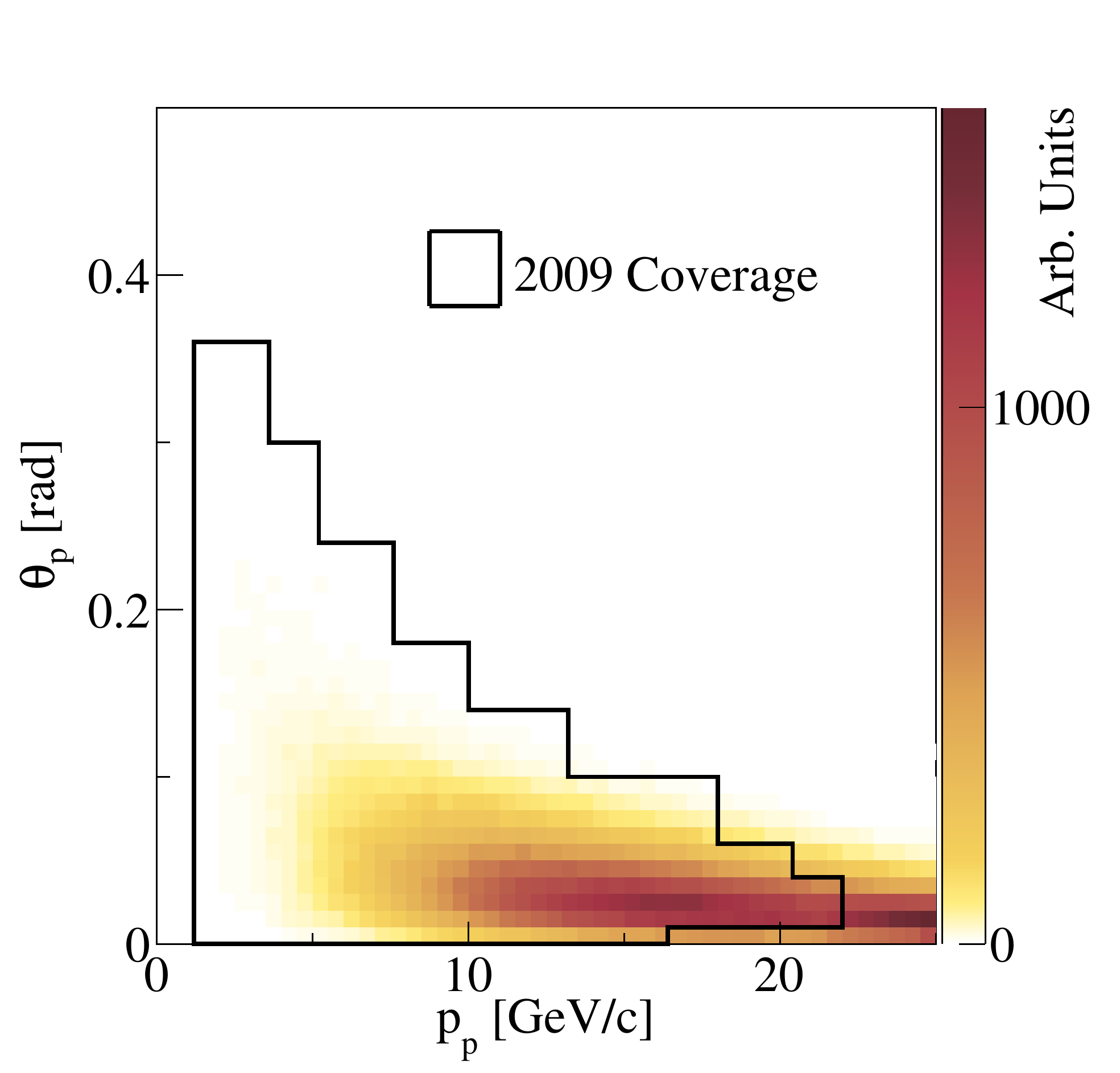}} 
\caption{
(Colour online)
The \pth phase space of $\pi^\pm$, $K^\pm$, $K^0_S$
and protons contributing 
to the predicted neutrino flux at SK 
in the ``positive'' focusing configuration, 
and the regions covered by the previously published \NASixtyOne 
measurements~\cite{pion_paper,kaon_paper} and by the new results presented 
in this article. 
Note that the size of the \pth bins used in the $K^0_S$ analysis of 
the 2007 data~\cite{V0_2007} is much larger compared to what is chosen
for the $K^0_S$ analysis presented here, see Section~\ref{Sec:v0}.
}
\label{fig:na61_coverage}
\end{figure*}

\begin{figure*}
\centering
\subfloat[$\pi^+$]{
\includegraphics[width=0.32\textwidth]{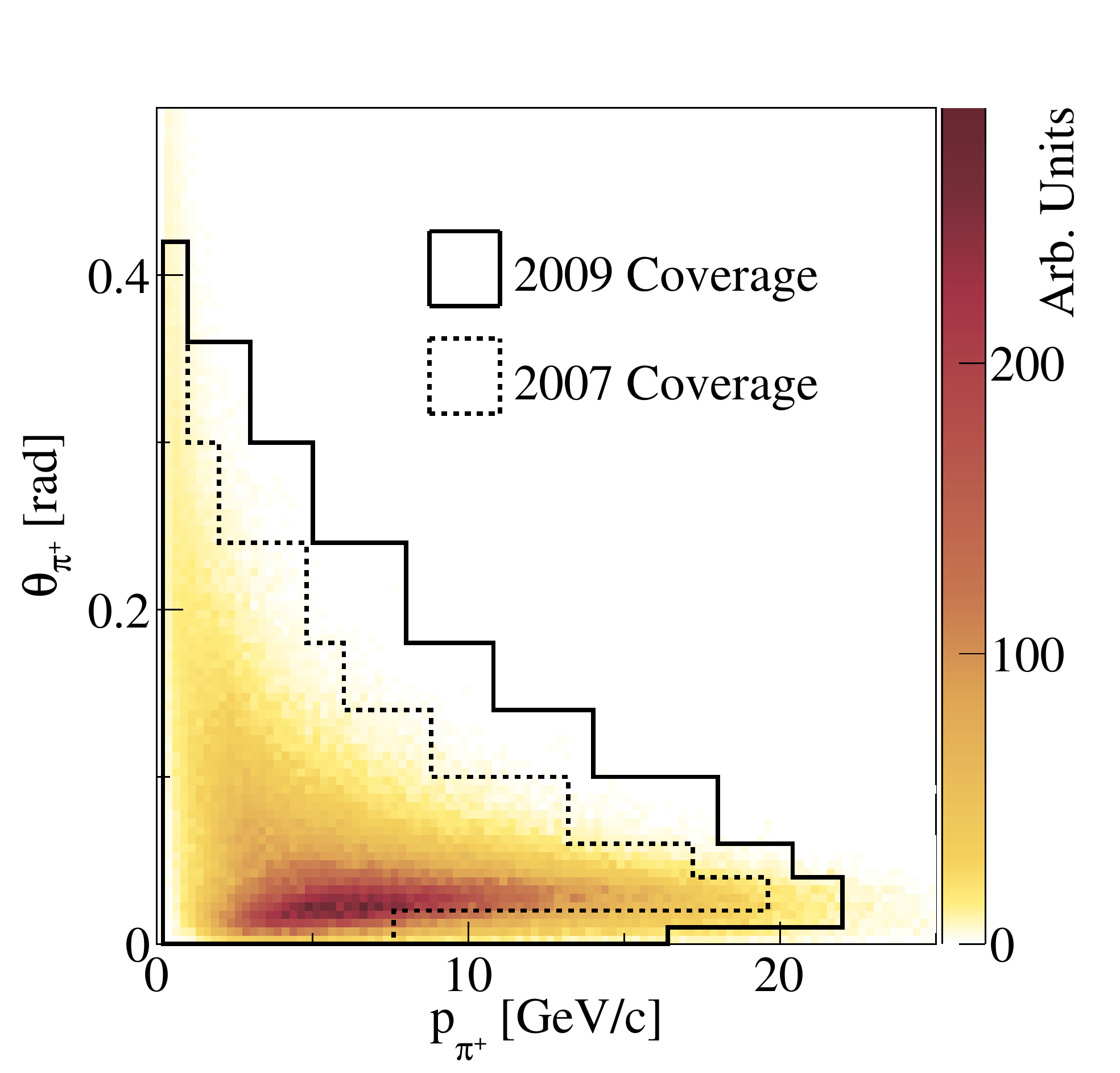}}
\subfloat[$\pi^-$]{
\includegraphics[width=0.32\textwidth]{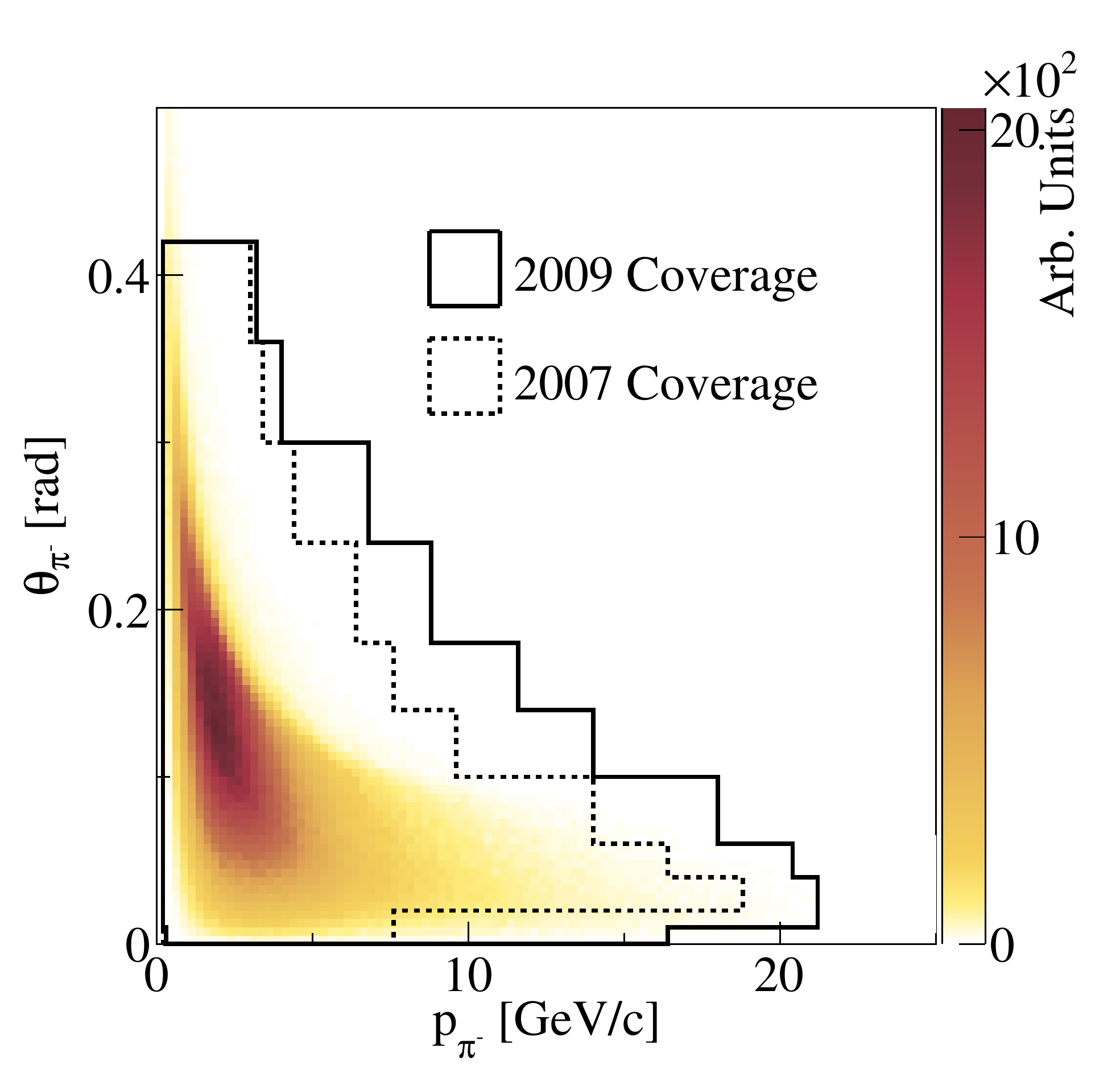}} 
\subfloat[$K^+$]{
\includegraphics[width=0.32\textwidth]{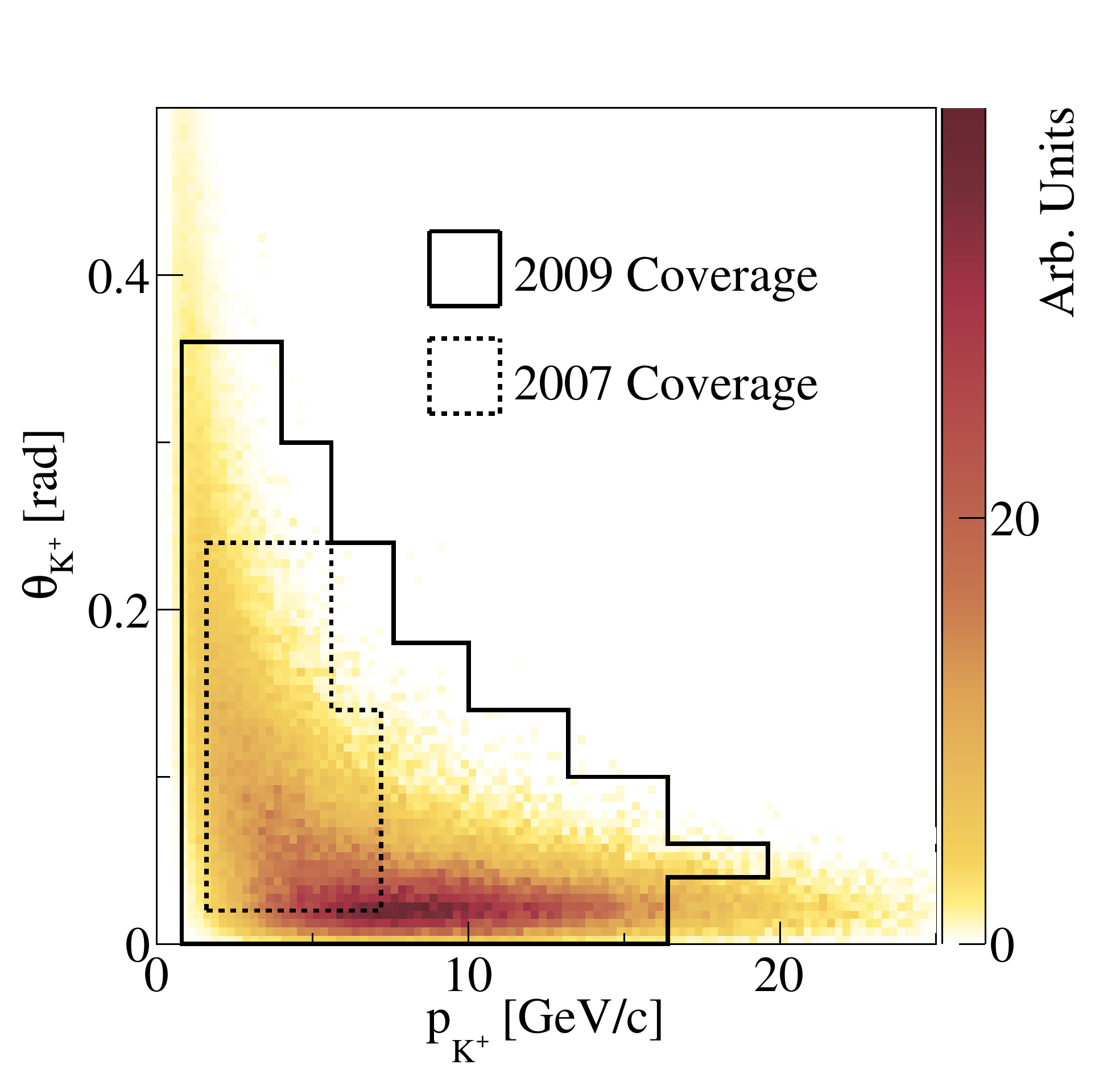}} \\
\subfloat[$K^-$]{
\includegraphics[width=0.32\textwidth]{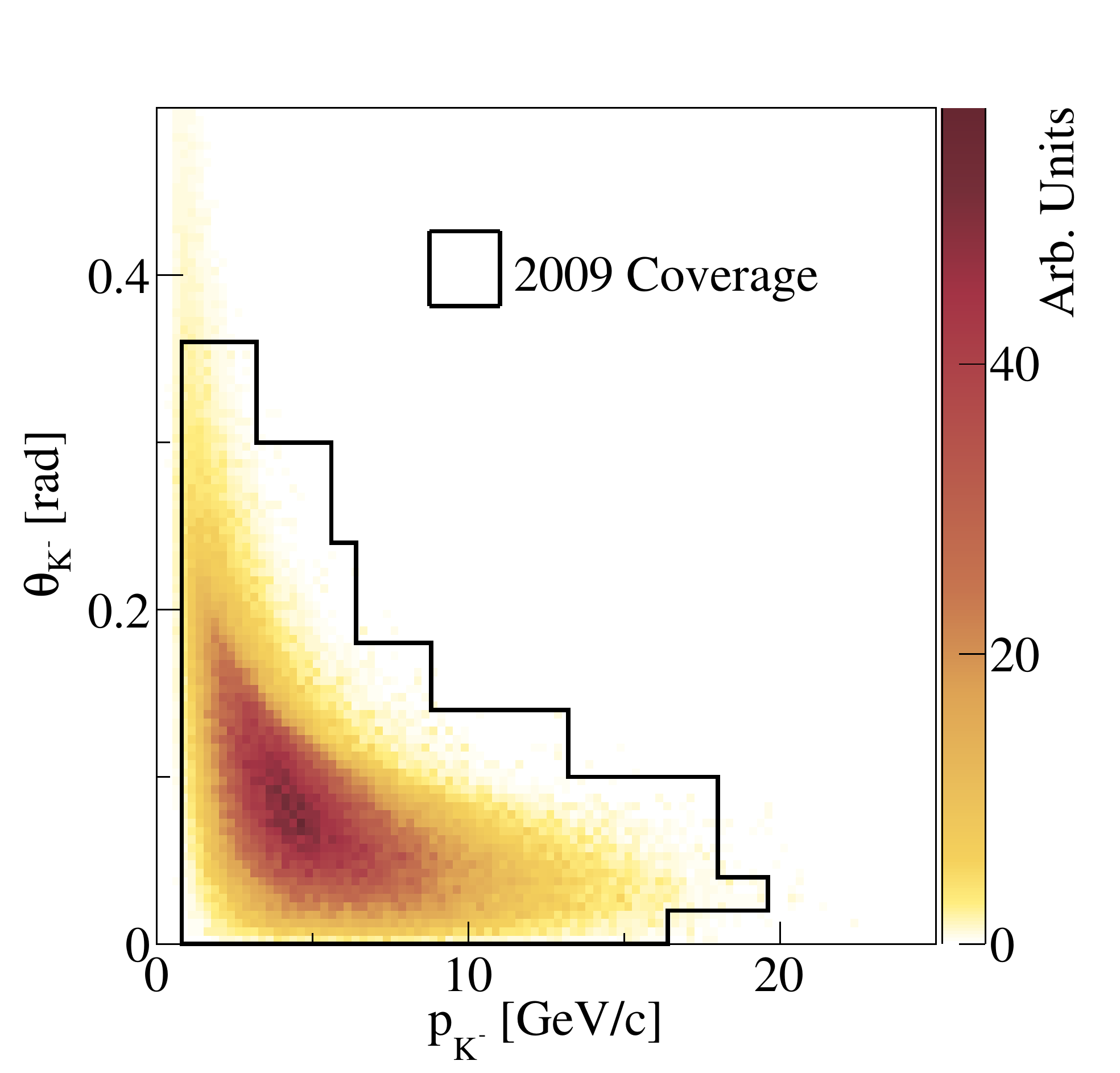}}
\subfloat[$K^0_S$]{
\includegraphics[width=0.32\textwidth]{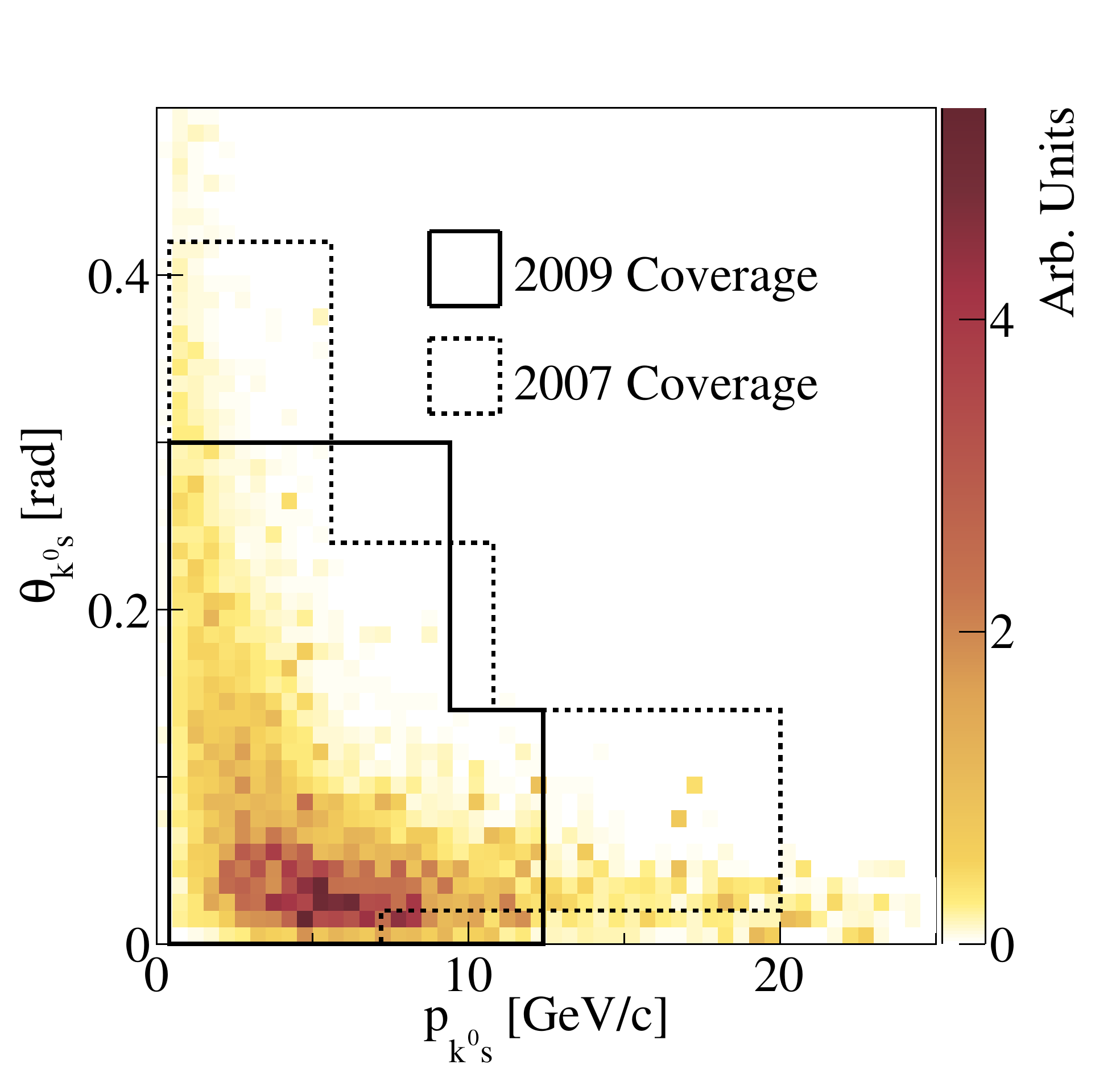}}
\subfloat[proton]{
\includegraphics[width=0.32\textwidth]{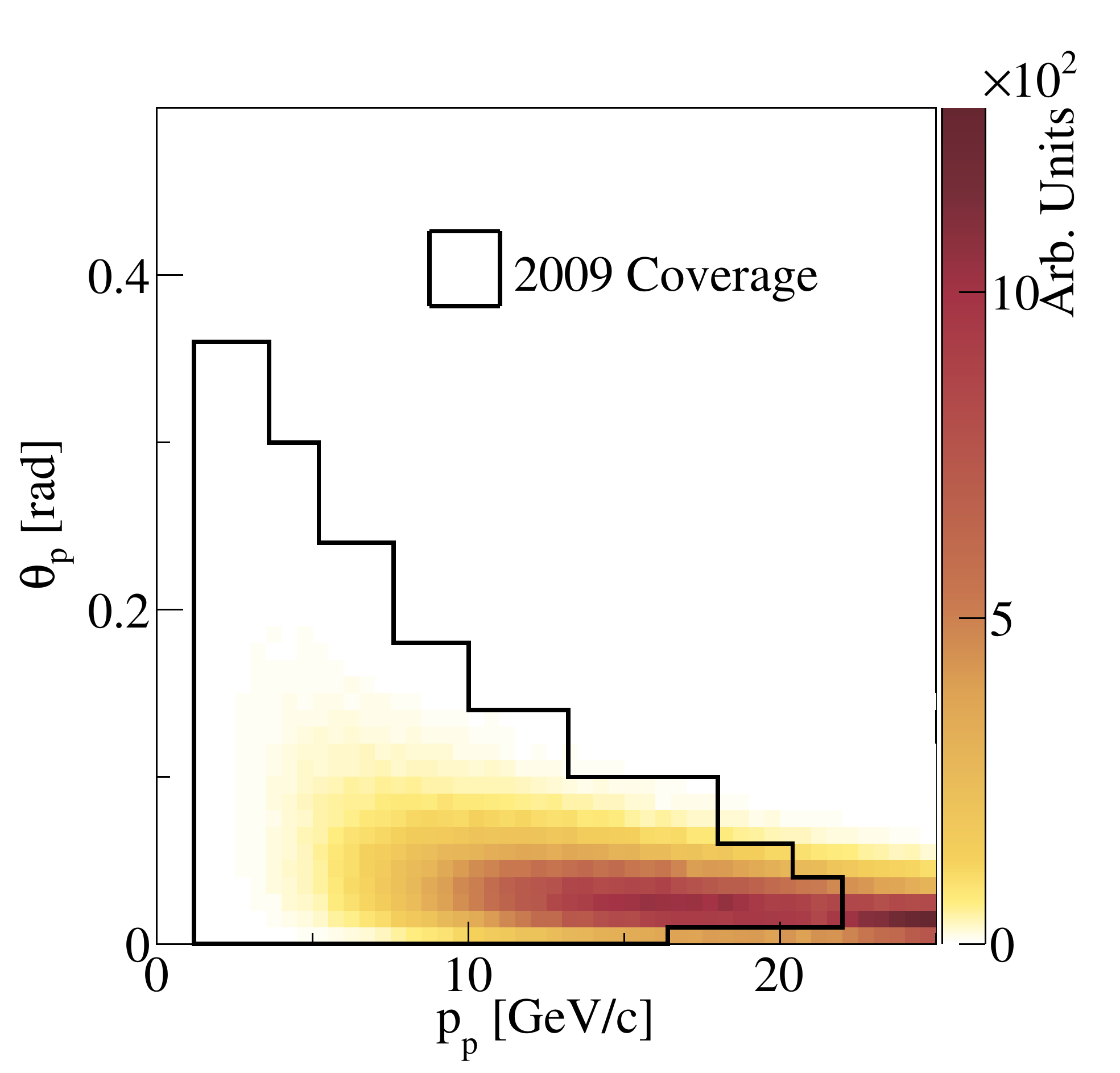}} 
\caption{
(Colour online)
The \pth phase space of $\pi^\pm$, $K^\pm$, $K^0_S$
and protons contributing 
to the predicted neutrino flux at SK 
in the ``negative'' focusing configuration, 
and the regions covered by the previously published \NASixtyOne 
measurements~\mbox{\cite{pion_paper,kaon_paper}} and by the new results presented 
in this article. 
Note that the size of the \pth bins used in the $K^0_S$ analysis of
the 2007 data~\cite{V0_2007} is much larger compared to what is chosen
for the $K^0_S$ analysis presented here, see Section~\ref{Sec:v0}.
}
\label{fig:na61_coverage_antinu}
\end{figure*}

 First measurements of 
 $\pi^\pm$ and $K^+$
 spectra
 in proton-carbon interactions 
 at 31\,GeV/$c$ 
 were already
 published by \NASixtyOne~\cite{pion_paper,kaon_paper} 
 and used for neutrino flux predictions 
 for the T2K experiment~\cite{T2K1stnue,T2K1stnumu,T2Kflux,T2Knueapp,T2Knumudisap,T2Knueapp_obs,T2Knumudisap_precise,T2Knumucs,Abe:2014nox,Abe:2014agb} 
 at J-PARC. 
 Yields of $K^0_S$ and $\Lambda$ were also published~\cite{V0_2007}.
 All of those measurements were performed using the data collected
 during the \NASixtyOne pilot run in 2007.
 A detailed description of the experimental apparatus and
 analysis techniques can be found in Refs.~\cite{NA61detector_paper,pion_paper}.

 This article presents new \NASixtyOne measurements of
 charged pion, kaon and proton as well as of $K^0_S$ and $\Lambda$
 spectra in p+C interactions at 31\,GeV/$c$,
 based on an eight times 
 larger dataset collected in 2009,
 after the detector and readout upgrades. 
 These results are important for achieving 
 the future T2K physics goals~\cite{T2Kfuture}. 

 T2K -- a long-baseline neutrino oscillation experiment from \mbox{J-PARC} 
 in Tokai to Kamioka (Japan)~\cite{T2K} -- aims to precisely measure
  $\nu_\mu\rightarrow\nu_e$ 
 appearance~\cite{T2K1stnue,T2Knueapp,T2Knueapp_obs} and
 $\nu_\mu$ disappearance~\cite{T2K1stnumu,T2Knumudisap,T2Knumudisap_precise}.
 The neutrino beam is generated by the \mbox{J-PARC}
 high intensity 30\,GeV (kinetic energy) proton beam interacting in a
 90\,cm long graphite target to produce $\pi$ and $K$ mesons, which decay
 emitting neutrinos. Some of the forward-going hadrons, mostly protons, 
 reinteract 
 in the target and surrounding material. To study and constrain
 the reinteractions in the long target, a special set of data
 was taken by \NASixtyOne with a replica of the T2K target:
 the first 
 study based on
 pilot data is presented in~Ref.~\cite{LTpaper},
 the analysis of the high-statistics 2009 dataset is finalized~\cite{Alexis},
 while the analysis of the last 2010 dataset
 is still on-going.

 The 
 T2K neutrino beam~\cite{T2Kflux} is aimed towards a near
 detector complex, 280\,m from the target, and towards the Super-Kamiokande
 (SK) far detector located 295\,km away at $2.5^\circ$ off-axis from
 the
 hadron beam. Neutrino oscillations are probed by comparing
 the neutrino event rates and spectra measured in SK to predictions of a
 Monte-Carlo (MC) simulation based on flux calculations and near detector
 measurements. Until the \NASixtyOne data were available, these flux
 calculations were based on hadron production models tuned to sparse
 available data, resulting in systematic uncertainties which are large
 and difficult to evaluate. Direct measurements of particle production
 rates in p+C interactions allow for more precise and
 reliable estimates~\cite{T2Kflux}. 
 Precise predictions of neutrino fluxes are also crucial for neutrino
 cross section measurements with the T2K near detector, 
 see e.g.~Refs.~\cite{T2Knumucs,Abe:2014nox,Abe:2014agb}.

 For the first stage of the experiment,
 the T2K neutrino beamline was set up
 to focus positively charged hadrons (the so-called ``positive'' focusing), 
 to produce a $\nu_\mu$ enhanced
 beam. While charged pions generate most of the low energy neutrinos,
 charged kaons generate the high energy tail of the T2K
 beam, and contribute substantially to the intrinsic $\nu_e$
 component in the T2K beam. See Ref.~\cite{T2Kflux} for more details.
 An anti-neutrino enhanced beam can be produced by
 reversing the current direction in the focusing elements of the beamline
 in order to focus negatively charged particles (``negative'' focusing).

 Positively and negatively charged pions and kaons whose daughter neutrinos 
 pass through the SK
 detector constitute the kinematic region of interest, shown in
 Figs.~\ref{fig:na61_coverage} and~\ref{fig:na61_coverage_antinu} 
 in the kinematic variables $p$ and
 $\theta$ -- the momentum and polar angle of particles in the
 laboratory frame for ``positive'' and ``negative'' focusing, 
 respectively. See Refs.~\cite{pion_paper,kaon_paper,T2Kflux}
 for additional information. 
 The much higher statistics available in the 2009 data
 makes it possible to use finer $\{p,\theta\}$ binning 
 (especially for charged kaons, $K_S^0$ and $\Lambda$)
 compared to previously published 
 results~\cite{pion_paper,kaon_paper,V0_2007} from the 2007 data.  
 The improved
 statistics of the 2009 data also allows for the first measurements of
 negatively charged kaons within \NASixtyOne.  

 The \NASixtyOne results on hadron production are also extremely important for
 testing and improving existing hadron production models in an energy
 region which is not well constrained by measurements at present.

 The paper is organized as follows: A brief description of the experimental
 setup, the collected data and their processing is presented 
 in Section~\ref{sec:setup}. 
 Section~\ref{Sec:norm} is devoted to the analysis technique
 used for the measurements of the inelastic and production cross
 sections in proton-carbon interactions at 31\,GeV/$c$ and 
 presents the obtained results. A detailed description 
 of the procedures used to obtain the differential inclusive spectra of hadrons
 is presented in Section~\ref{Sec:ana}. 
 Results on spectra are reported in Section~\ref{sec:Results}.
 A comparison of these results with the
 predictions of different hadron production models is discussed 
 in Section~\ref{Sec:models}. A summary and conclusions are given 
 in Section~\ref{Sec:summary}.


\section{The experimental setup, collected data and their processing} 
\label{sec:setup}

\begin{figure*}
\centering
  \includegraphics[width=0.89\textwidth]{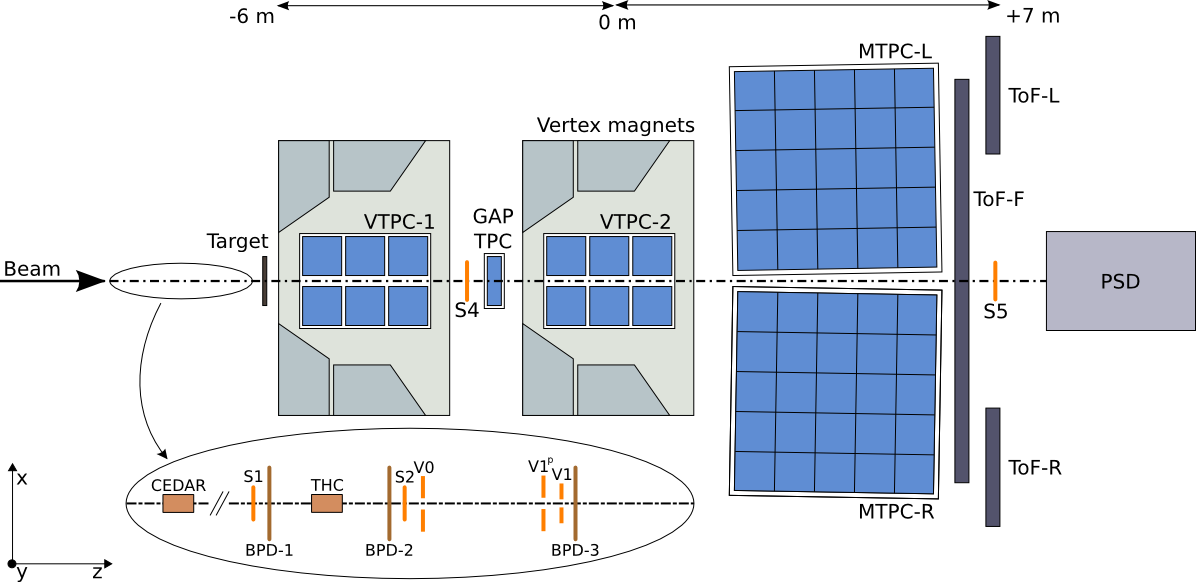}
  \caption{ 
  (Colour online)
  The schematic layout of the \NASixtyOne spectrometer 
  (horizontal cut, not to scale).
  The beam and trigger detector configuration used for data
  taking in 2009 is shown in the inset.
  The chosen coordinate system is drawn on the plot: 
  its origin lies in the middle of the VTPC-2, on the beam axis.
  The nominal beam direction is along the $z$ axis.
  The magnetic field bends charged particle trajectories
  in the $x$--$z$ (horizontal) plane. Positively charged particles are bent
  towards the top of the plot.
  The drift direction in the TPCs is along the $y$ (vertical) axis.
  }
\label{fig:detector}
\end{figure*}

The \NASixtyOne apparatus is a wide-acceptance hadron spectrometer 
at the CERN SPS. 
A detailed description of the \NASixtyOne setup is presented
in Ref.~\cite{NA61detector_paper}. Only parts relevant for the 2009 running
period are briefly described here.
The \NASixtyOne experiment has greatly profited from the long development of the CERN
proton and ion sources, the accelerator chain, as well as the H2 beamline
of the CERN North Area.
Numerous components of the \NASixtyOne setup were inherited from its
predecessors, in particular, the last one --  
the NA49 experiment~\cite{NA49-NIM}. 

The detector is built arround
five
Time Projection Chambers (TPCs), as shown in Fig.~\ref{fig:detector}. Two Vertex TPCs
(\mbox{VTPC-1} and \mbox{VTPC-2}) are placed in the magnetic field produced by two
superconducting dipole magnets and two Main-TPCs (\mbox{MTPC-L} and 
\mbox{MTPC-R}) are located
downstream symmetrically with respect to the beamline. 
An additional small TPC is placed between VTPC-1 and VTPC-2,
covering the very-forward region, and is referred to as the GAP TPC (GTPC).
The GTPC allows to extend the kinematic coverage at forward production
angles compared to the previously published results from the 2007 pilot run.

The TPCs are filled with Ar:CO$_2$ gas mixtures in proportions 90:10 
for the VTPCs and the GTPC, and 95:5 for the MTPCs.

\begin{figure}
\centering
\includegraphics[width=0.9\linewidth]{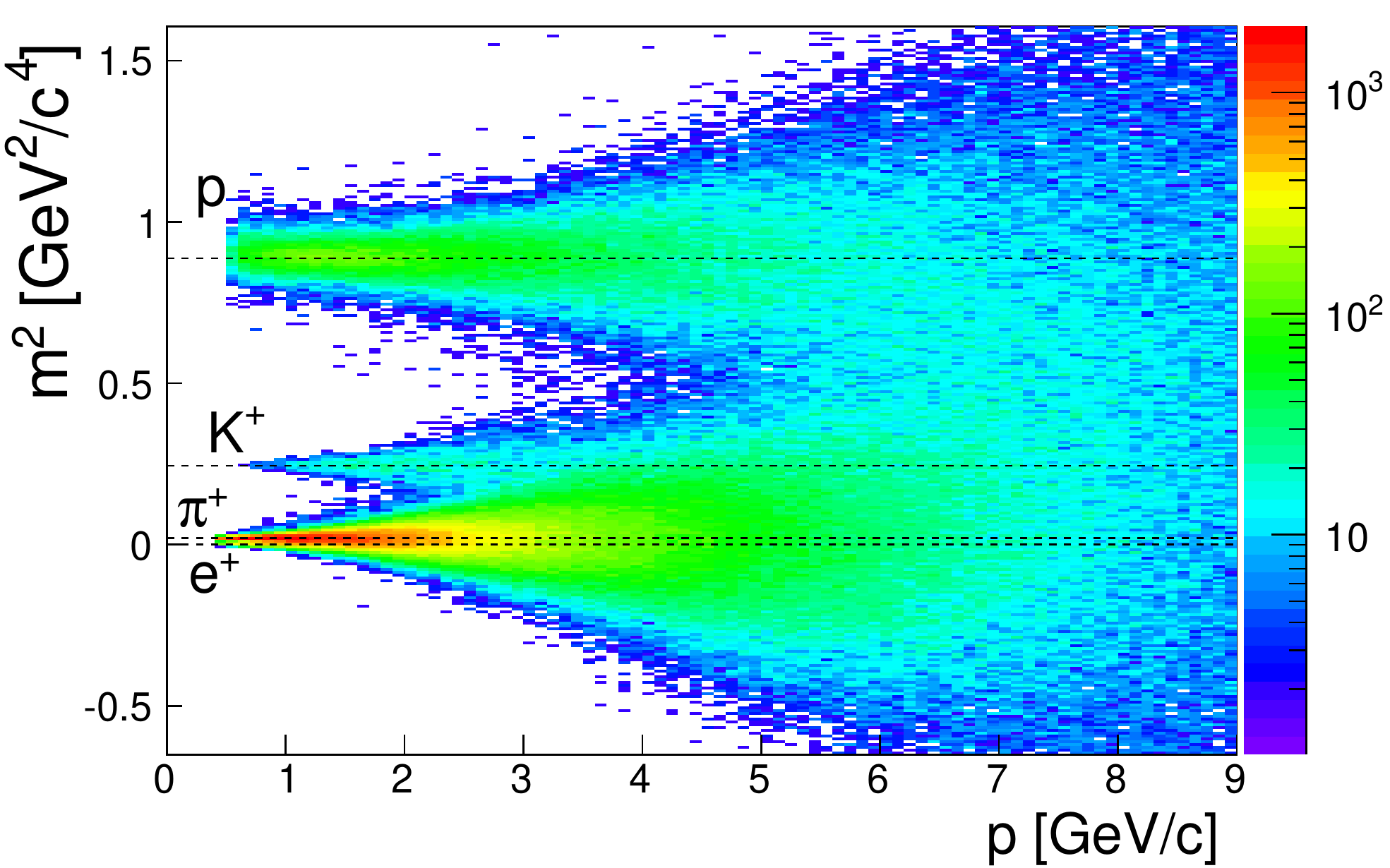}
\caption{(Colour online)
  Mass squared of positively charged particles,
  computed from the \mbox{ToF-F} measurement and the fitted 
  track parameters, as a function of momentum. The lines show the expected 
  mass squared values for different particles species.
}
\label{m2_vs_p}
\end{figure}

In the forward region, the experimental setup is
complemented by a time-of-flight (\mbox{ToF-F}) detector array
horizontally segmented into 80 scintillator bars, read out at both ends
by photomultipliers~\cite{NA61detector_paper}. 
Before the 2009 run, the \mbox{ToF-F} detector was upgraded with
additional modules 
placed on both sides of the beam in order 
to extend the acceptance for the analysis described here.
The intrinsic time resolution of each scintillator is
about 110\,ps~\cite{NA61detector_paper}.
The particle identification capabilities of the \mbox{ToF-F} are
illustrated in Fig.~\ref{m2_vs_p}.

For the study presented here the
magnetic field of the dipole magnets was set to a bending power of 1.14\,Tm.
This leads to a
momentum resolution $\sigma(p)/p^2$ in the track reconstruction of
about $5{\times}10^{-3}$\,(GeV/$c$)$^{-1}$ for long tracks reaching 
the \mbox{ToF-F}. 

Two scintillation counters, S1 and S2, provide the beam definition,
together with the three veto counters V0, V1 and V1$^p$,
which define the beam upstream of the target.
The S1 counter provides also the start time for all counters.          
The beam particles are identified by a CEDAR~\cite{CEDAR} and 
a threshold Cherenkov (THC) counter.  
The selection of beam protons (the beam trigger, $T_{beam}$) is then defined 
by the coincidence 
$\text{S1}\land\text{S2}\land\overline{\text{V0}}
\land\overline{\text{V1}}\land\overline{\text{V1p}}\land\text{CEDAR}\land 
\overline{\text{THC}}$.
The interaction trigger $T_{int} = T_{beam} \land \overline{\text{S4}}$
is given by a beam proton and the absence of a signal in S4, 
a scintillation counter, with a 2\,cm diameter, placed
between the VTPC-1 and VTPC-2 detectors along the  beam trajectory
at about 3.7~m from the target,
see Fig.~\ref{fig:detector}.
Almost all beam protons that interact in the target do not reach S4.
The interaction and beam triggers are run simultaneously.
The beam trigger events were recorded with a frequency
by a factor of about 10
lower than the frequency of interaction trigger events.

The incoming beam particle trajectories are precisely measured by a 
set
of three Beam Position Detectors (BPDs), placed along the beamline 
upstream of the target, as shown in the insert in Fig.~\ref{fig:detector}.
These detectors are $4.8 \times 4.8$\,cm$^2$ proportional chambers 
operated with an Ar:CO$_2$ (85:15) gas mixture. 
Each BPD measures the position of the beam particle in the plane 
transverse to the beam direction with a resolution of 
${\sim}100$\,$\upmu$m (see Ref.~\cite{NA61detector_paper} for more details). 

The same target was used in 2007 and 2009 -- 
an isotropic graphite sample with a thickness along 
the beam axis of 2\,cm, equivalent to about 4\% of a nuclear interaction length,
$\lambda_\text{I}$. During the data taking the target was placed
80\,cm upstream of the VTPC-1.

TPC readout and data acquisition (DAQ) system upgrades were performed before the 2009 run. 
To utilize the new DAQ capability, a much higher intensity beam was used 
during the 2009 data taking compared to the 2007 running period.
To cope with the high beam intensity in 2009
the passing times of individual beam particles before and after the event were recorded.
These are later used in the analysis to study and remove possible pileup effects. 

Reconstruction and calibration algorithms applied to the 2007 data are 
summarized in~Ref.~\cite{pion_paper}. Similar calibration procedures 
were applied to the 2009 data resulting in good data quality
suitable for the analysis (see e.g.~Ref.~\cite{NA61_2009_pp_EPJC}).

\begin{figure}
\includegraphics[width=0.9\linewidth]{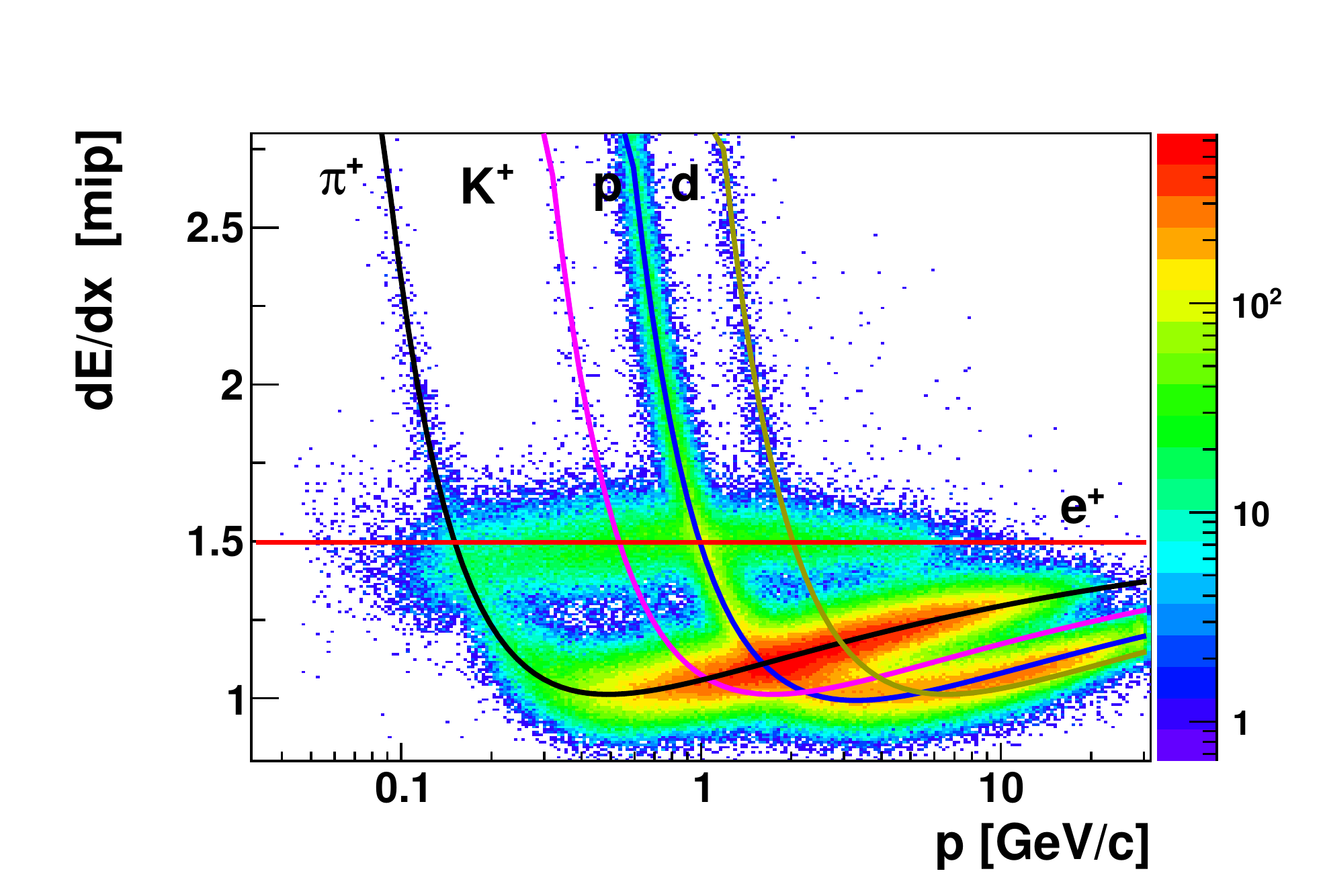}
\includegraphics[width=0.9\linewidth]{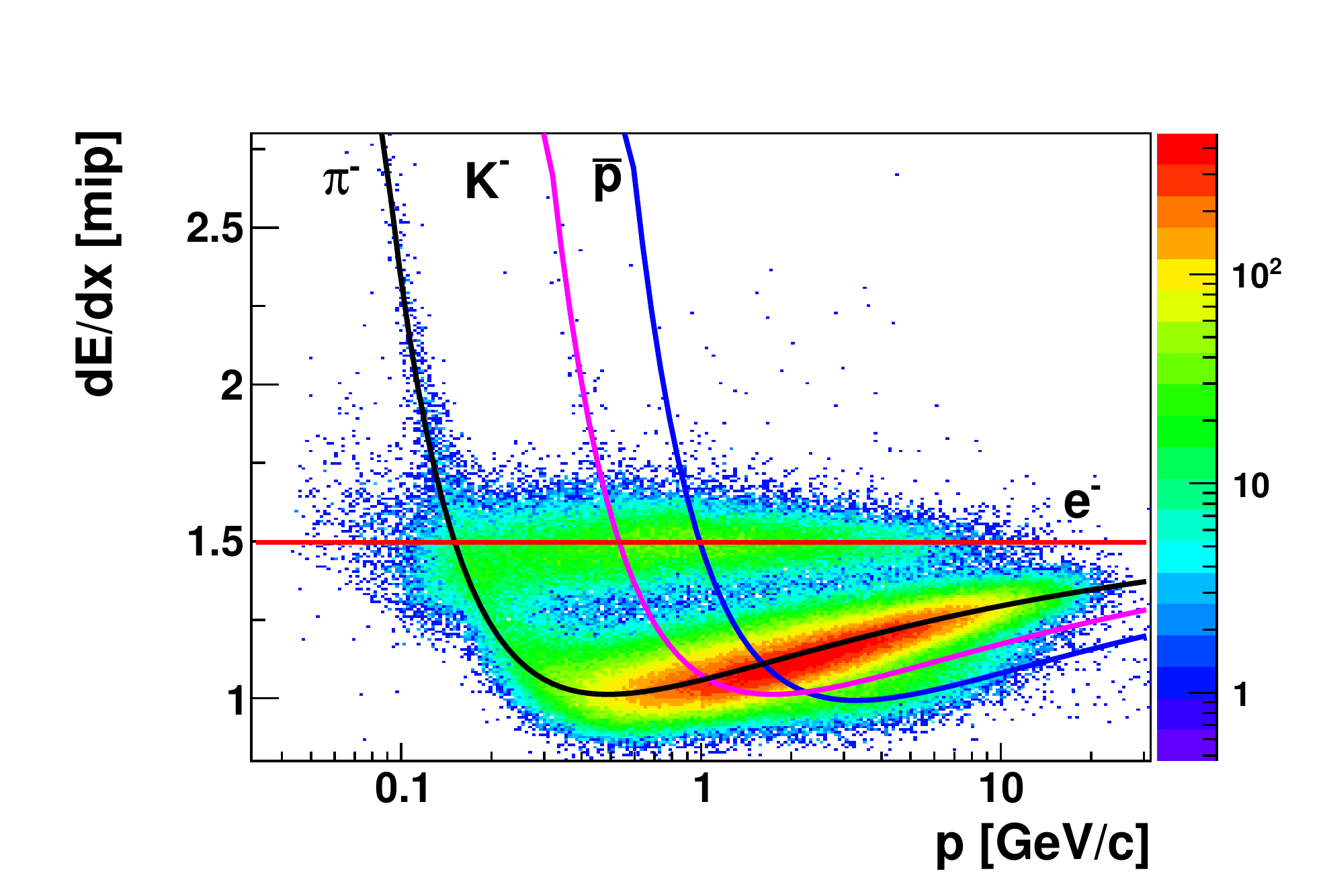}
\caption{(Colour online)
 Specific energy loss $\dd E/\dd x$ in the TPCs for positively (\emph{top}) and 
negatively (\emph{bottom}) charged particles as a function of momentum. 
Curves show the Bethe-Bloch (BB) parameterizations of the mean $\dd E/\dd x$ 
calculated 
for different hadron species. In the case of  electrons and positrons, which 
reach the Fermi plateau, the mean  $\dd E/\dd x$ is parameterized  
by a constant. }
\label{dedx_pos_neg}
\end{figure}

\begin{table*}
\centering
  \begin{tabular}{|c|c|c|c|c|c|}
    \hline
    beam & target & year & triggers\,$\times 10^6$ & status of the \NASixtyOne analysis & usage in the T2K beam MC \\ 
    \hline
    \hline
    & thin target& 2007 & 0.7 &  published: $\pi^\pm$ \cite{pion_paper}, $K^+$ \cite{kaon_paper}, $K^0_S$, $\Lambda$ \cite{V0_2007} & has already been used \cite{T2Kflux} \\
    \cline{3-6} 
    protons & 2\,cm ($0.04\lambda_\text{I}$)  & {\bf 2009} & 5.4 &  $\pi^\pm$, $K^\pm$, p, $K^0_S$, $\Lambda$ (this article) & currently being used \\
    \cline{2-6}
    at 31\,GeV/$c$ & T2K replica  & 2007 & 0.2 & published: $\pi^\pm$ \cite{LTpaper} & method developed \\
    \cline{3-6} 
    & target  & 2009 & 2.8 & analysis being finalized \cite{Alexis} & to be integrated \\
    \cline{3-6} 
    &  90\,cm ($1.9\lambda_\text{I}$) & 2010 & $7.2$ & analysis currently on-going & --- \\
    \hline
  \end{tabular} 
  \caption{A summary of the \NASixtyOne data collected for the T2K physics goals.
  }
  \label{tab_stat_NA61}
\end{table*}

Measurements of the specific energy loss $\dd E/\dd x$ of charged particles by ionisation in the TPC gas
are used for their identification. The $\dd E/\dd x$ of a particle is calculated as the 50\% truncated
mean of the charges of the clusters (points) on the track traversing the TPCs.
The calibrated $\dd E/\dd x$ distributions as a function of particle momentum
for positively and negatively
charged particles are presented in Fig.~\ref{dedx_pos_neg}.
The Bethe-Bloch (BB) parametrization of the mean energy loss, scaled to
the experimental data~(see Section~\ref{Sec:dedx}), is
shown by the curves for positrons (electrons),
pions, kaons, (anti)protons, and deuterons.
The typical achieved $\dd E/\dd x$ resolution is about 4\%.

Simulation of the \NASixtyOne detector response, used to correct
the raw data, is described in~Ref.~\cite{pion_paper} and 
additional details can be found in~Ref.~\cite{Nicolas}.

The particle spectra analysis described in Section~\ref{Sec:ana} 
is based on $4.6{\times}10^6$ reconstructed events 
with the target inserted (I) and $615{\times}10^3$ reconstructed
events with the target removed (R)
collected during the 2009 data-taking period
with a beam rate of about 100~kHz, much higher than in 2007 (15~kHz).
Only events for which a beam
track is properly reconstructed are selected for analysis.

A summary of the \NASixtyOne data collected for T2K is presented 
in Table\,\ref{tab_stat_NA61}.


\section{Inelastic and production cross section measurements}
\label{Sec:norm}

This section discusses the procedures used to obtain the inelastic and production cross 
section for p+C interactions at 31~GeV/$c$ and presents the results.
The inelastic cross section $\sigma_\text{inel}$ is defined as the difference between 
the total cross section $\sigma_\text{tot}$ and the coherent elastic cross section
$\sigma_\text{el}$~(see e.g.~\cite{Bobchenko}):
\begin{equation}
\sigma_\text{inel} = \sigma_\text{tot} - \sigma_\text{el}~.
\label{sigma_inel_abs}
\end{equation}
Thus it comprises every reaction which occurs with desintegration of 
the carbon nucleus.

The production processes are 
defined as those in which new hadrons are produced. Thus
the production cross section $\sigma_\text{prod}$ is the difference between
$\sigma_\text{inel}$ and the quasi-elastic cross section $\sigma_\text{qe}$
where the incoming proton scatters off
an individual nucleon which, in turn, is ejected from the carbon nucleus:
\begin{equation}
\sigma_\text{prod} = \sigma_\text{inel} - \sigma_\text{qe}~.
\label{sigma_prod_abs}
\end{equation}

Since many improvements were made to the trigger logic 
and a much higher beam rate was used during the 2009
data taking compared to the 2007 run (see Section~\ref{sec:setup})
the normalization analysis of the 2009 data~\cite{Davide} 
differs from the one used for the 2007 data~\cite{pion_paper,claudia}.


\subsection{Interaction trigger cross section}
\label{Sec:trigxsec}

The simultaneous use of the beam  and interaction
triggers allows a direct determination of
the interaction trigger probability, $P_\text{Tint}$:
\begin{equation}
P_\text{Tint} =
\frac{N(T_\text{beam} \wedge T_\text{int})}{N(T_\text{beam})} \; ,
\label{eq:ptint}
\end{equation}
where  $N(T_\text{beam})$ is the number
of events which satisfy the beam trigger condition and
$N(T_\text{beam} \wedge T_\text{int})$ is the number of events which satisfy
both the beam trigger and interaction trigger conditions.
The interaction trigger probability was measured for
the target inserted, $P_\text{Tint}^I$, and target removed, $P_\text{Tint}^R$, 
configurations.
Table \ref{tab:eventcut} summarizes the number of beam and interaction trigger 
events before and after the event selection.

\begin{table}
\begin{center}
\begin{tabular}{|l|c|c|} 
\hline 											& Target inserted 					& Target removed			\\
\hline N($T_{beam}$) before cuts						& 577894  						& 257430 					\\ 
\hline N($T_\text{beam} \wedge T_\text{int}$) before cuts		& 39644							& 3705					\\
\hline N($T_{beam}$) after cuts							& 331735  						& 145682 					\\ 
\hline N($T_\text{beam} \wedge T_\text{int}$) after cuts		& 20578							& 1110					\\
\hline 
\end{tabular}
\caption{Number of beam trigger, N($T_{beam}$), and interaction trigger, N($T_\text{beam} \wedge T_\text{int}$), 
events before and after the event selection. Note that beam trigger events  were recorded with a frequency
by a factor 10 lower than interaction trigger events (see Section~\ref{sec:setup}).
}		
\label{tab:eventcut}	
\end{center}
\end{table}

The interaction probability in the carbon target
was calculated  as follows:
\begin{equation}  
  P_\text{int} =  \frac{ P_\text{Tint}^I  - P_\text{Tint}^R }{ 1 - P_\text{Tint}^R  }~. 
\label{eq:pint}
\end{equation}
$P_\text{int}$ is used to obtain the interaction trigger cross section $\sigma_\text{trig}$ from the formula: 
\begin{equation}
\sigma_\text{trig} = \frac{1}{\rho~L_\text{eff}~N_\text{A} / A} P_\text{int}~,
\label{eq:trigxsec_def}
\end{equation}
where $N_\text{A}$ is Avogadro's number and $\rho$, $A$ and $L_\text{eff}$ are  
the density, atomic mass and effective length of the target, respectively.
The effective target length accounts for the exponential beam attenuation
and can be computed according to 
\begin{equation}
L_\text{eff} = \lambda_\text{abs} \left( 1 - \exp^{- L / \lambda_\text{abs}} \right)~,
\label{eq:leff}
\end{equation}
where the absorption length is:
\begin{equation}
\lambda_\text{abs} = \frac{A}{ \rho~N_\text{A}~\sigma_\text{trig} }~.
\label{eq:lambda}
\end{equation}

Substituting Eqs.~\eqref{eq:leff} and \eqref{eq:lambda} 
into Eq.~\eqref{eq:trigxsec_def}, the formula for the interaction 
trigger cross section is obtained:
\begin{equation}
\sigma_\text{trig} = \frac{ A }{ L~\rho~N_\text{A} } \, \ln \left( \frac{1}{1 - 
P_\text{int}} \right)~.
\label{eq:sigma_trig_anal}
\end{equation}


\subsection{Event selection}
\label{Sec:evtsel}

An event selection was applied to improve the rejection of out-of-target 
interactions. The following two quality cuts based on the measurements of 
the beam position and the beam proton passage times were imposed:
\begin{enumerate}[(i)]
\item Requirement to have
both the $x$ and $y$ positions of the beam particle measured by all three BPDs. 
This selection is referred to later on as the {\it standard} BPD selection.
\item 
Rejection of events in which one or more additional beam particles
are detected in the time window $t=[-2,0]\,\upmu$s before the 
triggering beam particle. This avoids pileup in the BPDs due the
long signal shaping time.

\end{enumerate}

After applying these cuts 
the amount of out-of-target interactions decreased by about 45\% 
(see Table \ref{tab:eventcut}).


\subsection{Study of systematic uncertainty on $\sigma_\text{trig}$}
\label{Sec:trigxsec_syst}

The first component of systematic uncertainty was evaluated 
by varying the event selection criteria described in the previous subsection.
It amounts to 1.0\,mb.

Elastic scattering of the beam along the beamline was considered as a systematic bias. 
However, the fraction of events for which the beam extrapolation falls outside of the 
carbon target and that pass all the event selections, was evaluated
and found to be negligible.

Another source of potential systematic uncertainty relates to pileup
in the trigger system. The trigger logic has a time resolution of about 9\,ns.
If a pileup particle arrives within this time window
it can not be distinguished from the one which caused the trigger.
The measured $P_\text{Tint}$ and corrected unbiased 
$P_\text{Tint}^\text{corr}$
interaction trigger probabilities  
are related as: 
\begin{equation}
  P_\text{Tint}^\text{corr} = \frac{ P_\text{Tint} }{ 1 - P_\text{2beam} } \ ,    
\label{eq:ptrig_corr}
\end{equation}
where 
$P_\text{2beam}$ is the probability that a pile-up beam is within 
the trigger logic time window.
This probability was found to be $(0.18 \pm 0.07)$\%, thus
the correction to $\sigma_\text{trig}$ is negligible 
and no corresponding systematic uncertainty was assigned.

\begin{figure}
  \centering
  \includegraphics[width=.45\textwidth]{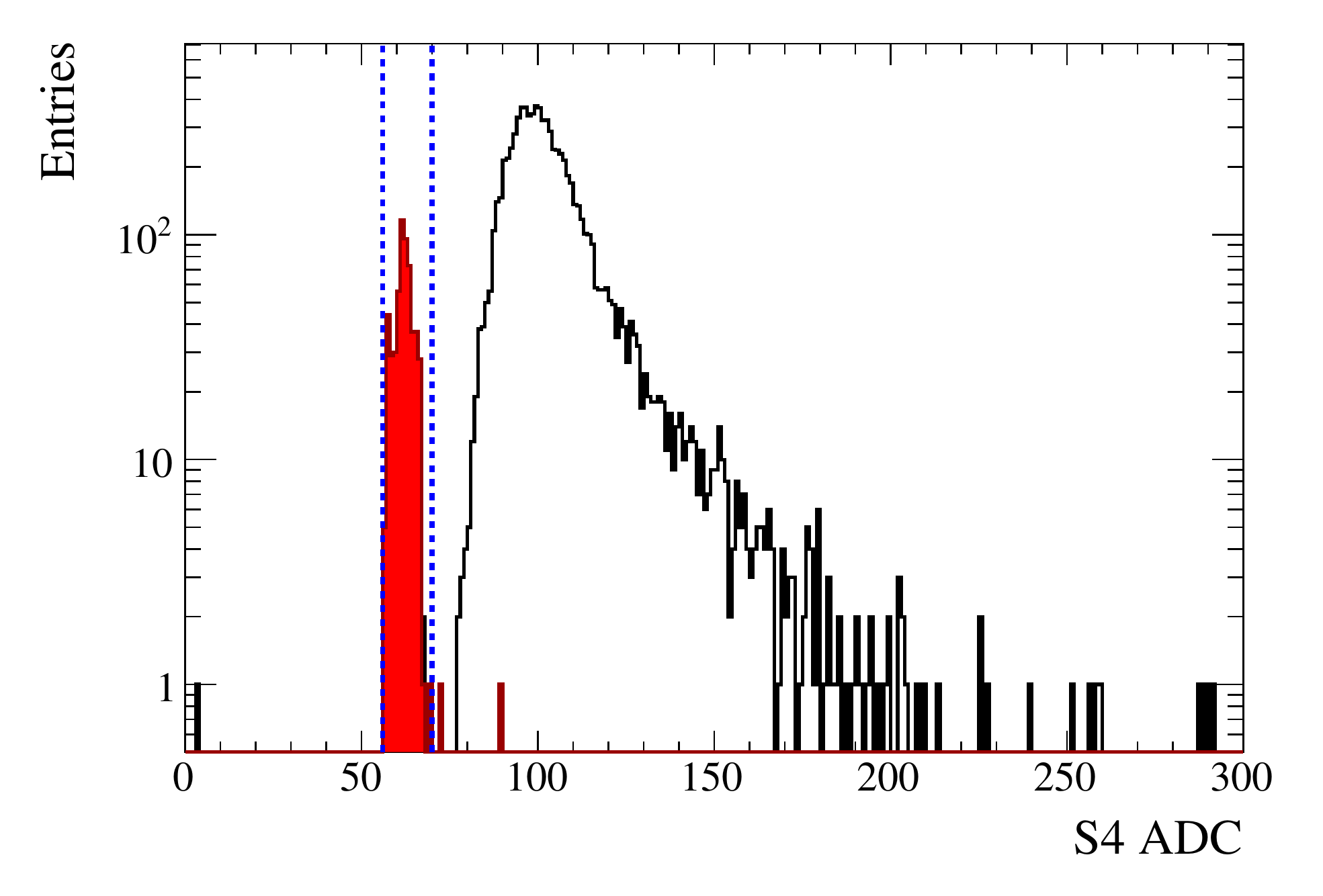}
\caption{ (Colour online)
    Distribution of S4 ADC counts for the beam trigger $T_\text{beam}$. 
    Red histogram corresponds to the interaction trigger subsample 
    \mbox{$T_\text{beam} \wedge T_\text{int}$}.}	
  \label{fig:adc_s4}
\end{figure}

The beam composition at 31\,GeV/$c$, measured with the CEDAR and the THC, 
is 84\% pions, 14\% protons and about 2\% kaons.
The proton component of the beam was selected by requiring respectively 
the coincidence and the anti-coin\-cidence of the CEDAR and THC counters
(see Section~\ref{sec:setup}).
In order to check the purity of the identified proton beam,  
the beam was deflected into the TPCs with the maximum magnetic field (9\,Tm)
and its composition was determined using
the energy loss measurements in the TPCs.
The fraction of misidentified particles in the proton beam
was found to be lower than 0.2\% and was considered negligible.

The efficiency 
of the interaction trigger was estimated using the ADC information 
from the S4 scintillator counter.
The ADC signal of S4 can be distorted if pileup beam particles 
are close in time to the triggering beam. 
To avoid this effect all the events with at least one pileup beam particle 
within $\pm4\,\upmu$s around the triggering beam particle were rejected.
In Fig.~\ref{fig:adc_s4} the distribution of the ADC signal is shown 
for a sample of events tagged by the beam trigger $T_\text{beam}$.
If a beam proton does not interact in the target, and thus hits S4, 
the ADC counts will be larger than 70. 
If both the beam and interaction trigger conditions are satisfied 
\mbox{$T_\text{beam} \wedge T_\text{int}$}, 
the  ADC signal corresponds to the pedestal and 
will be distributed between 56 and 
70  counts ($\Delta_{adc}$).
The  efficiency of the S4 counter as a part of the $T_\text{int}$ trigger
is defined as the ratio between the number of ADC counts in the $\Delta_\text{adc}$ interval 
for \mbox{$T_\text{beam} \wedge T_\text{int}$} events 
and the total number of ADC counts in the $\Delta_\text{adc}$ interval
for $T_\text{beam}$ events. 
The measured ratio is $(99.8 \pm 0.2) \%$.
This estimate of the S4 efficiency was cross-checked using the GTPC.
Beam track segments reconstructed in the GTPC were extrapolated to the $z$ position 
of S4. The fraction of the number of extrapolations hitting S4 
and also satisfying $T_\text{int}$ provides another estimate of the efficiency of S4.
Both results were found to be in agreement. Thus a possible bias caused
by inefficiency of S4 is considered negligible.


\subsection{Results on $\sigma_\text{trig}$}
\label{Sec:trigger_cross_section}

\begin{figure*}
	\centering
	\subfloat[]{

	\includegraphics[width=.45\textwidth]{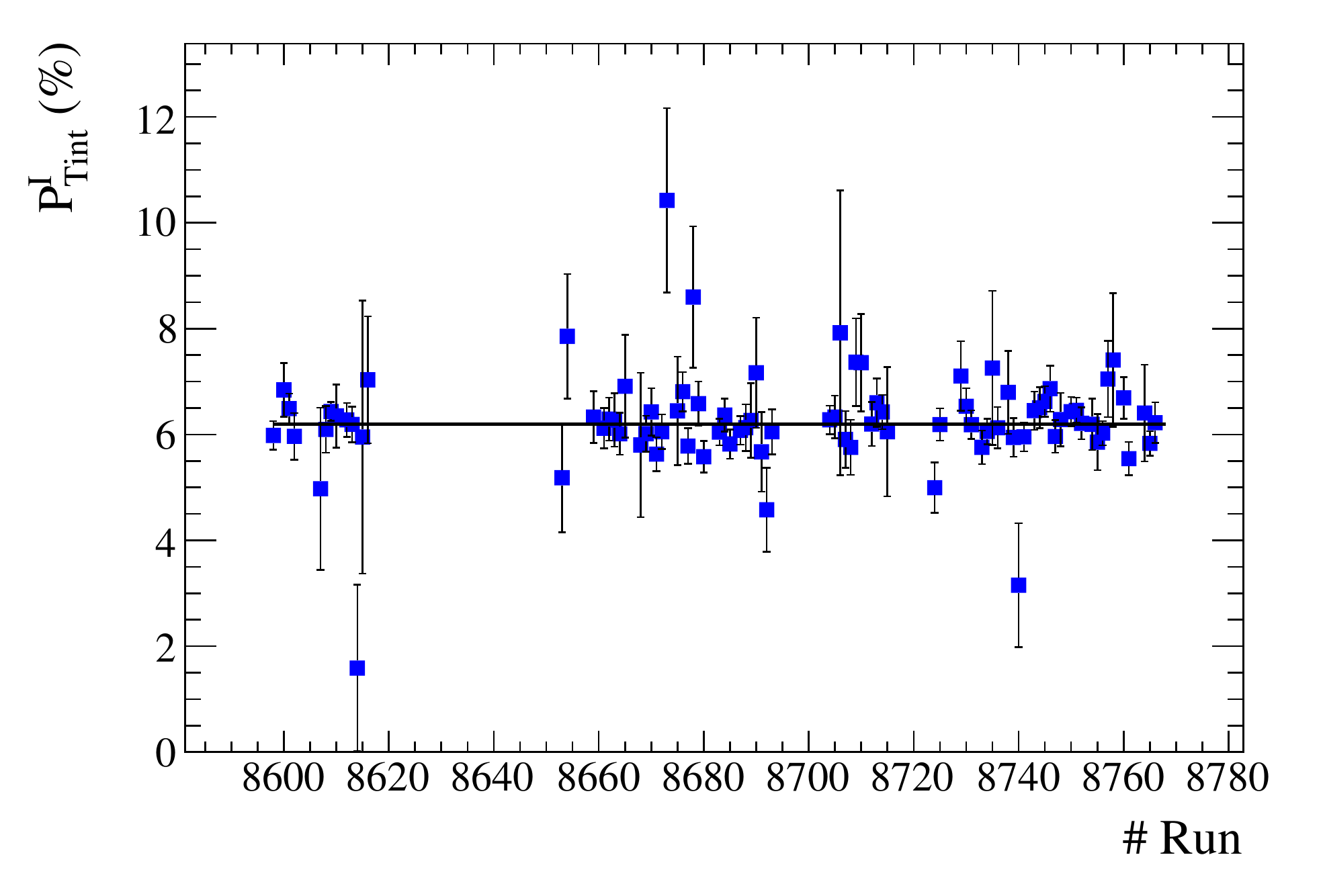}}
	\subfloat[]{

	 \includegraphics[width=.45\textwidth]{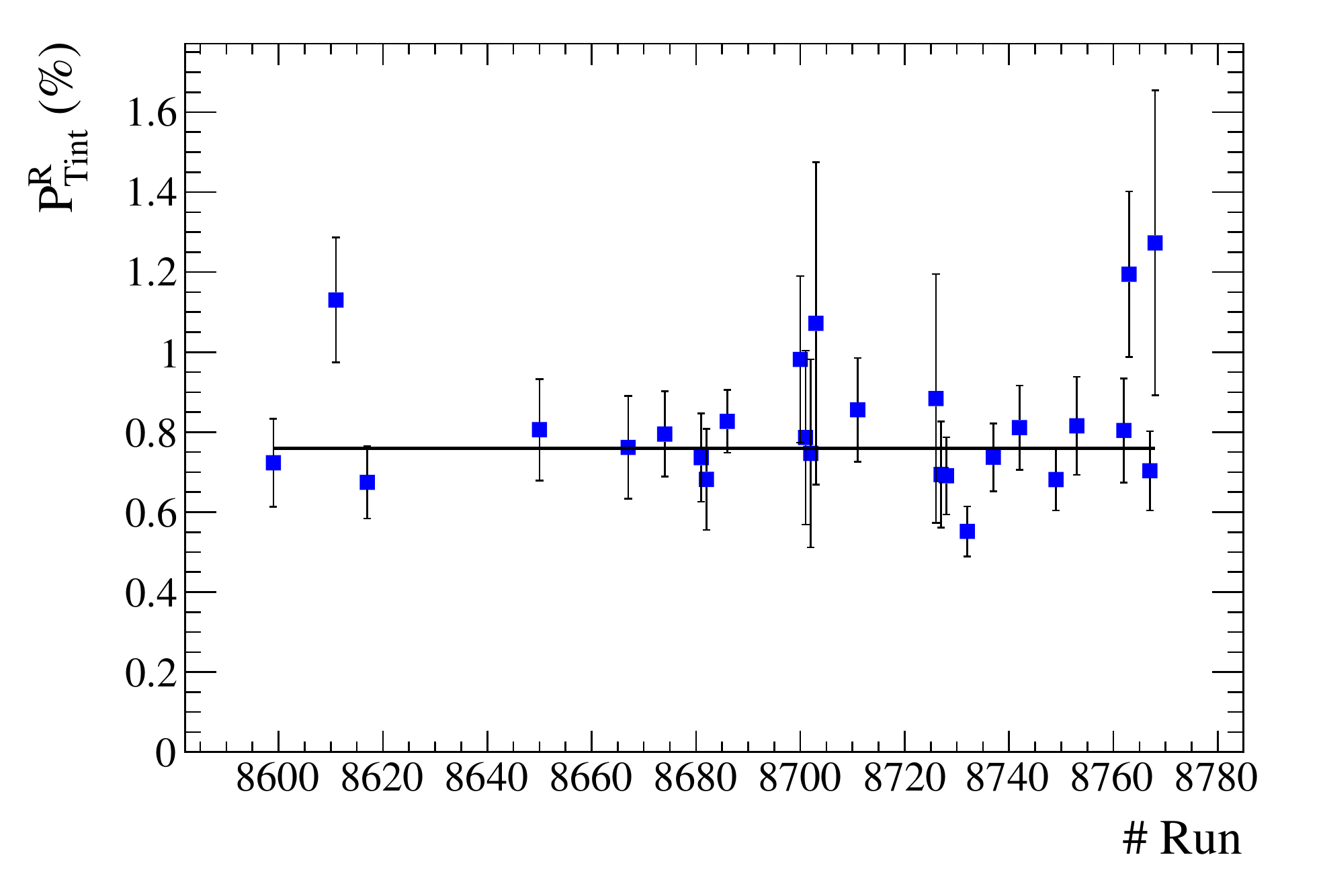}} \\
	\caption{
(Colour online)
The interaction trigger probability as a function of the run number 
for target inserted (a) and target removed (b) runs. 
The solid lines correspond to the measured mean values of 
the interaction trigger probabilities presented 
in Section~\ref{Sec:trigger_cross_section}.
Points far away from the measured values correspond 
to the runs with low number of events.
}
	\label{fig:TrigProbRun}
\end{figure*}

The interaction trigger probabilities for both the target inserted and target removed 
samples are time independent, 
as shown in Fig.~\ref{fig:TrigProbRun}.
The mean values of trigger probabilities were found to be
$P_\text{Tint}^\text{I} = (6.20 \pm 0.04) \%$ and 
$P_\text{Tint}^\text{R} = (0.76 \pm 0.02) \%$.
Insertion into the Eq.~\eqref{eq:pint} gives the interaction trigger probability
$P_\text{int} = (5.48 \pm 0.05) \%$. 
Finally, the corresponding trigger cross section is:

\begin{align}
\sigma_\text{trig} = 305.7 \pm 2.7 (\text{stat}) \pm 1.0 (\text{det}) \,\text{mb} ~,
\label{eq:trigxsec_meas}
\end{align}
where ``stat'' is the statistical uncertainty and ``det'' is the detector systematic uncertainty.
This measurement of $\sigma_\text{trig}$ is more precise than 
the result obtained with the 2007 data:
$298.1 \pm 1.9 (\text{stat}) \pm 7.3 (\text{det})$\,mb~\cite{pion_paper}. 
The detector systematic uncertainty of 2009 is significantly smaller. 
It is a consequence of the fact that in 2009 the beam triggers were recorded 
by DAQ simultaneously with physics triggers. 
Thus the same selection cuts (see Section~\ref{Sec:evtsel}) could be applied to all triggers.
In contrast, during the 2007 run the beam information was recorded by scalers which were not 
read out by the main DAQ.  
As a consequence, event-by-event quality selection was not possible~\cite{claudia}.
Instead, special runs were taken with the beam trigger to estimate biases. 
It had to be  assumed that the effect of the selection was stable over 
the whole data taking period. 
On the other hand, the statistical uncertainty on $\sigma_\text{trig}$ from
the 2009 data (see Eq.\,(\ref{eq:trigxsec_meas}))
is slightly larger because of the smaller number of recorded $T_\text{beam}$ triggers.

The fraction of out-of-target interaction background in the sample of 
the target inserted events is
\begin{align}
\epsilon = \frac{P_\text{Tint}^\text{R}}{P_\text{Tint}^\text{I}} = (12.3 \pm 0.4)\% .
\label{eq:frac_out}
\end{align}
The uncertainty of $\epsilon$ is larger for the 2009 than the 2007 data 
because of a larger statistical uncertainty of $P_\text{Tint}^\text{R}$.


\subsection{Results on inelastic and production cross section}
\label{Sec:inelprodxsec}

\begin{figure*}[t]
	\centering
	\subfloat[]{
	\includegraphics[width=.45\textwidth]{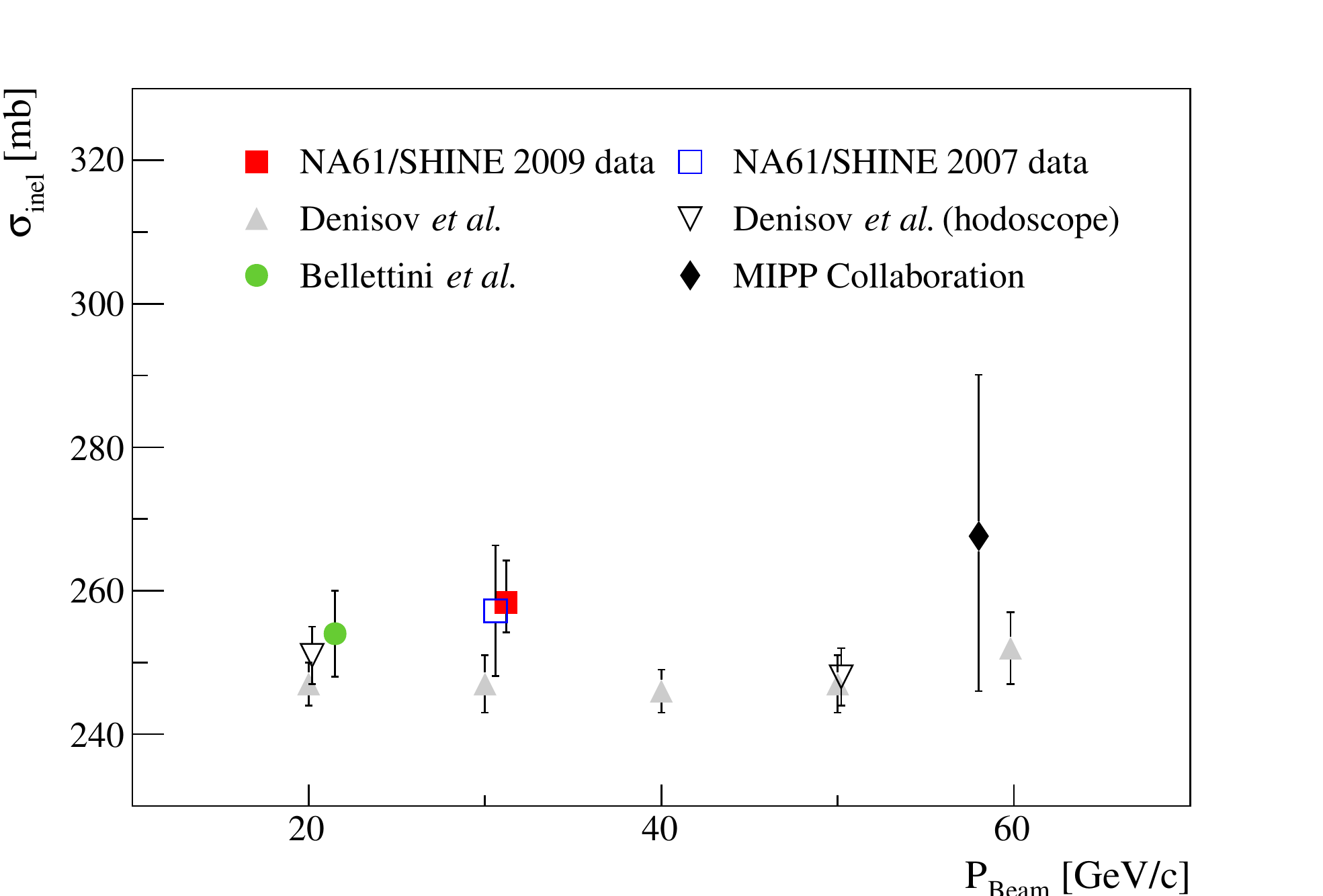}}  
	\subfloat[]{
	\includegraphics[width=.45\textwidth]{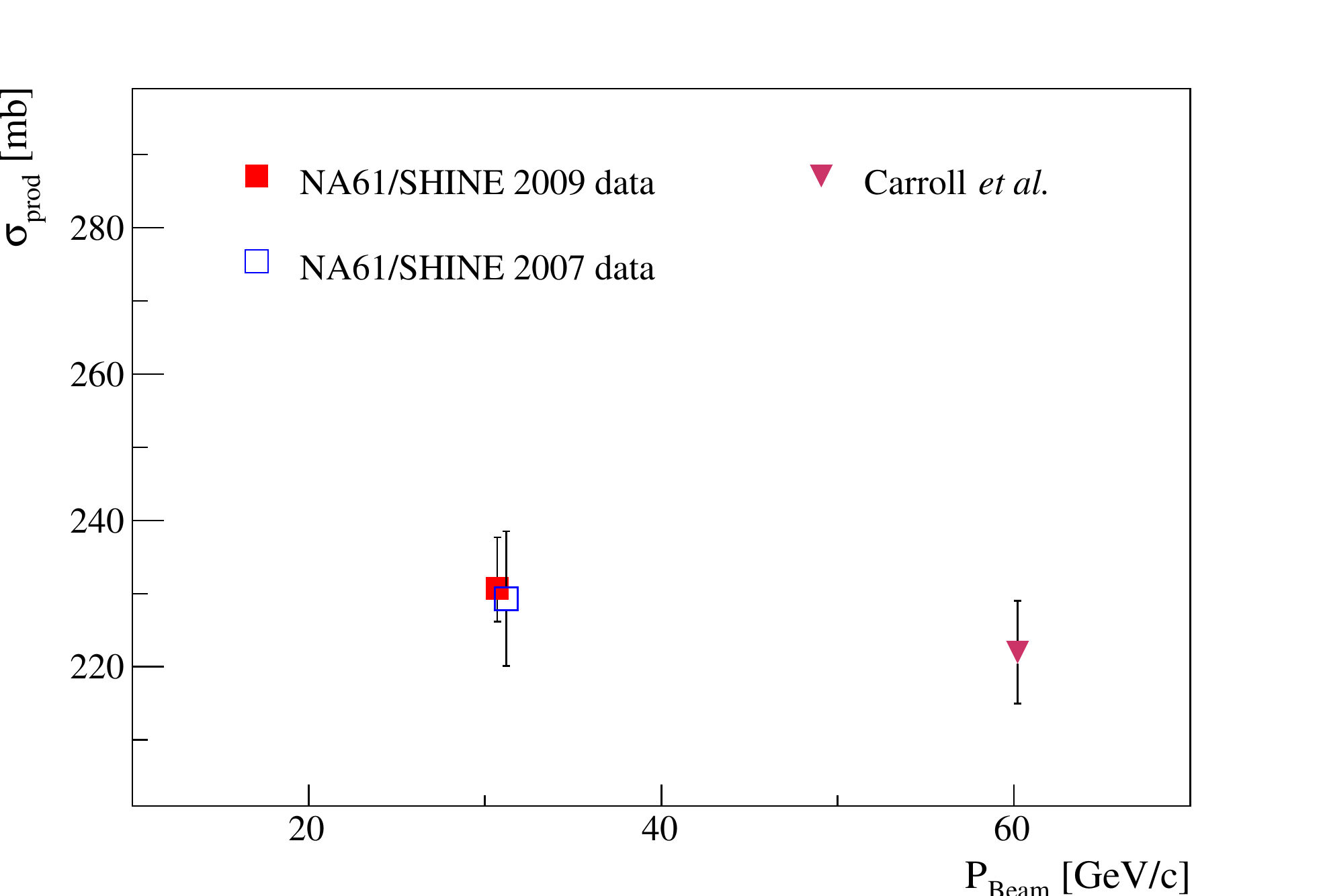}}  
	\caption{
(Colour online)
A comparison of the measured inelastic ($left$) and production ($right$)
cross sections at different momenta with previously published results.
Bellettini \emph{et al.}\ (green full circle)~\cite{Bellettini}, 
Denisov \emph{et al.}\ (grey full triangles)~\cite{Denisov}
and MIPP (black full diamond)~\cite{MIPP} measured 
the inelastic cross section
while Carroll \emph{et al.}\ (pink full inverted triangle)~\cite{Carroll} 
result corresponds to the production cross section. 
Inelastic cross section measurements performed by Denisov \emph{et al.}\  
with the hodoscope method are shown as well (open inverted triangles).
The \NASixtyOne measurements with 2007 (blue open square) 
and 2009 (red full square) data samples are shown. 
	}	
	\label{fig:XSecLiterature}
\end{figure*}

Using Eqs. (\ref{sigma_inel_abs}) and (\ref{sigma_prod_abs})
one can calculate the inelastic and production cross sections 
by representing them in the following way:
\begin{align}
&\sigma_\text{inel} = \left( \sigma_\text{trig} - \sigma^\text{f}_\text{el} \right) \frac{1}{f_\text{inel}} 
\label{sigma_inel_rel}
\\
&\sigma_\text{prod} = \left( \sigma_\text{trig} - \sigma^\text{f}_\text{el} - \sigma^\text{f}_\text{qe} \right) \frac{1}{f_\text{prod}} ~,
\label{sigma_prod_rel}
\\
&\sigma^\text{f}_\text{el} = \sigma_\text{el} f_\text{el} ~,
\\
&\sigma^\text{f}_\text{qe} = \sigma_\text{qe} f_\text{qe} ~,
\end{align}
where $f_\text{el}$, $f_\text{qe}$, $f_\text{inel}$ and $f_\text{prod}$ are the fractions of 
elastic, quasi-elastic, inelastic and production events, respectively, in which all charged
particles miss the S4 counter and which are therefore accepted as interactions by the $T_\text{int}$ trigger.
The values of $\sigma^\text{f}_\text{el}$ and $\sigma^\text{f}_\text{qe}$ are the contributions of elastic
and quasi-elastic interactions to $\sigma_\text{trig}$ which have to be subtracted to obtain $\sigma_\text{inel}$ 
and $\sigma_\text{prod}$.
The values of $f_\text{inel}$ and $f_\text{prod}$ depend upon the efficiency 
of $T_\text{int}$ for selecting inelastic and production events.
In order to take into account the correlations between $\sigma_\text{el}$, $f_\text{el}$
and $\sigma_\text{qel}$, $f_\text{qel}$, estimates of systematic uncertainties are based on those 
for $\sigma^\text{f}_\text{el}$ and $\sigma^\text{f}_\text{qe}$.

This method differs from the one used in the analysis of the 2007 data \cite{pion_paper,claudia}.
Here the simulation (see below) is basically only required to extract the magnitudes of the fractions $f$. 
For the absolute values of $\sigma_\text{inel}$ and $\sigma_\text{prod}$
one can use the results of experimental measurements, if available. 
In addition, in the approach used for the 2007 data the simulated values of 
inelastic and production cross sections
were part of the corrections of $\sigma_\text{inel}$ and $\sigma_\text{prod}$.
This is not the case for the method applied in the present analysis.

The corrections to $\sigma_\text{trig}$ as well as the corresponding uncertainties   
were estimated with GEANT4.9.5~\cite{GEANT4,GEANT4bis} using the FTF\_BIC 
physics list (see Section~\ref{Sec:models} 
for more detailed comparisons of the spectra measurements reported in this paper with the GEANT4 physics lists),
except for the elastic cross section, for which large uncertainties were found
in the GEANT4 simulation\footnote{For $\sigma_\text{el}$ 
significant differences were found between various releases of GEANT4 
-- 4.9.5, 4.9.6 and 4.10 -- for different physics lists. 
The obtained values of $\sigma_\text{el}$ were in the range from 78~mb to 88~mb.}. 
Since the total elastic cross section decreases
in good approximation linearly with  proton beam momentum
in the range 20-70~GeV/$c$,
the measurements performed by Bellettini \emph{et al.} at 21.5~GeV/$c$~\cite{Bellettini}
and Schiz \emph{et al.} at 70~GeV/$c$~\cite{Schiz}
were used to estimate the elastic cross section at 31~GeV/$c$ beam momentum.
The elastic cross section measured by Bellettini \emph{et al.} is  
\begin{align*}
\sigma_\text{el} (21.5~\text{GeV/}c) &= 81.00 \pm 5.00 \text{(sys)} \,\text{mb}
\end{align*}
Schiz \emph{et al.} reported the measured differential cross section $\frac{d\sigma}{dt}$ as a function 
of the momentum transfer $t$ with a parametrization.
The total elastic cross section can be obtained by integrating the differential 
cross section over the whole range of momentum transfer and is equal to:
\begin{align*}
\sigma_\text{el} (70~\text{GeV/}c) &= 76.6 \pm 6.9 \text{(sys)} \,\text{mb}
\end{align*}
The corresponding elastic cross section at the \NASixtyOne momentum 
was obtained by linear interpolation  between these two measurements:
\begin{align*}
\sigma_\text{el} (30.92~\text{GeV/}c) &= 80.1 \pm 5.4 \text{(sys)} \,\text{mb}
\end{align*}
The $\pm 1 \sigma$ range covers the interval $\left[74.8, 85.5 \right]$~mb.
The deviations from the extremes of the interval and the nominal value of $\sigma_{el}$
estimated with GEANT4 are taken into account as a model systematic uncertainty.

The values for the elastic and quasi-elastic cross section, estimated with GEANT4, are  
(see~Ref.~\cite{Laura_PhD} for more details):
\begin{align*}
\sigma^\text{f}_\text{el} &= 50.4 \, ^{+0.6}_{-0.5} \text{(det)} \,  ^{+4.9}_{-2.0} \text{(mod)} \,\text{mb}
\\
\sigma^\text{f}_\text{qe} &= 26.2 \, ^{+0.4}_{-0.3} \text{(det)} \,  ^{+3.9}_{-0.0} \text{(mod)} \,\text{mb} 
\end{align*}
The fractions estimated to be accepted by the interaction trigger are:
\begin{align*}
&f_\text{prod} = 0.993 \pm 0.000 \text{(det)} \, ^{+0.001}_{-0.012} \text{(mod)}
\\
&f_\text{inel} = 0.988 ^{+0.001}_{-0.008} \text{(det)} \, ^{+0.000}_{-0.008} \text{(mod)}
\end{align*}
where ``det'' is the detector systematic uncertainty obtained by
performing the simulation for different positions and sizes of S4,
taking also into account the beam divergence measured from the data.
The  uncertainty  ``mod'' resulting from the choice of  physics model
was calculated as the largest difference between the contributions estimated 
for $\sigma^\text{f}_\text{qe}$ with different GEANT4 physics models
(FTFP\_BERT, QBBC and QGSP\_BERT, as well as FTF\_BIC physics list)
and from measured data for $\sigma^\text{f}_\text{el}$ as described above.

Inserting these values of the elastic and quasi-elastic cross sections, 
and of the fractions accepted by the trigger into
Eqs.~\eqref{sigma_inel_rel} and \eqref{sigma_prod_rel}, 
one obtains the final results:
\begin{align}
&\sigma_\text{inel} = 258.4 \pm 2.8 \text{(stat)} \, \pm 1.2 \text{(det)} \, 
^{+5.0}_{-2.9} \text{(mod)} \,\text{mb}~, 
\label{eq:sigma_inel_res}
\\
&\sigma_\text{prod} = 230.7 \pm 2.8 \text{(stat)} \, \pm 1.2 \text{(det)} \, 
^{+6.3}_{-3.5} \text{(mod)} \,\text{mb}~, 
\label{eq:sigma_prod_res}
\end{align}
where ``stat'' is the statistical uncertainty, ``det'' is the total detector systematic uncertainty 
and ``mod'' is the uncertainty caused by the choice of physics model.
The total uncertainty of $\sigma_\text{prod}$ is $^{+7.0}_{-4.6}$\,mb, 
which is significantly
smaller than that of the \NASixtyOne result obtained from the 2007 data. 
The dominant uncertainty comes from the choice of physics model used to derive 
the production cross section from the 
trigger cross section.

The new \NASixtyOne results on inelastic and production cross section 
agree, in general,  
with the previously published measurements as shown in Fig.~\ref{fig:XSecLiterature}.
A possible tension with measurements by Denisov \emph{et al.}~\cite{Denisov}, 
which are assigned a rather small systematic uncertainty of 1\%,
could be due to different experimental techniques used to extract 
$\sigma_\text{inel}$. 
As discussed in Ref.~\cite{Bobchenko}, various approaches 
to define and measure $\sigma_\text{inel}$ could lead
to differences of up to 8~mb for proton-carbon interactions.

\section{Spectra analysis techniques and uncertainties}
\label{Sec:ana}

This section presents analysis techniques developed for the measurements
of the differential inclusive spectra of hadrons.
Details are shown on data selection and binning, 
on particle identification (PID) methods as well as on the 
calculation of correction factors and the estimation of
systematic uncertainties.

The data analysis procedure consists of the following steps:
\begin{enumerate}[(i)]
  \item application of event and track selection criteria,
  \item determination of spectra of hadrons
        using the selected events and tracks,
  \item evaluation of corrections to the spectra based on
        experimental data and simulations,
  \item calculation of the corrected spectra.
\end{enumerate}  

Corrections for the following biases were evaluated and applied:
\begin{enumerate}[(i)]
 \item geometrical acceptance, 
 \item reconstruction efficiency,
 \item contribution of off-target interactions,
 \item contribution of other (misidentified) particles,
 \item feed-down from decays of neutral strange particles,
 \item analysis-specific effects (e.g. ToF-F efficiency, PID, 
   $K^-$ and $\bar{p}$ contamination, etc.).
\end{enumerate}

All these steps are described in the following subsections
for each of the employed identification technique separately.

 The \NASixtyOne measurements refer to 
 hadrons
 (denoted as \textit{primary 
 hadrons})
 produced in p+C interactions at 31~GeV/\textit{c}
 and in the electromagnetic decays of
 produced hadrons (e.g. $\Lambda$ from $\Sigma^0$ decay). Contributions from
 products of weak decays and secondary interactions are corrected for.

\subsection{Event and track selection}\label{sec:track_sel}

Events recorded with the ``interaction'' ($T_\text{int}$) trigger
were required to have a well-reconstructed incoming beam trajectory 
(the {\it standard} BPD selection).

Several criteria were applied to select well-measured tracks 
in the TPCs in order to ensure high reconstruction efficiency 
as well as to reduce the contamination of tracks from secondary interactions:
\begin{enumerate}[(i)]
\setlength{\itemsep}{1pt}
\item track momentum fit at the
interaction vertex should have converged,
\item the total number of reconstructed points on the track
should be greater than $N_\text{point}$, 
\item at least $N_\text{field}$ reconstructed points in the three TPCs 
  were used for momentum measurement 
  (\mbox{VTPC-1}, \\ 
  \mbox{VTPC-2} and GTPC), 
\item distance of closest approach of the fitted track to the
  interaction point (impact parameter) smaller than
  $D_x$ ($D_y$) in the horizontal (vertical) plane. 
\end{enumerate}
A summary of cut values used with the different identification techniques
described in the following sections,
is given in Table~\ref{tab_cuts}.

The adopted \pth binning scheme is chosen based on the available statistics
and the kinematic phase-space of interest for T2K.
The  highest $\theta$ limit
is analysis-dependent.
The polar angular region down to $\theta=0$ is covered.


\begin{table}
\centering
  \begin{tabular}{|c|c|c|c|c|}
    \hline
    Identification method & $N_\text{point}$ & $N_\text{field}$ & $D_x$  & $D_y$  \\ 
    \hline
    \hline
    $V^0$ & 20 & -- & 4~cm & 2~cm \\
    \hline
    $tof$-$dE/dx$ & 30 & 6 & 4~cm & 4~cm \\
    \hline
    $dE/dx$ & 30 & 12 & 4~cm & 4~cm \\
    \hline
    $h^-$ & 30 & 12 & 4~cm & 4~cm \\
    \hline
  \end{tabular} 
  \caption{A summary of cuts used with different identification  techniques.
  }
  \label{tab_cuts}
\end{table}

\subsection{Derivation of spectra}
\label{Sec:norma}


The raw number of hadron candidates has to be corrected for various effects 
such as the loss of particles due to selection cuts, reconstruction inefficiencies 
and acceptance.
In each of the \pth bins, the correction factor was computed
from a GEANT3 based detector simulation using the \VenusLong model as 
primary event generator and applying the same event and track selections as
for the data (for description see Ref.~\cite{pion_paper}):
\begin{equation}
\label{eq:corr_factor}
C_h(p, \theta) = 
\left(\frac{\Delta n_h^\text{rec,fit}}{\Delta n_h^\text{sim,gen}}\right)_\text{MC} / 
 \left(\frac{N^\text{acc}}{N^\text{gen}}\right)_\text{MC} ~,
\end{equation}
The numerator corrects for the loss of candidates of particle type $h$, 
where $\Delta n_h^\text{sim,gen}$ is the number of true $h$ particles generated in 
a specific \pth bin and $\Delta n_h^\text{rec,fit}$ is the number of $h$
candidates extracted from the reconstructed tracks in the simulation. 
The denominator accounts for 
the events lost due to the trigger bias with $N^\text{gen}$ 
the number of generated and $N^\text{acc}$ 
the number of accepted  inelastic events. 
The corrected number of hadron candidates $\Delta n_h^\text{corr}$ was then obtained
from the raw number $\Delta n_h^\text{raw}$ by:
\begin{equation}
\label{eq:corr_nb}
\Delta n_h^\text{corr}(p, \theta) = \frac{\Delta n_h^\text{raw}(p, 
\theta)}{C_h(p, \theta)} ~.
\end{equation}

The  procedures presented in 
Sections~\ref{Sec:v0} --
\ref{Sec:h-} below
were used to analyze events with the carbon target inserted (I) as well as 
with the carbon target removed (R).
The corresponding
corrected numbers of particles in \pth bins
are denoted as
$\Delta n^\text{I}_h$ and
$\Delta n^\text{R}_h$, where $h$ stands for the particle type
(e.g.\ $\pi^-$).
Note that the same event and track selection criteria as well as the
same corrections discussed in 
the previous sections
were used in the analysis of events
with the target inserted and removed.
The latter events allow to correct the measurements for the contribution of
out-of-target interactions.

The double differential inclusive
cross section was calculated from the formula:
\begin{equation}
    \label{eq:xsecmeas3}
\frac {\dd^2 \sigma_h}{\dd p \dd\theta} = \frac {\sigma_\text{trig}} {1-\epsilon}
\left( \frac {1} {N^\text{I}}  \frac { \Delta n^\text{I}_h } { \Delta p \Delta \theta} -
\frac {\epsilon} {N^\text{R}}  \frac { \Delta n^\text{R}_h } { \Delta p \Delta \theta}\right)~,
\end{equation}
where
\begin{enumerate}[(i)]
\setlength{\itemsep}{1pt}

  \item $\sigma_\text{trig} = (305.7 \pm 2.7 \pm 1.0)$\,mb is
  the ``trigger'' cross section
  as given in Eq.~(\ref{eq:trigxsec_meas}) of Section~\ref{Sec:trigger_cross_section},

  \item $N^\text{I}$ and $N^\text{R}$ are the numbers of events 
    with the target inserted and removed, respectively,
    selected for the analysis
    (see Subsection~\ref{sec:track_sel}),
    
  \item $\Delta p$ ($\Delta \theta$) is the bin size in momentum (polar angle), and

  \item $\epsilon = 0.123 \pm 0.004$ is the ratio of the interaction probabilities 
  for operation with the target removed and inserted as given in 
  Eq.~(\ref{eq:frac_out}).

\end{enumerate}

The overall uncertainty on the inclusive cross section due to the normalization 
procedure amounts to 1\%. 

The particle spectra normalized to the mean particle multiplicity in
production interactions was calculated as
\begin{equation}
    \label{eq:xsecprod}
\frac {\dd^2 n_{h}}{\dd p \dd\theta} = \frac{1}{\sigma_\text{prod}} \,
\frac {\dd^2 \sigma_h}{\dd p \dd\theta}~,
\end{equation}
where $\sigma_\text{prod}$ is the cross section for production processes given in 
Eq.~(\ref{eq:sigma_prod_res}).

The normalization uncertainty on the multiplicity spectra is $^{+2.8\%}_{-1.6\%}$.

The statistical uncertainty of the measured spectra receives contributions from the finite
statistics of both the data and the simulated events
used to obtain the correction factors.
The dominant contribution is the uncertainty of the data which was 
calculated assuming a Poisson probability distribution for the number of entries
in a \pth bin. The simulation  statistics was about ten times higher than the data 
statistics, thus the uncertainty of the corrections is
neglected
for all bins.

The systematic uncertainty of the measured spectra was calculated taking into
account various contributions discussed in detail in the following sections
separately for each identification technique.

\subsection{$V^0$ analysis}
\label{Sec:v0}

Understanding the neutral strange particle  $V^0$ 
(here $V^0$ stands for $K^0_S$ and $\Lambda$) production in 
p+C interactions at 31\,GeV/$c$ is of  interest for T2K for two reasons. 
First, it allows to decrease the systematic uncertainties 
on measurements of charged  
pions and protons, since a data-based feed-down correction can be used. 
Second, measurements of $K^0_S$ production  improve 
the knowledge of the $\nu_e$ and $\bar \nu_e$ fluxes  
coming from the three-body  
$K^0_L \to \pi^0 e^\pm \nu_e (\bar\nu_e)$ decays.\\
The following $V^0$ decay channels were studied:

\begin{align}
K^0_S &\to \pi^+ + \pi^- \quad \text{B.R. = (69.20 $\pm$ 0.05)\%}~,\\
\Lambda &\to p \phantom{^+} + \pi^- \quad \text{B.R. = (63.9 $\pm$ 0.5)\%}~.
\end{align}

Fits to the invariant mass distributions of $V^0$ 
candidates were used to extract measured numbers of $K^0_S$
and $\Lambda$ decays in momentum and polar angle bins.
These numbers were corrected for acceptance and other experimental biases using
simulated events.

\subsubsection{Event and track selection for the $V^0$ analysis}
\label{Subsec:v0_sel}

\begin{figure*}
\centering
\includegraphics[width = 0.3\textwidth]{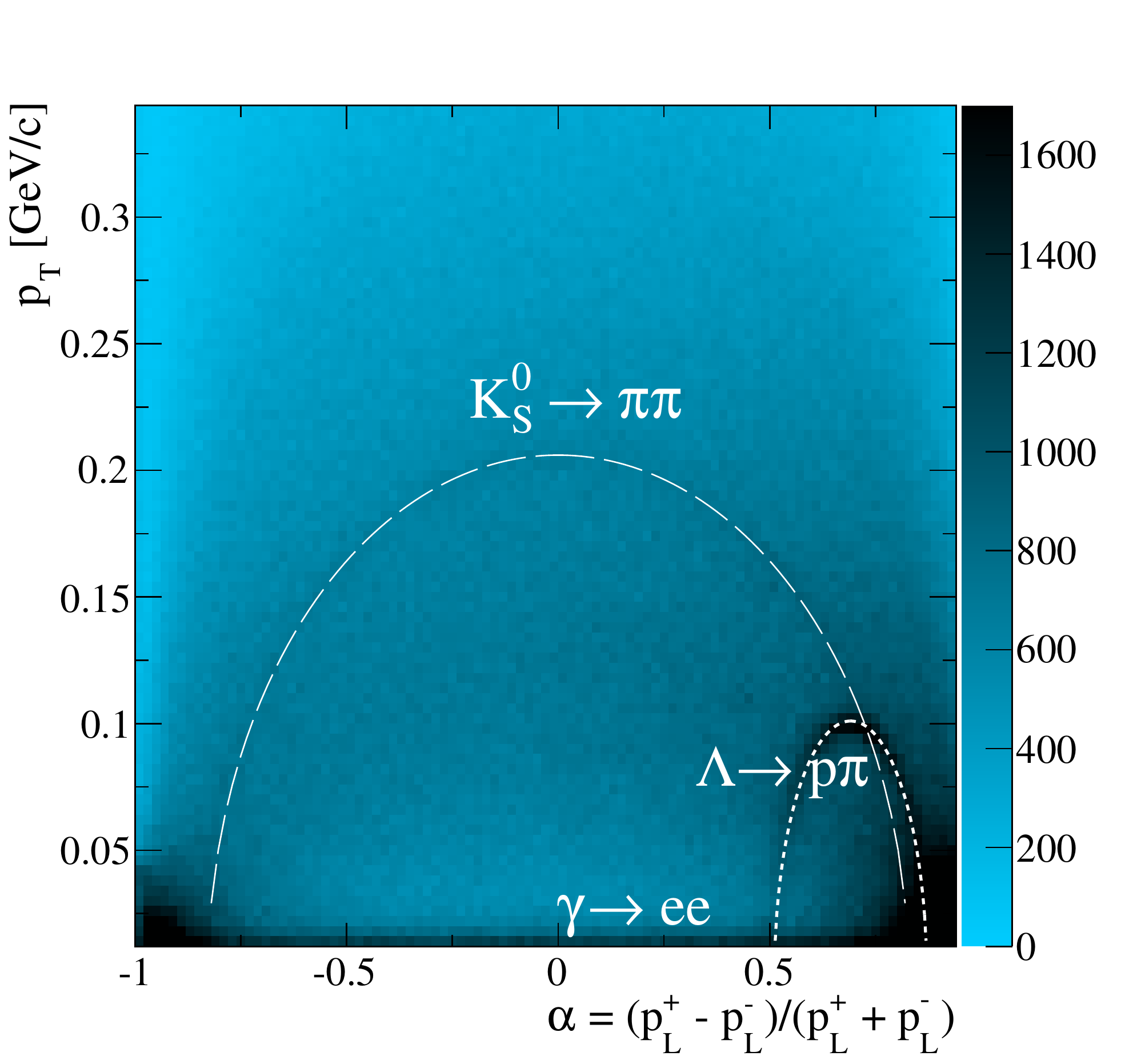}
\includegraphics[width = 0.3\textwidth]{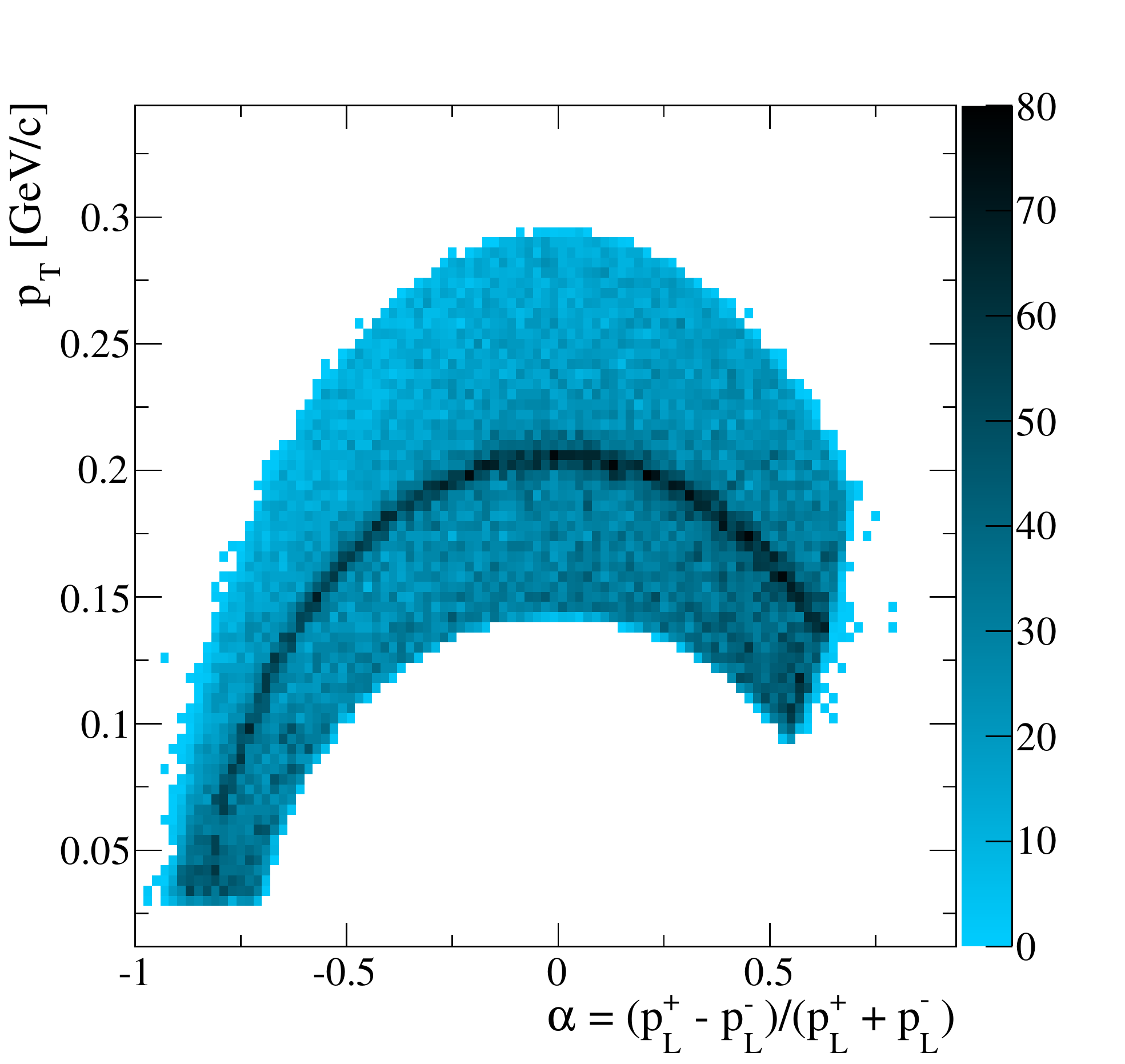}
\includegraphics[width = 0.3\textwidth]{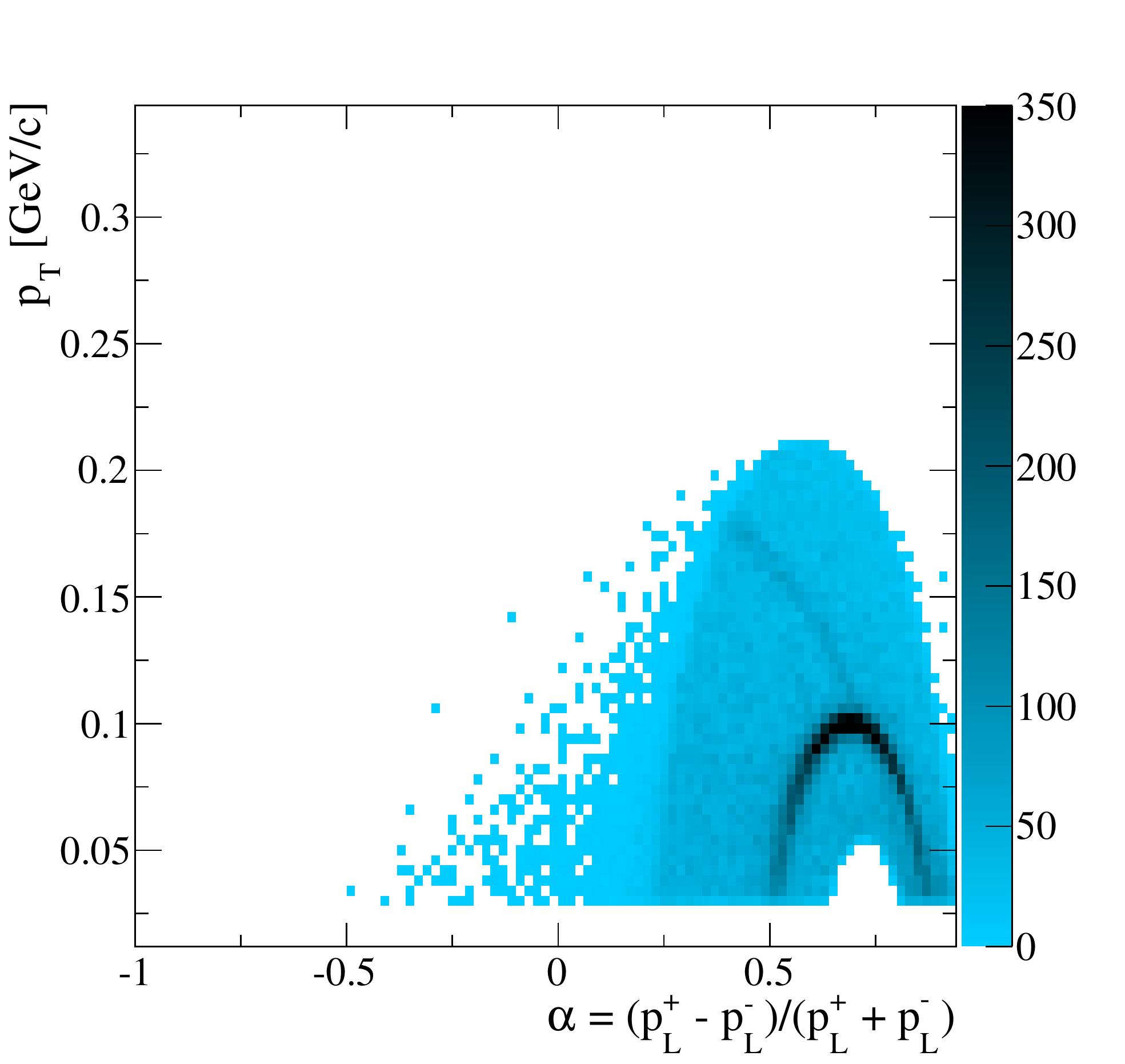}

\caption{\label{fig:v0_pa}
(Colour online)
Distributions of the $V^0$ candidates in the Podolanski-Armenteros 
variables~\cite{PAplots} after 
the event and topological cuts ($left$) and after the additional kinematic cuts 
($K^0_S$: $middle$, $\Lambda$: $right$). 
Ellipses showing the expected positions of $K^0_S$ and $\Lambda$ are also drawn.}
\end{figure*}

The track and $V^0$ candidate selections can be separated 
into three categories: 
event and track quality selections, 
topological selections aimed at finding $V^0$-type candidates, 
and, finally, kinematic selections to
separate $K^0_S$ and $\Lambda$ candidates. 

The standard quality selections for events and tracks were applied
(see  Section~\ref{sec:track_sel}). 

The $V^0$ topological criteria require that a fitted secondary vertex, 
located downstream of the interaction, is built out of two tracks with opposite 
electric charges.
Moreover, the distance of closest approach between the daughter tracks 
and the secondary vertex had to be smaller than 0.5\,cm.
The same quality and topological criteria were applied for 
selecting $\Lambda$ and $K^0_S$ candidates.

In order to extract the $K^0_S$ candidates, 
the following kinematic cuts were applied to the selected $V^0$  
candidates:
\begin{enumerate}[(i)]
\setlength{\itemsep}{1pt}
\item The transverse momentum  of the daughter tracks 
      relative to the $V^0$ momentum
      must be greater than 
      0.03\,GeV/$c$ 
      in order to remove converted photons,
\item The cosine of the angle $\theta^*$ between the momentum of the $V^0$ candidate 
      and the momentum of the daughter in 
      the center of mass  must be smaller than  0.76. 
      This cut allows the rejection of  most of the $\Lambda$ candidates for
       which the distribution of $\cos\theta^*$ computed under the $K^0_S$ hypothesis 
      is concentrated in the region $\cos\theta^* >$~0.8,
\item The candidates must have an invariant mass for the $K^0_S$ hypothesis within 
      the range of $[0.4,0.65]$\,GeV/$c^2$, 
\item The reconstructed proper decay length should be greater than 
      a quarter of the mean proper
      decay length~\cite{PDG} of $K^0_S$ mesons ($c\tau > 0.67$\,cm). 
\end{enumerate}

The kinematic cuts used to extract the $\Lambda$ candidates are the following:
\begin{enumerate}[(i)]
\setlength{\itemsep}{1pt}
\item Transverse momentum of the daughter tracks must be greater than 
0.03\,GeV/$c$,
\item The candidate must have an invariant mass for the $\Lambda$ hypothesis within the range of $[1.09,1.215]$\,GeV/$c^2$, 
\item The reconstructed proper decay length should be greater than 
      a quarter of the mean proper
      decay length~\cite{PDG} of $\Lambda$ hyperons ($c\tau > 1.97$\,cm).
\end{enumerate}

Figure~\ref{fig:v0_pa} shows the Podolanski-Armenteros~\cite{PAplots}
plots once the event and topological selections, as well as 
all $K^0_S$ and $\Lambda$ kinematic cuts have been applied. One can see that this set of cuts 
allows an efficient selection of  the desired $V^0$ candidates. 

\begin{figure*}
\centering
\begin{overpic}
[width = 0.48\textwidth]{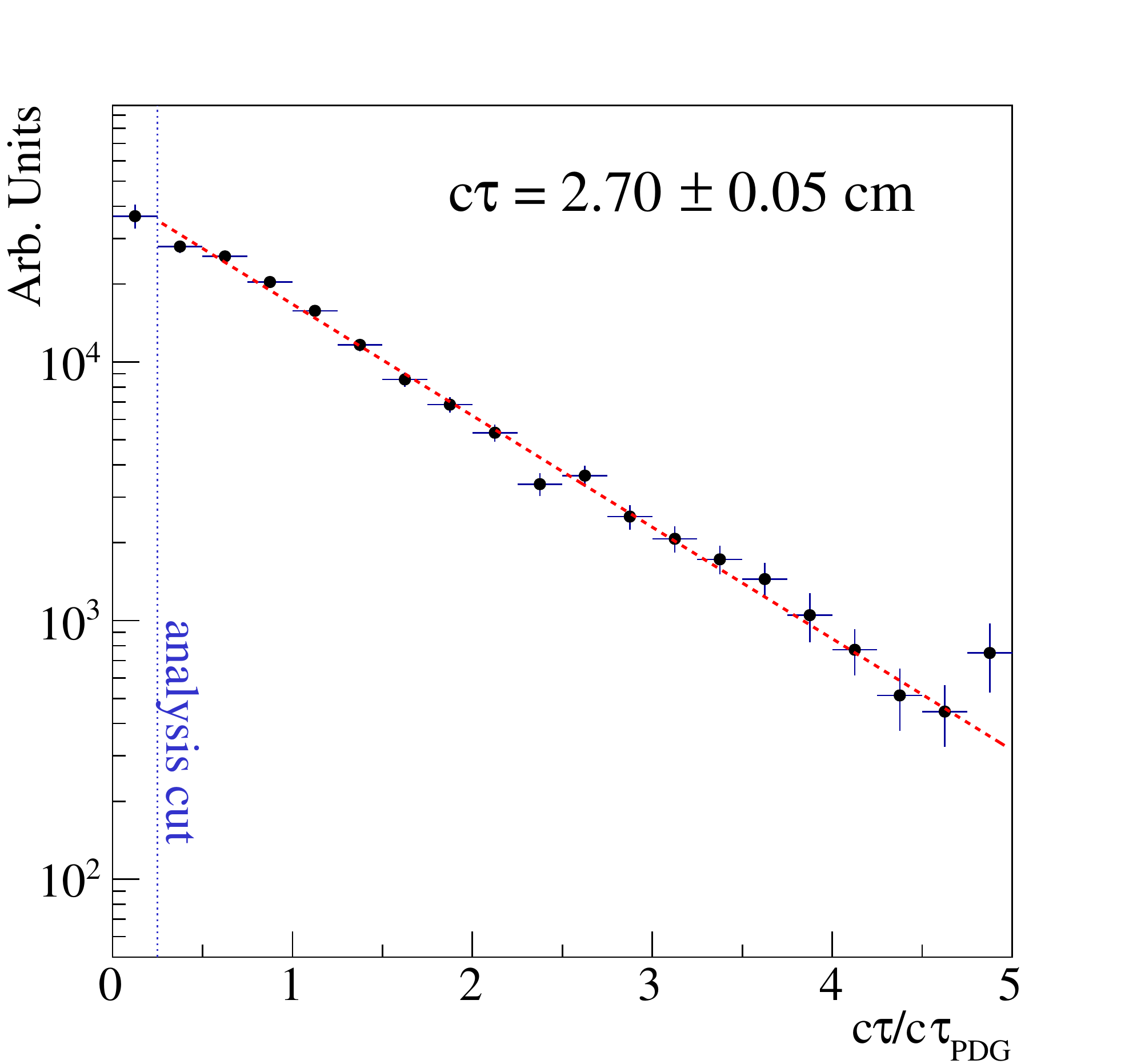}
\put(150, 150){\Large{$K^0_S$}}
\put(95.5, 188){\line(1,0){10}}
\end{overpic}
\begin{overpic}
[width = 0.48\textwidth]{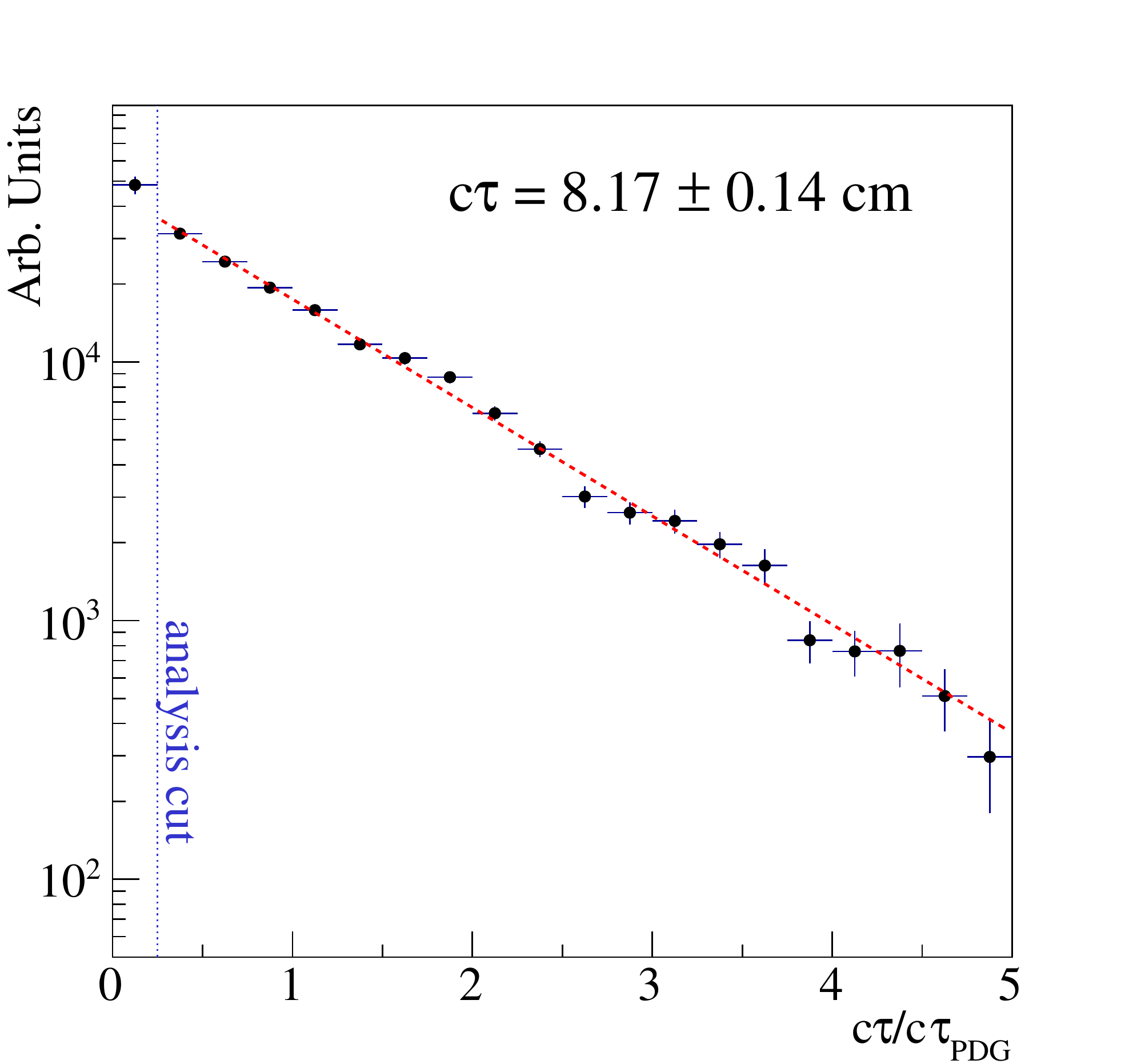}
\put(150, 150){\Large{$\Lambda$}}
\put(95.5, 188){\line(1,0){10}}

\end{overpic}
\caption{\label{fig:lifetime}
(Colour online)
The measured proper decay length ($c\tau$) distributions for 
$K^0_S$ ($left$) and $\Lambda$ ($right$).}
\end{figure*}

The measured proper decay length distributions corrected for all  
experimental biases\footnote{The numbers of $K^0_S$ and $\Lambda$ as well as the correction factores
were obtained in bins of proper decay length.} are shown in Fig.~\ref{fig:lifetime} 
for both $K^0_S$ and $\Lambda$. 
The fitted mean proper decay lengths are in reasonable agreement with 
the PDG values~\cite{PDG}. 

\begin{figure*}
\centering
\includegraphics[width = 0.48\textwidth]{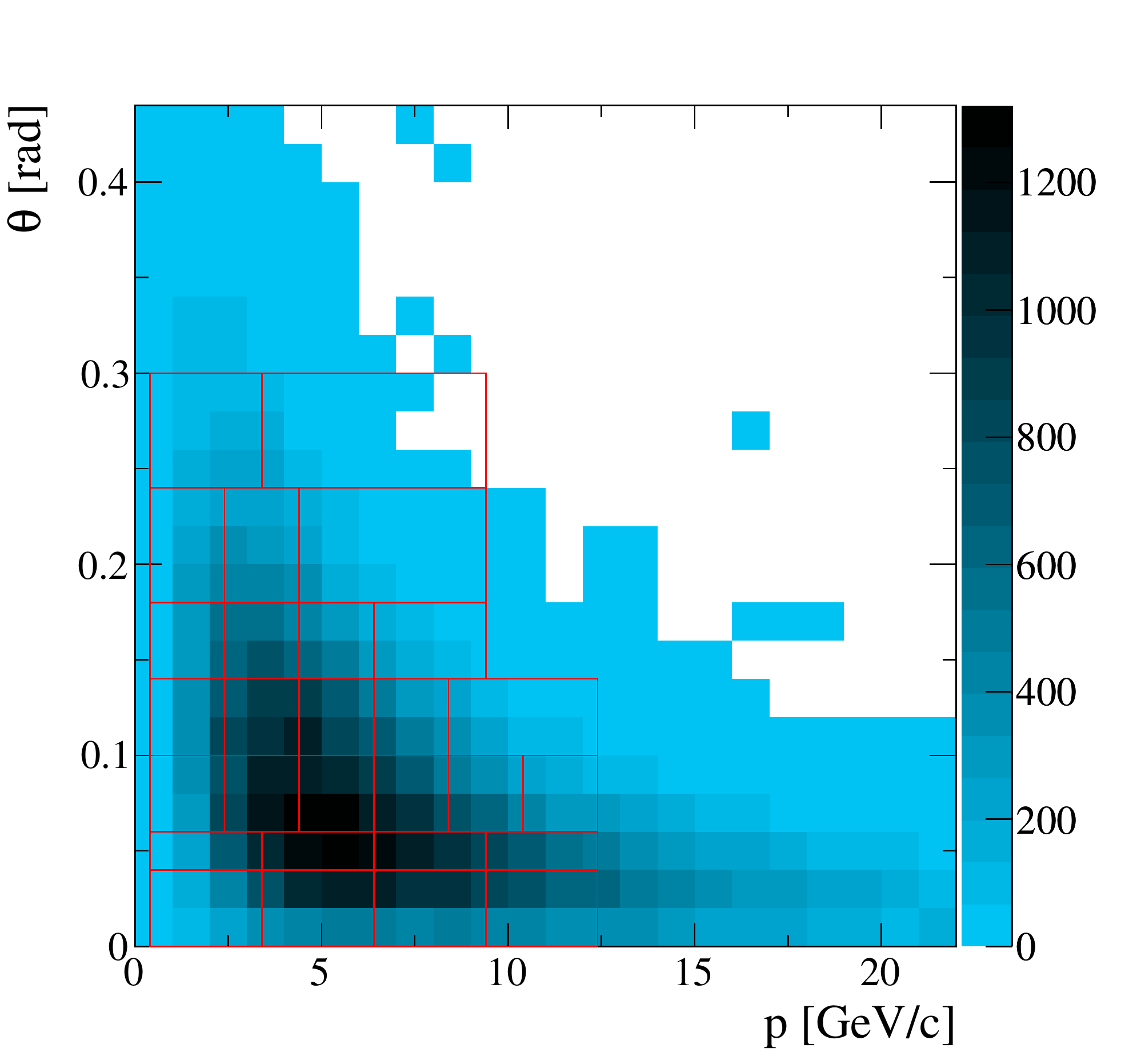}
\begin{overpic}
[width = 0.48\textwidth]{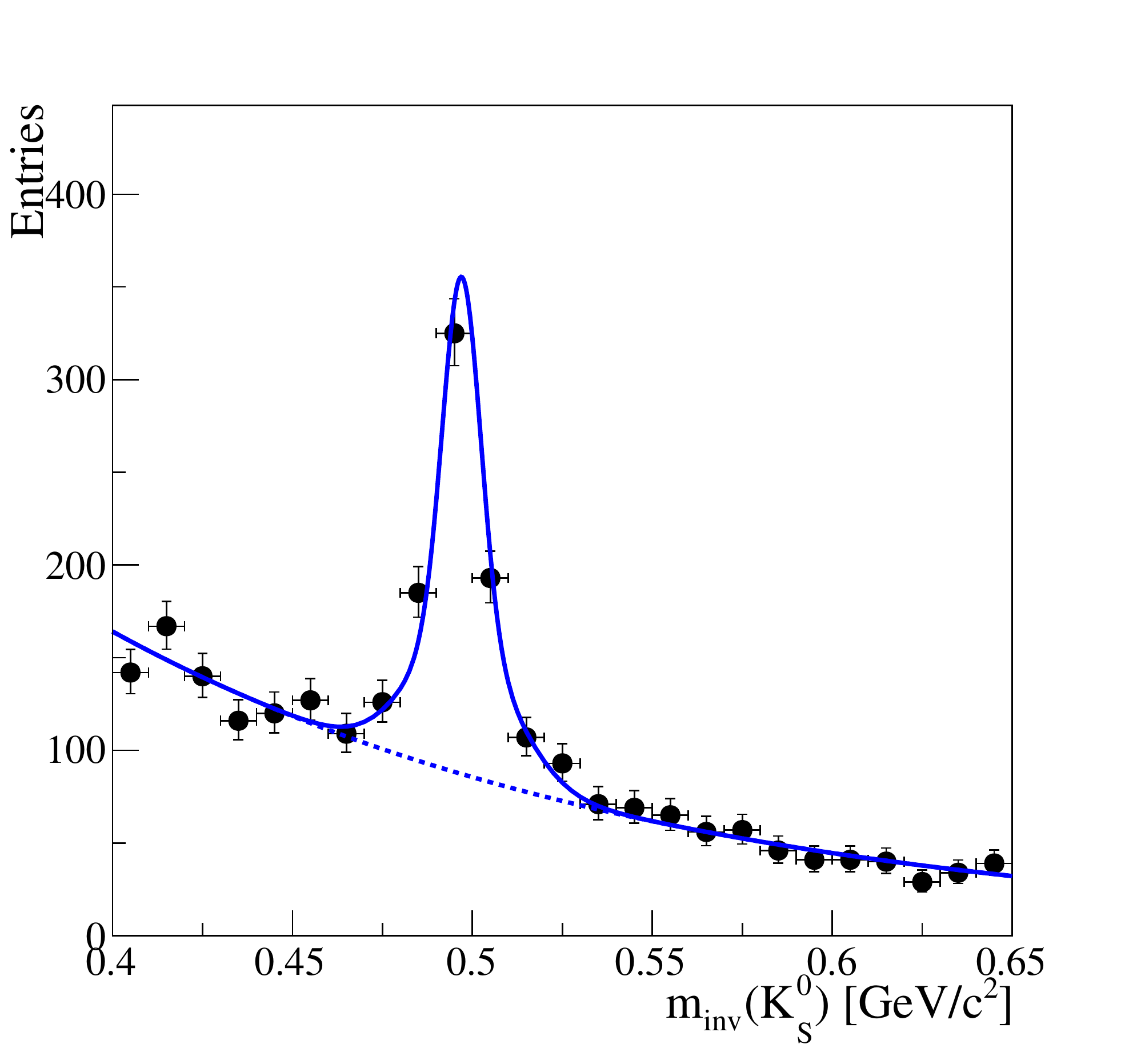}
\put(138, 180){$0.14 < \theta[\text{rad}] < 0.18$}
\put(135, 160){$2.4 < p[\text{GeV}/c] < 4.4$}
\end{overpic}

\caption{\label{fig:k0_ps_fit}
(Colour online)
$Left$: The \pth phase space of 
interest for T2K is shown by the shaded area.
The binning used for the $K^0_S$ analysis is indicated by the outlined boxes. 
$Right$: An example of the $K^0_S$ invariant mass distribution 
in a selected \pth bin with the fit result overlaid.}
\end{figure*}

\begin{figure*}
\centering
\includegraphics[width = 0.48\textwidth]{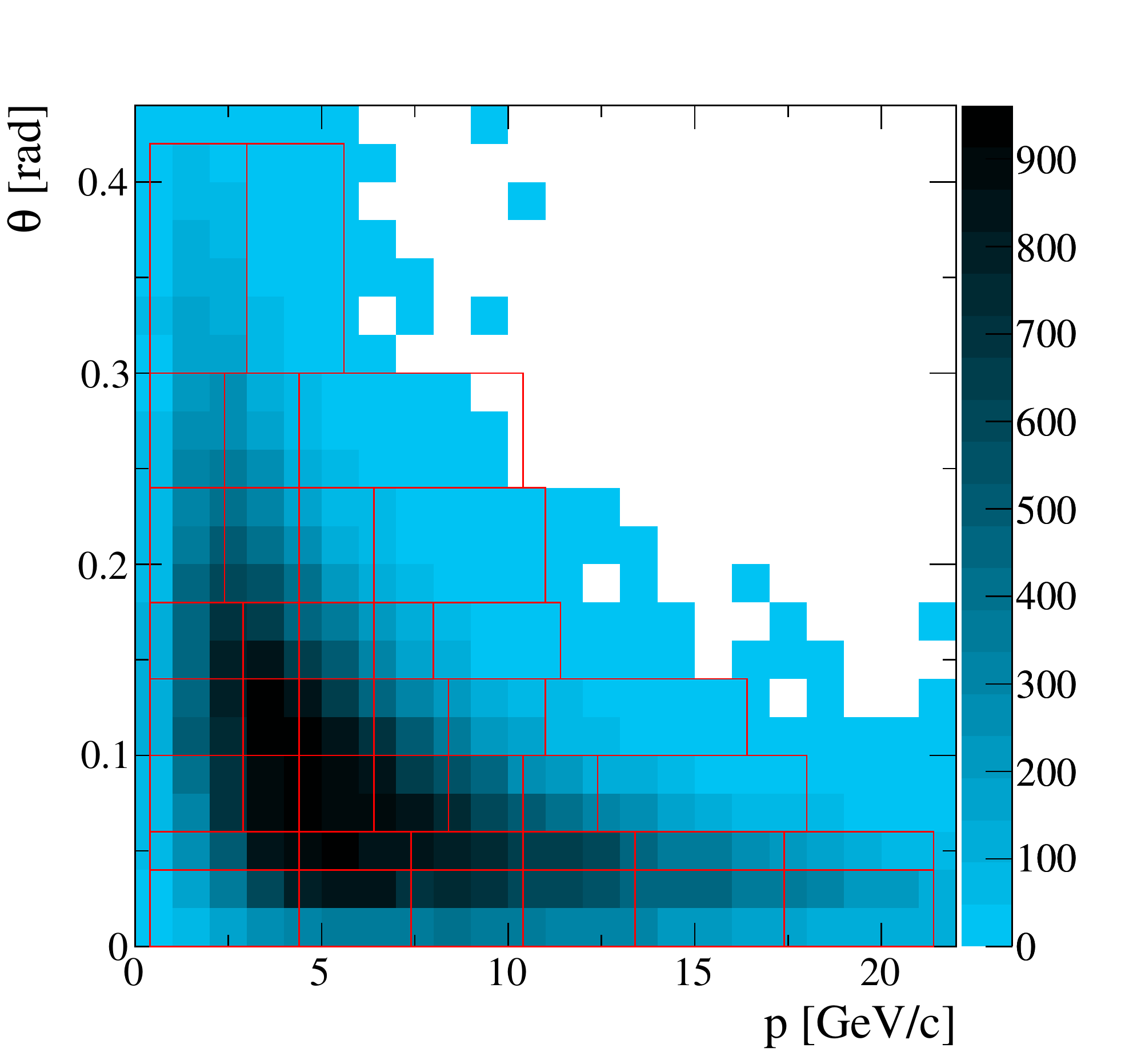}
\begin{overpic}
[width = 0.48\textwidth]{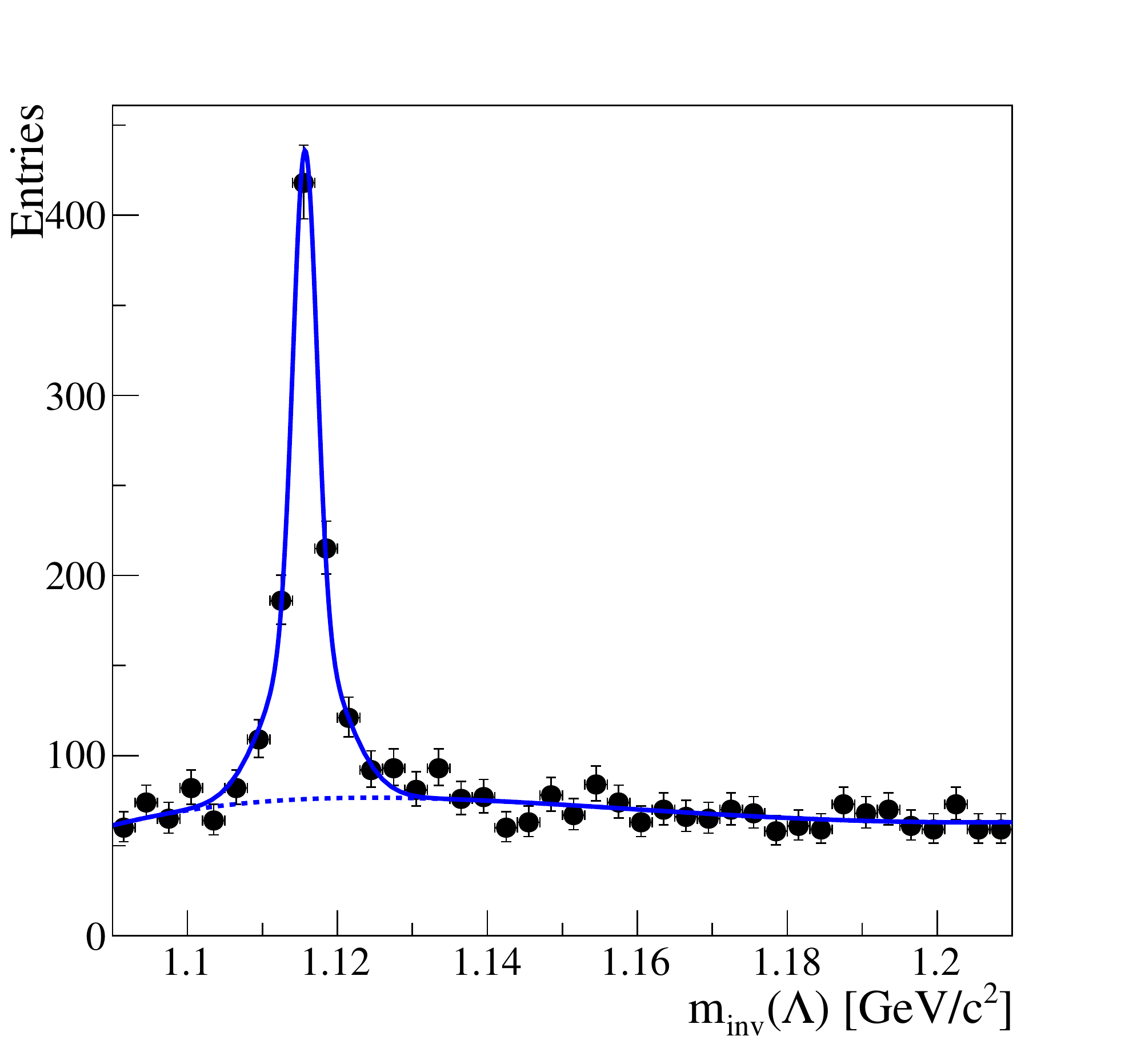}
\put(138, 180){$0.06 < \theta[\text{rad}] < 0.10$}
\put(135, 160){$4.4 < p[\text{GeV}/c] < 6.4$}
\end{overpic}

\caption{\label{fig:lambda_ps_fit}
(Colour online)
$Left$: The \pth phase space of 
interest for T2K is shown by the shaded area.
The binning used for the $\Lambda$ analysis is indicated by the outlined boxes. 
$Right$: An example of the $\Lambda$ invariant mass distribution 
in a selected \pth bin with the fit result overlaid.}
\end{figure*}

\subsubsection{Binning, fitting, corrections}
\label{Subsec:v0_fit}

The selected $V^0$ candidates are binned in \pth phase space. 
Due to detector acceptance and reconstruction efficiency, the momentum range 
of reconstructed charged particles starts from 0.4\,GeV/$c$.
For the $K^0_S$ analysis, 28 bins are used, 
whereas the $\Lambda$ candidates are divided into 39 bins.
The choice of the binning scheme is driven by the available statistics. 

A fit of the invariant mass distribution was performed in each of these \pth bins. 
The shape of the $K^0_S$ and $\Lambda$ signal was parametrised by
a sum of two Gaussians
\begin{equation}
L_{sig}(m) = f_{sig}G_W(m; \sigma_W, m_{PDG}) + (1-f_{sig})G_N(m; \sigma_N, m_{PDG})
\end{equation}
where $m$ stands for the reconstructed invariant mass and subscripts 
$W$ and $N$ refer to the wide and narrow Gaussian, respectively. 
The parameter $f_{sig}$ describes the fraction contributed by each Gaussian.
In order to reduce the statistical uncertainty on the fitted signal, 
which is closely related to the number of free parameters, $f_{sig}$ and $\sigma_W$ were fixed. 
Also, the central position of the Gaussians was fixed to the well-known PDG value, $m_{PDG}$.

For the $K^0_S$ analysis
the background shape was modeled by an exponential function, 
while a 3$^{rd}$ order Chebyshev function was used in the $\Lambda$ analysis.

The \pth binning scheme and an example of
a fit in one \pth bin is presented in Figs.~\ref{fig:k0_ps_fit} 
and~\ref{fig:lambda_ps_fit} for $K^0_S$ and $\Lambda$, respectively. 

The same analysis procedure was applied to 
simulated events, for which the \VenusLong~\cite{Venus1,Venus2} model
was used as the primary event generator. The fitted numbers of $K^0_S$ and $\Lambda$ in the data
were then corrected as described in Section~\ref{Sec:ana}.2 taking also into account 
$V^0$ decay channels that cannot be reconstructed with the detector.

The number of $V^0$ candidates from interactions in the material surrounding the target was estimated 
by analysing the special runs taken with the target removed. After applying the same selections as used for 
the target inserted data a negligible number of candidates remained. 

Spectra were then derived from the corrected number of fitted $K^0_S$ and $\Lambda$ 
using the standard \NASixtyOne procedure (Eqs.~\ref{eq:corr_nb},~\ref{eq:xsecmeas3},~\ref{eq:xsecprod}).

\subsubsection{Systematic uncertainties of the $V^0$ analysis}
\label{Subsec:v0_syst}

The contributions from five sources of systematic uncertainties associated with this analysis were studied adopting
the same procedure in all cases. Namely, the relative difference between the standard analysis and the one 
in which the respective source was varied was taken as an estimate of the systematic uncertainty. The 
following contributions were considered:
\begin{enumerate}[(i)]
\setlength{\itemsep}{1pt}
\item {\it Correction factors}: The dependence of the correction factors on the primary event generator was tested 
by performing two other simulations with the same statistics, based on \FlukaNew~\cite{Fluka_CERN,Fluka,Fluka_new} 
and \EposLong~\cite{EPOS} as generators for primary interactions. The systematic uncertainty associated with this source 
is from 7\% to 10\% for both $K^0_S$ and $\Lambda$.
\item {\it Fitting procedure}: Several alternative fitting functions were tested on the invariant mass distributions:
a bifurcated Gaussian for the peak signal and a 3$^{rd}$ order Chebyshev polynomial or a 4$^{th}$ 
order polynomial function for the background for $K^0_S$ and $\Lambda$, respectively. The contribution to the systematic uncertainty
associated with this source is up to 12\% (7\%) for $K^0_S$ ($\Lambda$).
\item {\it Reconstruction algorithm}: Two different primary interaction vertex reconstruction algorithms 
were used, either fitting all three coordinates or fixing the z-coordinate to the survey position.
The uncertainty connected to the algorithm is from 5\% to 8\% for both analyses. 
\item {\it Quality cuts}: All quality cuts were varied independently within a range given in Ref.~\cite{Laura_PhD}
and the uncertainty connected to this source was found to reach up to 10\% (5\%) for $K^0_S$ ($\Lambda$). 
\item {\it Kinematic cuts}: As for the previous source of uncertainty, all cuts were varied and the 
resulting uncertainty is up to 10\% (7\%) for $K^0_S$ ($\Lambda$).
\end{enumerate}
The uncertainty estimates of items (iv) and (v) are strongly correlated since the same datasets and analysis techniques 
were used. Hence, only the maximum deviation due to these cut variations was taken into account. The total systematic error
was taken as the sum of all contributions added in quadrature.

Comparison of the final corrected $K^0_S$ and $\Lambda$ spectra to the 2007 measurements~\cite{V0_2007}
shows that the results are compatible within the attributed uncertainties.

Tables~\ref{tab:k0_results} and~\ref{tab:l_results} present 
the final double differential cross section, 
$d^2 \sigma /(d p d \theta)$, for $K^0_S$ 
and $\Lambda$ production in p+C interactions at 31\,GeV$/c$, 
with statistical and systematic uncertainties. 
Fig.~\ref{fig:k0_09DataMixVenusEpos} 
and Fig.~\ref{fig:lambda_data_vs_models} 
show the spectra of $K^0_S$ and $\Lambda$ yields.

\subsection{The $tof$-$dE/dx$ analysis method}
\label{Sec:dedxtof}

Depending on the momentum range and charged particle species different
particle identification (PID) techniques need to be applied. 
The method described in this section utilizes
the measurements of the specific energy loss $dE/dx$ in the TPCs and the measurements 
of the time-of-flight ($tof$) by the time-of-flight ToF-F detector.
The energy loss information can be used in the full momentum range of \NASixtyOne.
The \mbox{ToF-F} detector, which is installed about 13\,m downstream of
the target (see Fig.~\ref{fig:detector}),
contributes to particle identification up to 8\,GeV/$c$.
The distribution of $m^2$, the square of the mass calculated
from the ToF-F measurement and the fitted track parameters,
is shown for positively charged particles as a function of momentum in Fig.\,\ref{m2_vs_p}.

Simultaneous use of these two sources of PID information is particularly important
in the momentum range from 1 to 4\,GeV/$c$ where the $\dd E/\dd x$ bands 
of charged hadrons cross over (see Fig.\,\ref{dedx_pos_neg}).
Thus the $\dd E/\dd x$ measurement alone would not be enough to identify 
particles with sufficient precision
and $tof$ is especially important to resolve this ambiguity.

The combined \emph{tof}-$\dd E/\dd x$ analysis technique was employed 
to determine yields of $\pi^\pm$, K$^\pm$ and protons
in the momentum region above 1\,GeV/$c$.
For lower momenta the $\dd E/\dd x$-only approach (see Section~\ref{Sec:dedx} below) 
provides better statistical precision.
The spectra of $\pi^-$ can also be obtained precisely with the so-called
$h^-$ analysis technique (see Section~\ref{Sec:h-} below).

The standard event and track selection procedures common to all charged hadron analyses are described in 
Section~\ref{sec:track_sel}. 
The following additional cuts were applied in the \emph{tof}-$\dd E/\dd x$ analysis:
\begin{enumerate}[(i)]
\setlength{\itemsep}{1pt}
\item exclusion of kinematic regions where the spectrometer acceptance changes rapidly and
  a small mismatch in the simulation can have a large effect on the corrected hadron spectra.
  Basically, these are regions where the reconstruction capability is limited by the ToF-F acceptance,
  by the magnet aperture or by the presence of uninstrumented regions in the VTPCs. 
  To exclude these regions a cut on the hadron azimuthal angle $\phi$ was applied.
  Since the spectrometer acceptance drops quickly with increase of the polar angle 
  $\theta$ of the particle, typical cut intervals for $\phi$ at low and high $\theta$ were  
  $\pm 60^\circ$  and $\pm 6^\circ$, respectively.
  
\item the track must have an associated ToF-F hit.
  Since particles can decay or interact before reaching the  detector, 
  the $z$ position of the last reconstructed TPC cluster of a track 
  should be reasonably close to the \mbox{ToF-F} wall, thus a cut
  $z_\text{last}>6$\,m was applied to all tracks.
  This requirement is especially important for $K^\pm$ ($c\tau\approx 3.7$\,m)
  many of which decay before reaching \mbox{ToF-F}.
  Pions have a higher chance to reach the \mbox{ToF-F} ($c\tau\approx 7.8$\,m)
  also due to their lower mass and thus larger Lorentz factor. 
  Moreover a muon produced in the decay of a pion follows
  the parent pion trajectory. 
  Such a topology is in general  reconstructed as a single track.
  
\end{enumerate}

Having ToF-F slabs oriented vertically, the position of a hit 
is measured only in the $x$ direction.
The precision is determined by the width of the scintillator slab
producing the signal.  
A ToF-F hit is associated with a track if the trajectory can be
extrapolated to the pertaining slab.

Another important reason for using the ToF-F information is the time tag which 
it provides and which ensures that all associated tracks originated from the triggered event. 
The ToF-F time resolution of 110 ps \cite{NA61detector_paper} 
guarantees an unambiguous discrimination against tracks from out-of-time events for
the used beam rate of 100 kHz (one beam particle on average each $10\, \mu$s).

\begin{figure*}
\centering
\includegraphics[width = 0.99\textwidth]{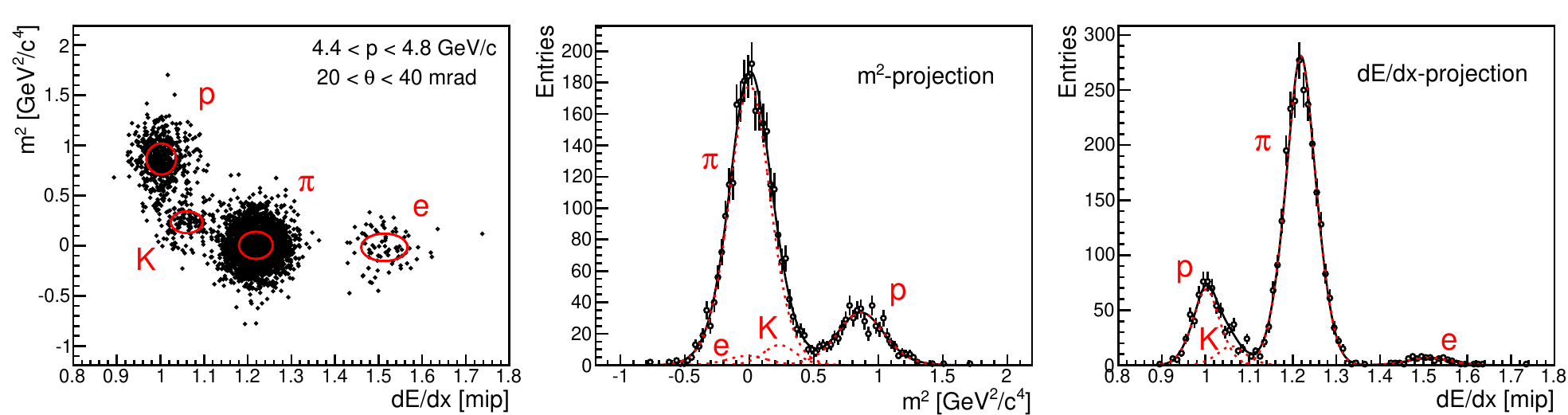}
\caption{
  (Colour online)
  Example of a two-dimensional fit to the $m^2 - dE/dx$ distribution of 
  positively charged particles ($left$). 
  The $m^2$ ($middle$) and $dE/dx$ ($right$) projections are 
  superimposed with the results of the fitted functions.
  Distributions correspond to the \pth bin:
  $4.4 < p < 4.8$ GeV/$c$ and $20<\theta<40$ mrad.
\label{fig:ToF-dEdx_fit}
}
\end{figure*}

The $tof$-$dE/dx$ analysis was performed separately for positively and
negatively charged particles following the  procedure  described 
in detail in Refs.~\cite{pion_paper,kaon_paper,Sebastien}.
A two-dimensional
histogram of $m^2$ versus $\dd E/\dd x$ was filled for every \pth bin.
An example of such a distribution is shown in Fig.\,\ref{fig:ToF-dEdx_fit}.
In this distribution particles of different types form regions
which are parametrized by a product of two one-dimensional Gaussian functions 
in $m^2$ and $\dd E/\dd x$, respectively.
The binned maximum likelihood
method is applied to fit the distribution with 20 parameters
(4 particle types $\times$ 5 parameters of the Gaussians).
Depending on the momentum range and particle species, 
some of these parameters were fixed or constrained.
In particular, this is important for $K^+$ which are difficult to separate from
protons in the projected $\dd E/\dd x$ distributions at higher momenta 
where the $tof$ information can no longer provide PID.
Therefore, the mean $\langle\dd E/\dd x\rangle$ position of kaons was fixed 
and the width of the $\dd E/\dd x$ peak was constrained from above by using 
information from pions and protons.

As a result of the fit one obtained raw yields of particles
($e^\pm$, $\pi^\pm$, $K^\pm$, p, $\bar{\rm p}$) in bins of \pth 
which were then corrected 
using the \NASixtyOne  
simulation chain with the \VenusLong \cite{Venus1,Venus2} model for primary interactions and 
a GEANT3-based part for tracking the produced particles through the detector.

\begin{figure*}
\centering
\includegraphics[width = 0.44\textwidth]{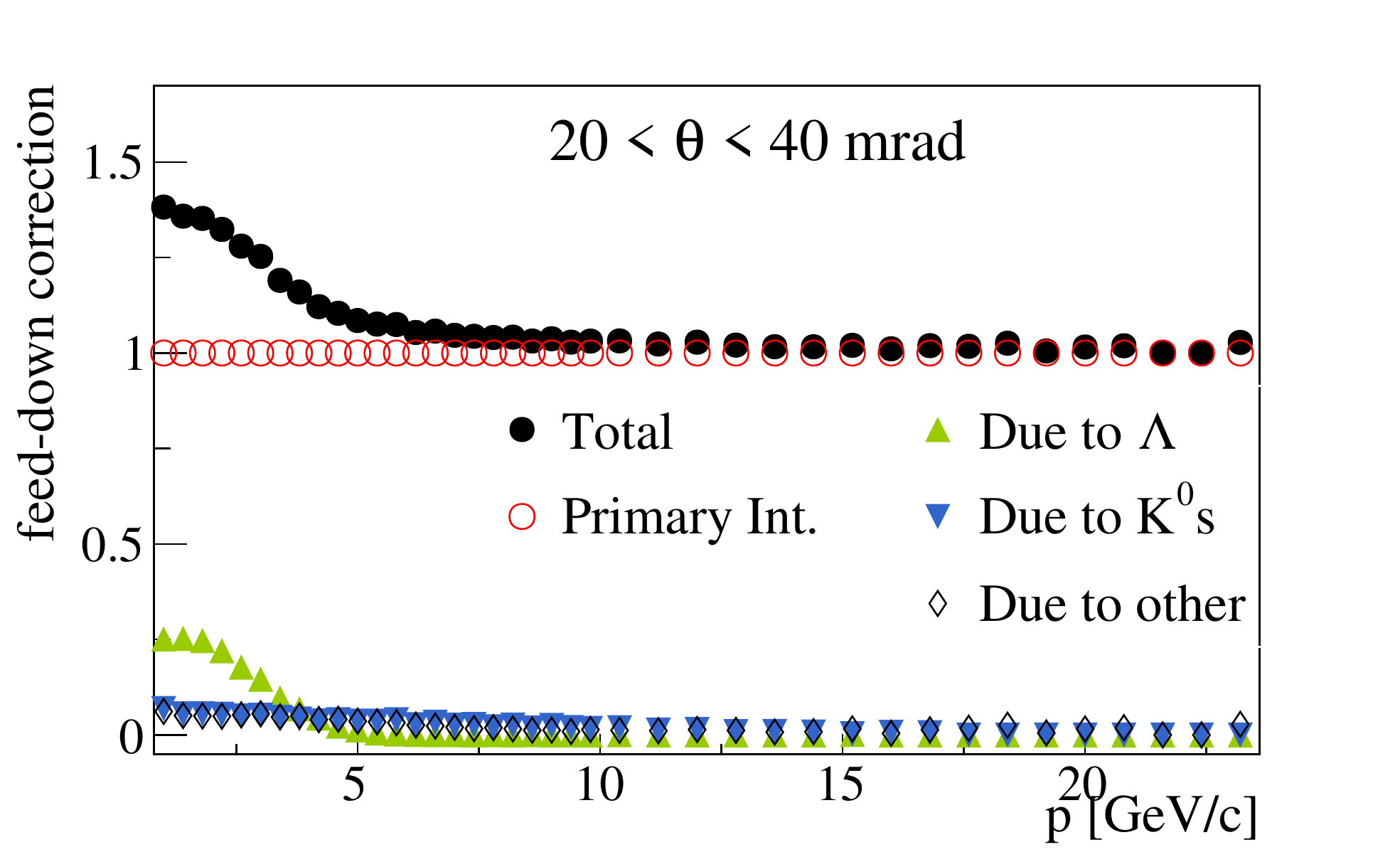}
\hspace{0.05\textwidth}
\includegraphics[width = 0.44\textwidth]{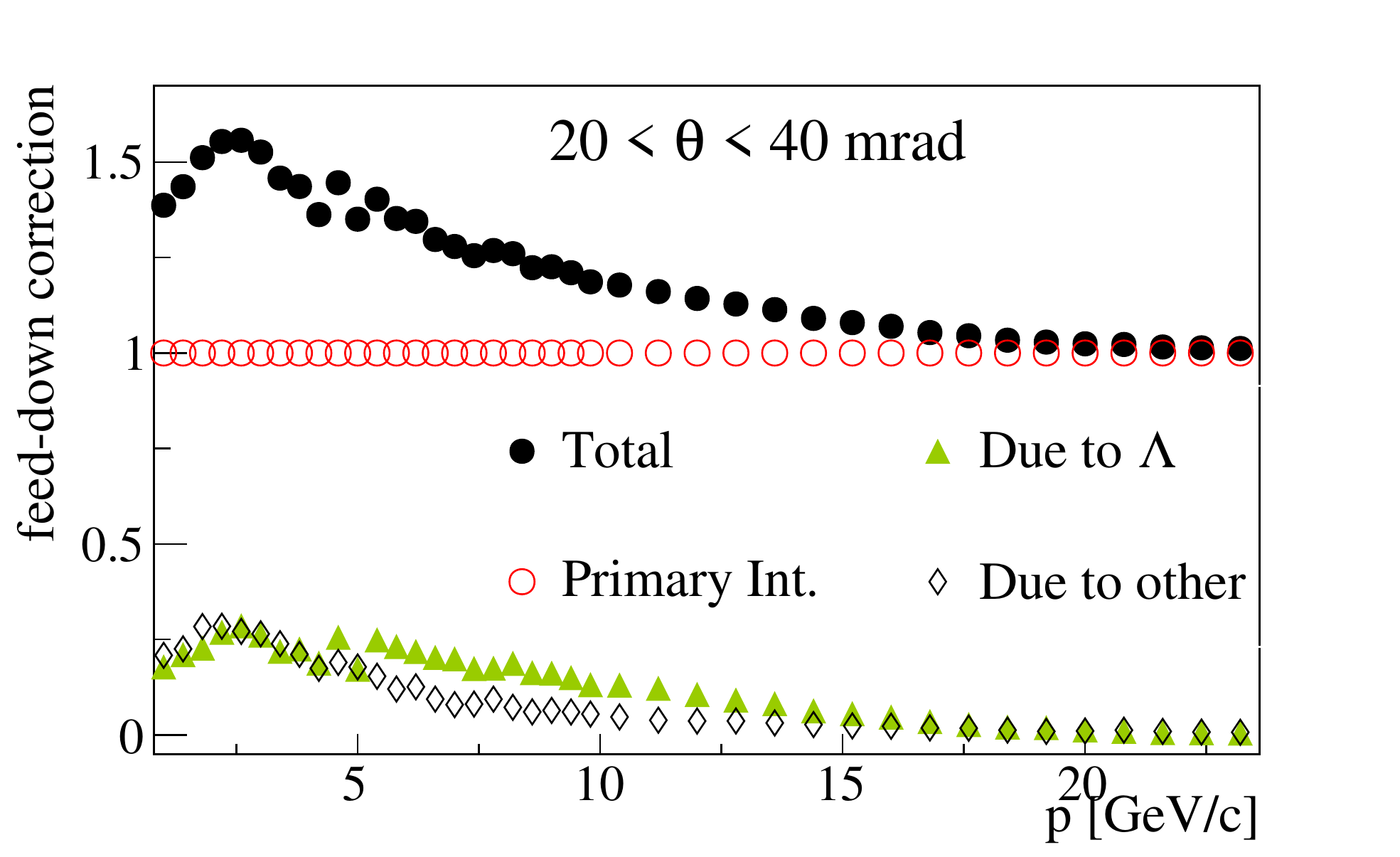}
\caption{
  (Colour online)
  An example of the dependence of the feed-down correction on momentum for
  $\pi^-$ mesons ($left$) and protons ($right$) for the [20,40]\,mrad 
  angular interval. Contributions due to $\Lambda$ and $K^0_S$ decays 
  are shown separately.
\label{fig:feed_down}
}
\end{figure*}

\begin{figure*}
\centering
\includegraphics[width = 0.44\textwidth]{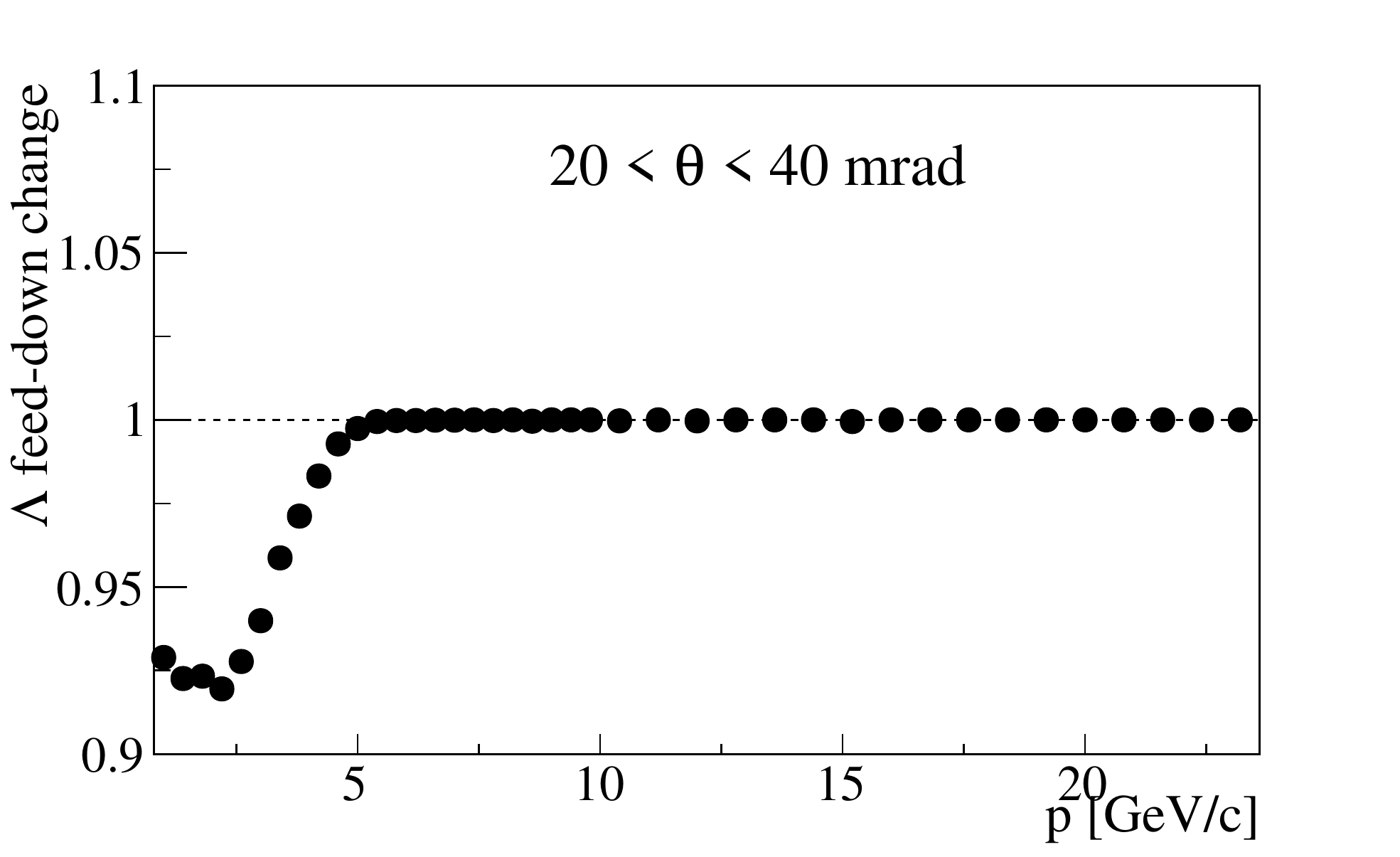}
\hspace{0.05\textwidth}
\includegraphics[width = 0.44\textwidth]{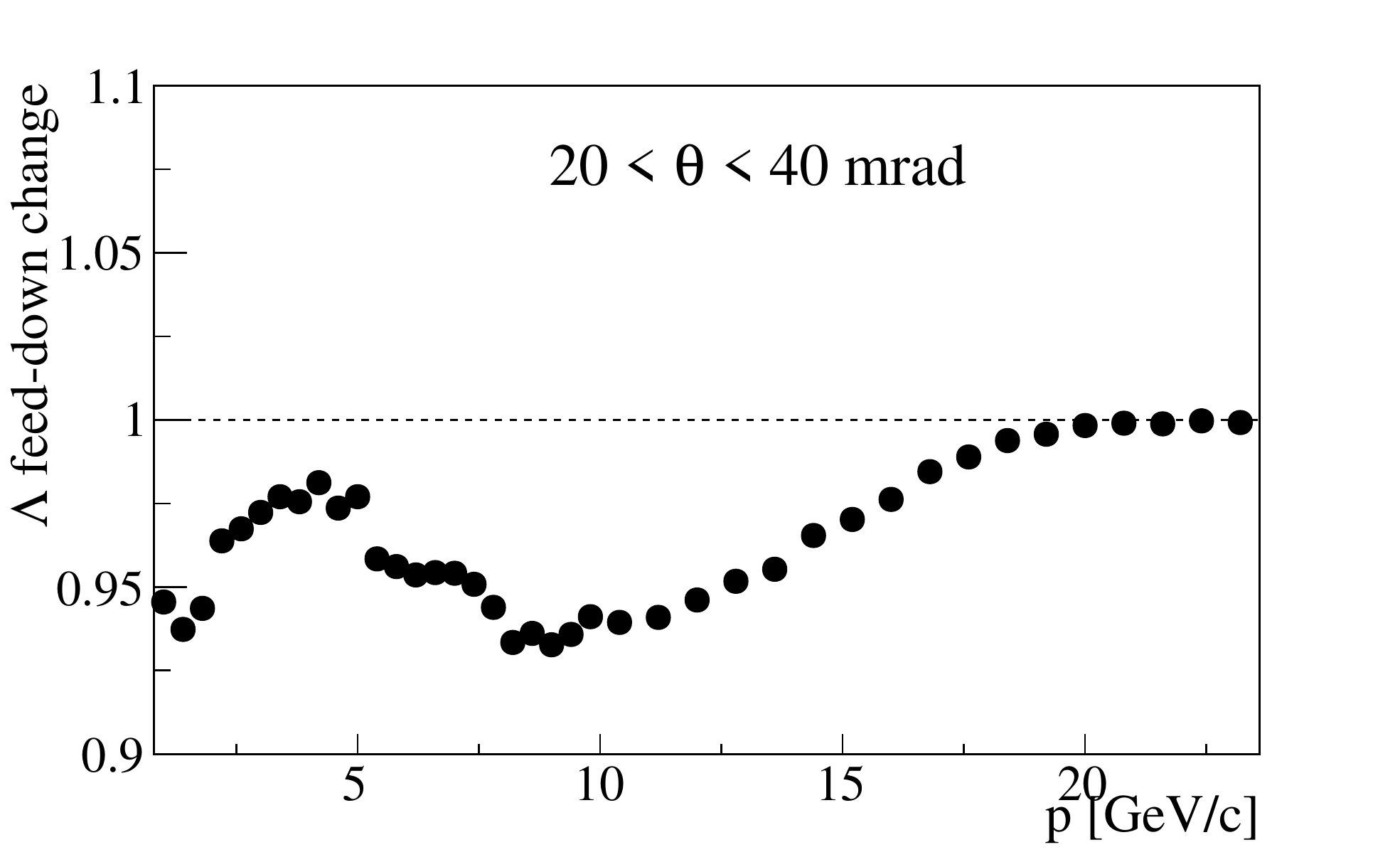}
\caption{
  (Colour online)
  The effect of the $\Lambda$ re-weighting. The ratio of multiplicities
  calculated using the \VenusLong
  model without modification and with the $\Lambda$ spectra reweighted to
  reproduce the \NASixtyOne measurements. Graphs are shown as a function of
  momentum for $\pi^-$ mesons ($left$) and protons ($right$).
\label{fig:reweighting}
}
\end{figure*}


\subsubsection{Feed-down corrections and $\Lambda$ re-weighting}
\label{subsec:lambda_reweighting}

Hadrons which were not produced in the primary interaction 
can amount to a significant fraction of the selected track sample.
Thus a special effort was undertaken to evaluate and subtract 
this contribution. 
Hereafter this correction will be referred to as feed-down.

According to the simulation with the \VenusLong model as the primary event generator,
the correction reaches  12\% for $\pi^+$, 
40\% for $\pi^-$ and up to 60\% for protons at low polar angles
and small momenta. For kaons it is not significant 
(${\lesssim}2\%$).
Figure~\ref{fig:feed_down} shows 
the feed-down correction for $\pi^-$ and protons
as a function of momentum for one of the $\theta$ bins.
Decomposition of the correction reveals that the main contribution  comes 
from $\Lambda$-hyperon decays:
\[
 \begin{array}{lcl}
 p + \text{C} & \rightarrow & \Lambda + \text{X} \\
       &             & ~ \stackrel{\rotatebox[origin=c]{180}{$\Lsh$}}{} \, \pi^- + p~.
 \end{array}
\]
For $\pi^-$ mesons these decays are responsible for about $2/3$ of 
the non-primary contribution at $p=1$\,GeV/$c$.
For protons, they amount to about $1/2$ for the whole momentum range.
The measurements of $\Lambda$ spectra described in Section~\ref{Sec:v0}
can be used to improve the precision of the correction.
Therefore the feed-down contribution of $\Lambda$ is calculated separately from 
the feed-down correction due to other weak decays.

Technically the $\Lambda$ feed-down correction based on data was evaluated
by weighting the \VenusLong generated spectra of $\Lambda$ to agree with the measurements.
The resulting change of the  $\pi^-$ and proton spectra is shown in 
Fig.~\ref{fig:reweighting}. 
Thus rescaling of the $\Lambda$ spectra reduces the feed-down
in $\pi^-$ spectra by a maximum of 8\% at $p=2$\,GeV/$c$.
For protons the reduction was found in most of the momentum 
range with a maximum at $p=9$\,GeV/$c$.

The reweighting  of the \VenusLong $K^0_S$ spectra
based on  the \NASixtyOne measurements impacts 
the corrected charged pion spectra by less than 2\%.
This refinement of the correction was neglected in the results presented here. 


\subsubsection{Systematic uncertainties of the $tof$-$d E/d x$ analysis} \label{sec:FA}

\begin{figure*}
\centering
\includegraphics[width = 0.9\textwidth]{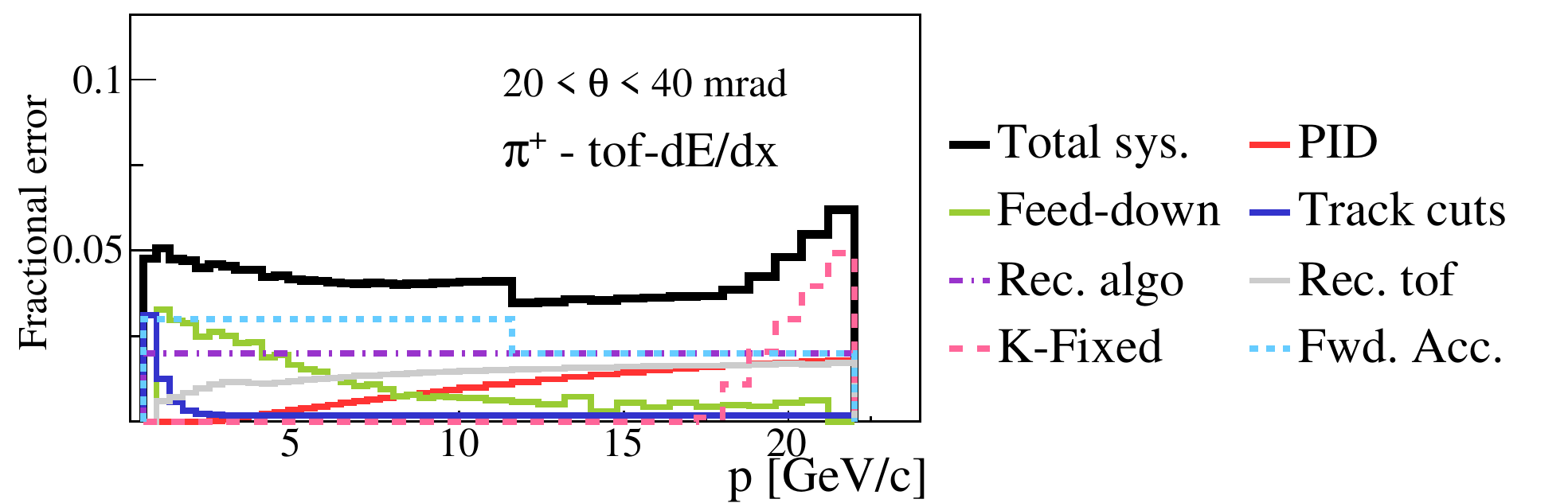}
\caption{
\label{fig:pip_err_breakdown_tofdedx} 
(Colour online)
Breakdown of systematic uncertainties of $\pi^+$ spectra from the \emph{tof}-$\dd E/\dd x$ analysis, presented as a function 
of momentum for the [20,40]~mrad angular interval.
}
\includegraphics[width = 0.9\textwidth]{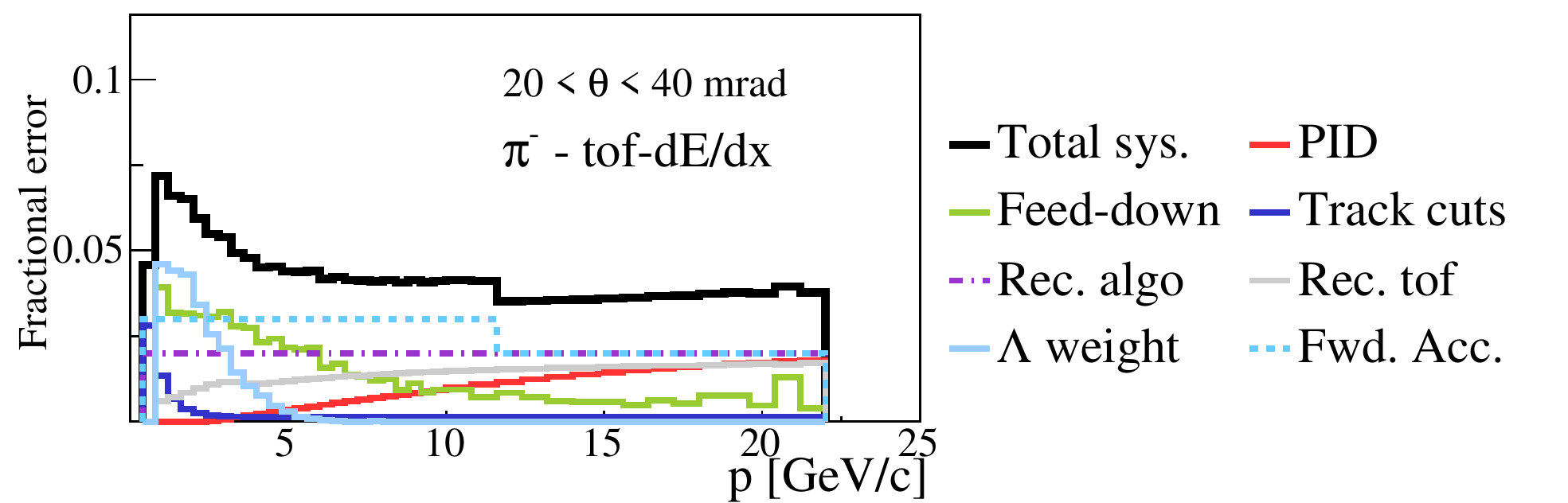}
\caption{
\label{fig:pim_err_breakdown_tofdedx} 
(Colour online)
Breakdown of systematic uncertainties of $\pi^-$ spectra from the \emph{tof}-$\dd E/\dd x$ analysis, presented as a function 
of momentum for the [20,40]~mrad angular interval.
}
\end{figure*}

Systematic uncertainties of the hadron spectra were estimated by
varying track selection and identification criteria as well as
the parameters used to calculate the corrections. 
The following sources of systematic uncertainties were considered:

\begin{enumerate}[(i)]

\item {\it PID ($\dd E/\dd x$).}
 A Gaussian function is used to parametrize the $\dd E/\dd x$ distribution 
 at a fixed value of momentum. 
 The width of the distribution 
 decreases with increasing
 number of TPC clusters used to determine the $\dd E/\dd x$
 value of a track.  
 Having in one \pth bin tracks 
 with different number of clusters
 would cause a deviation from the single-Gaussian shape.
 The contribution of this effect to the uncertainty of the fitted 
 number of particles of a certain species was estimated by 
 performing an alternative fit by a sum of two Gaussians.
 
 At low momenta the ToF-F resolution ensures an unambiguous
 particle identification,
 thus small deviations of the $\dd E/\dd x$ distribution from a single Gaussian
 in general are not important.  
 A deviation  has a significant effect only for momenta above 3\,GeV/$c$.
 For pions it steadily increases up to 2\% at $p=20$\,GeV/$c$.
 However for kaons,  
 the uncertainty is an order of magnitude larger: up to 20\% 
 for $K^+$ at high momenta.

\item {\it Hadron loss.} 
 To ensure a high quality match of tracks and ToF-F hits 
 the last point of a track should be within 1.6\,m 
 from the ToF-F wall ($z_\text{last}>6$\,m).
 This implies
 that a track segment is reconstructed in the 
 \mbox{MTPC-L} or \mbox{MTPC-R} detectors. 
 Possible imperfections in the description of the spectrometer
 can introduce a difference in the acceptance and reconstruction 
 efficiency (merging track segments between \mbox{VTPC-2} and \mbox{MTPC-L/R}) 
 between simulation and real data
 which can be 
 important  for the reconstruction of long tracks.
 To check how sensitive the results are to the $z_\text{last}$ cut, 
 it was relaxed down to $z_\text{last}>-1.5$\,m for pions and 
 $z_\text{last}>-3$\,m for kaons. 
 The difference in the resulting final spectra was assigned
 as the systematic uncertainty.
 It reaches up to 2\% at 1\,GeV/$c$ and drops quickly
 with increasing momentum.

\item {\it Reconstruction efficiency.}
 To estimate the uncertainty of the reconstruction efficiency the following track selection 
 criteria were varied: the minimum number of points measured on the track,
 the azimuthal angle and the impact parameter cuts.
 Also results were compared which were obtained with two independent track topologies,
 different algorithms for merging track segments from different TPCs
 into global tracks, and with
 two different algorithms for the primary vertex reconstruction.
 It was found that the influence of such changes is small compared to  
 the statistical uncertainties.
 The corresponding systematic uncertainty was estimated to be 2\%.

\item {\it Forward acceptance.}
As stated in Section~\ref{sec:setup}, the GTPC detector was used for the first time in the reconstruction algorithms within \NASixtyOne.
The GTPC increases the number of reconstructed tracks
mainly for smaller angles, $\theta<40$\,mrad, and thus allows
finer momentum binning in the forward region $\theta<20$\,mrad.
The estimate of the  systematic uncertainty in the forward region is based on the
comparison of results obtained with and without the GTPC included in the 
reconstruction algorithms
and by varying the required number of GTPC clusters in the analysis.
The latter takes into account inefficiencies of the GTPC 
electronic readout which were not included in the simulations.
The difference of spectra 
obtained with and without using the GTPC information in the reconstruction
was found
 to be  4\% up to $\theta<20$\,mrad and
 about 3\% for $20<\theta<40$\,mrad. 
The variation of the required number of  GTPC clusters  between 4 and 6 resulted in changes  
of up to  4\% for $0<\theta<10$\,mrad and  up to 2\%  for $10<\theta<20$\,mrad
for tracks with momentum $p>12$\,GeV/$c$. In this region 
the majority of tracks do not traverse the \mbox{VTPC-1/2} 
detectors and thus the reconstruction of the track segment in the magnetic field totally relies 
on the GTPC.

\item {\it ToF-F reconstruction.}
 The ToF-F reconstruction efficiency was estimated using a sample of events with very 
 strict selection requiring no incoming beam particle within  a $\pm20\,\upmu$s 
 time window around the triggered interaction.
 The inefficiency was found to vary from 4\% in the central region
 to a fraction of a percent 
 in the ToF-F slabs far away from the beamline.
 This variation correlates with the higher particle density
 in the near-to-beam region where 
 several particles can hit a single slab, thus contributing to inefficiency.
 A value of 50\% of the inefficiency correction was assigned as a conservative limit 
 to this source of systematic uncertainty.

 \item {\it Secondary interactions and non-$\Lambda$ feed-down corrections.}
 As in the case of the $\dd E/\dd x$ (Section~\ref{Sec:dedx}) and $h^-$ (Section~\ref{Sec:h-}) 
 approaches, the important contribution 
 to the systematic uncertainty at low momenta comes from the uncertainty of 
 the simulation-based correction for secondary interactions and weak decays of strange 
 particles (excluding $\Lambda$ hyperons). 
 Following arguments described in Ref.~\cite{pion_paper}
 an uncertainty of 30\% of the correction value was assigned for 
 both of these sources.

\item {\it $\Lambda$ feed-down correction.}
 The correction for the feed-down to pions and protons
 originating from $\Lambda$ decays was calculated separately
 based on measured $\Lambda$ spectra
 (see Section~\ref{subsec:lambda_reweighting}).
 The uncertainty assigned to this correction was estimated to be 30\% 
 which is an upper limit on the overall uncertainty of the measured $\Lambda$ spectra.

\end{enumerate}
Figures~\ref{fig:pip_err_breakdown_tofdedx} 
and~\ref{fig:pim_err_breakdown_tofdedx} show a breakdown of 
the total systematic uncertainty in the \emph{tof}-$\dd E/\dd x$ analysis 
for the example of the angular interval [20,40]~mrad.

\subsection{The $\dd E/\dd x$ analysis method} \label{Sec:dedx}

The analysis of charged pion production at low momentum was performed
using particle identification based only on measurements of specific energy loss in the TPCs.
For a large fraction of tracks \emph{tof} can not be measured
since the majority of low-momentum particles does not reach the ToF-F detector.
A reliable  identification of $\pi^+$ mesons was not possible
at momenta above 1\,GeV/$c$ where the BB
curves for pions, kaons, and protons cross each other 
(see Fig.~\ref{dedx_pos_neg}).
On the other hand, since the contamination from $K^-$ and antiprotons
is almost negligible for $\pi^-$ mesons, the $\dd E/\dd x$ analysis could be performed
for momenta up to 3\,GeV/$c$ allowing consistency checks
with the other identification methods in the region of overlap.

The procedure of particle identification, described below, is tailored
to the region where a rapid change of energy loss with momentum is observed.
This procedure was used already for the 2007 data and more details 
can be found in Ref.~\cite{Magda}. 
Here just the most important steps of the analysis are described.

In order to optimize the parametrization of the BB function,
samples of $e^\pm$, $\pi^\pm$, $K^\pm$, p, and $d$ tracks with
reliable particle identification were chosen in the $\beta \gamma$ 
range from 0.2 up to 100. The dependence of the BB function on
$\beta\gamma$ was then fitted to the data using the
Sternheimer and Peierls para\-metrization of Ref.~\cite{stern}.
This function was subsequently used to calculate for every track
of a given momentum the expected $\dd E/\dd x_\text{BB}$
values for all considered identity hypotheses
for comparison with the measured $\dd E/\dd x$.
A small (a few percent) dependence of the mean $\avg{\dd E/\dd x}_\text{data}$ 
on the track polar angle had to be corrected for.
	
\begin{figure}[!hb]
\begin{center}
\includegraphics[width=0.95\linewidth]{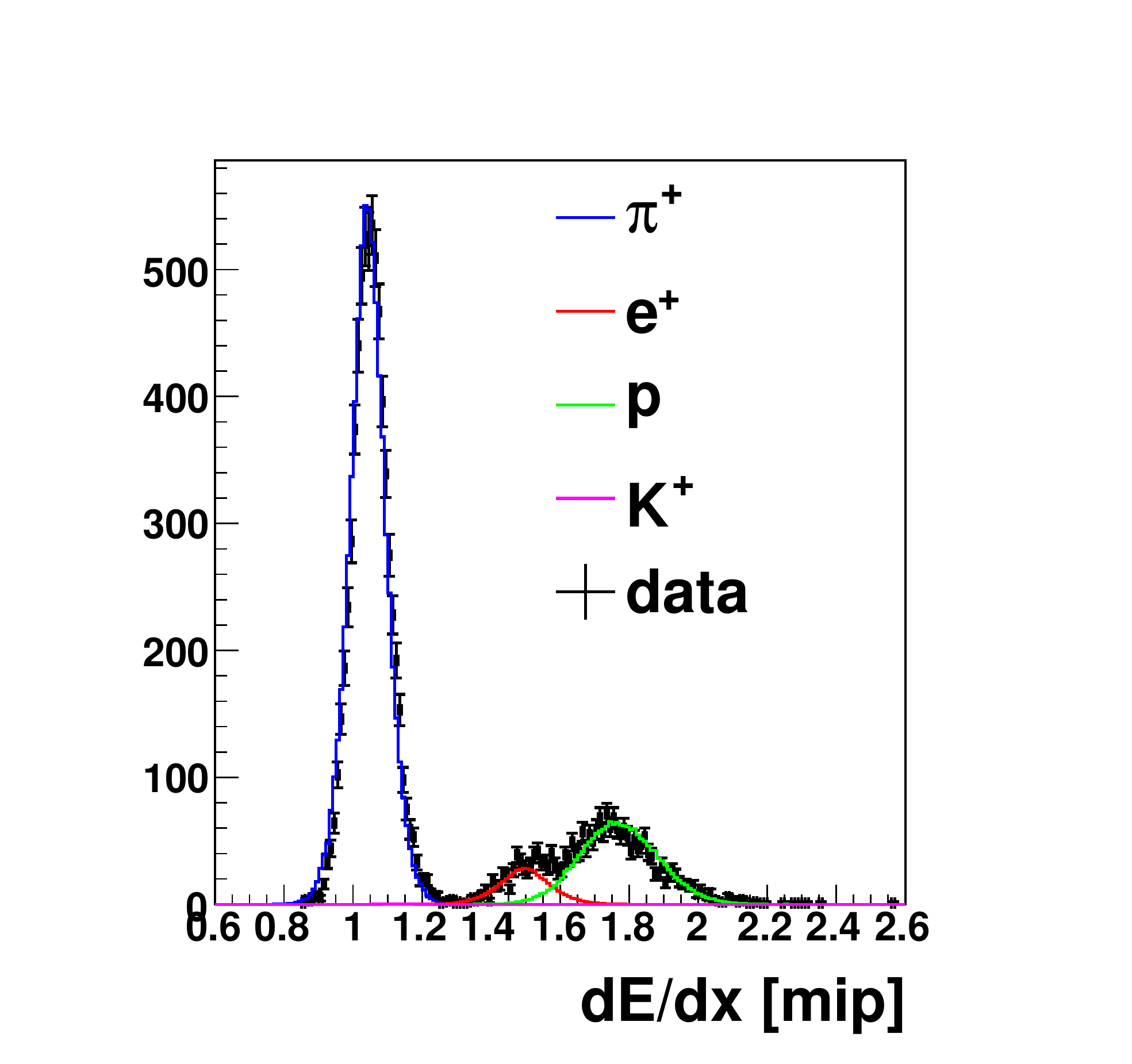}
\includegraphics[width=0.95\linewidth]{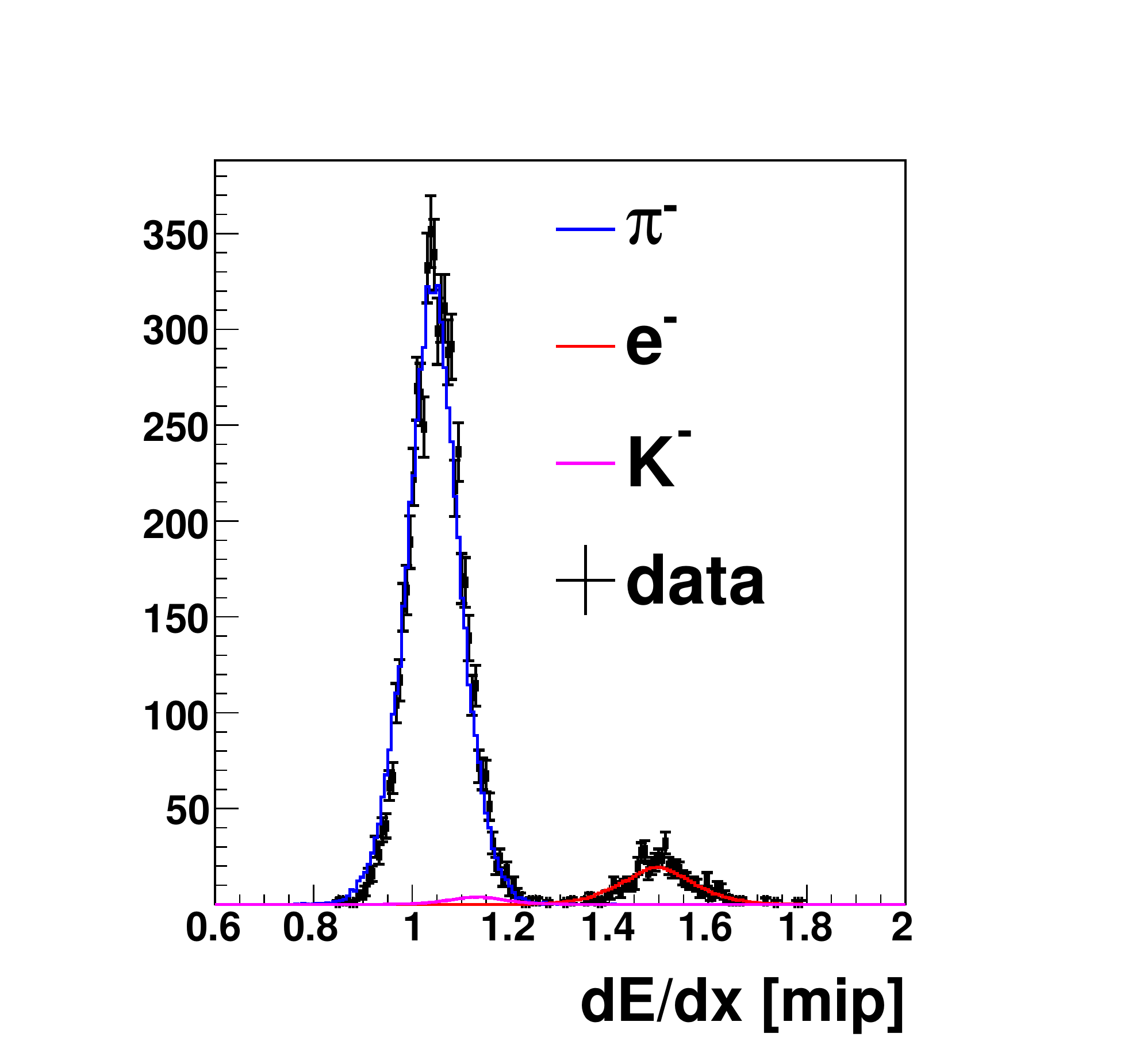}
\caption{(Colour online)
The $\dd E/\dd x$ distributions for positively ($top$) and negatively ($bottom$) charged
particles in the momentum interval [0.8,0.9]\,GeV/$c$ 
and angular bin [180,240]\,mrad compared
with the distributions calculated using the fitted relative abundances.}
\label{dedx_fit_example}
\end{center}
\end{figure}

The identification procedure was performed in \pth bins.
Narrow momentum intervals (of 0.1\,GeV/$c$ for $p<1$\,GeV/$c$ and
0.2\,GeV/$c$ for $1<p<3$\,GeV/$c$) were chosen because of the strong
dependence of $\dd E/\dd x$ on momentum.
The event and track selection criteria described in Section~\ref{sec:track_sel}
were applied.
In each \pth bin an unbinned
maximum likelihood fit (for details see Ref.~\cite{marek}) 
was performed  to extract yields
of $\pi^+$ and $\pi^-$ mesons. The probability density functions
were assumed to be a sum of Gaussian functions for each particle species,
centered on $\dd E/\dd x_\text{BB}$ with variances derived from the data.
The $\dd E/\dd x$ resolution is a function of
the number of measured points and the particle momentum.
In the $\pi^+$ analysis three independent abundances were fitted
($\pi^+$, $K^+$ and proton), in the $\pi^-$ analysis only two ($\pi^-$ and $K^-$). 
The $e^+$ and $e^-$ abundances
were determined from the total number of particles in the fit.
	
As an example, the $\dd E/\dd x$ distributions for positively and negatively charged
particles in the momentum interval [0.8,0.9]\,GeV/$c$ and angular bin [180,240]\,mrad
are shown in Fig.~\ref{dedx_fit_example} and compared with the distributions 
obtained from the fitted function.

Finally, the \VenusLong based simulation
was used to calculate bin-by-bin corrections for
pions from weak decays and interactions in the target and the detector material. The corrections include also track reconstruction efficiency and resolution as well as
losses due to pion decays.
Also $\mu$ tracks coming from $\pi$ decay were taken into account (not done for the 2007 data analysis). 
Simulation studies showed that this additional category of tracks has 
a non-negligible impact on the value of corrections increasing them by about ${\sim}8\%$. 
The $\mu$ tracks from $\pi \to \mu$ decays
can not be distinguished from $\pi$ candidates in the data, as both particles show  
similar values of $\dd E/\dd x$ due to the small difference in their masses.

\subsubsection{Systematic uncertainties of the $\dd E/\dd x$ analysis}

\begin{figure*}
\centering
\includegraphics[width = 0.9\textwidth]{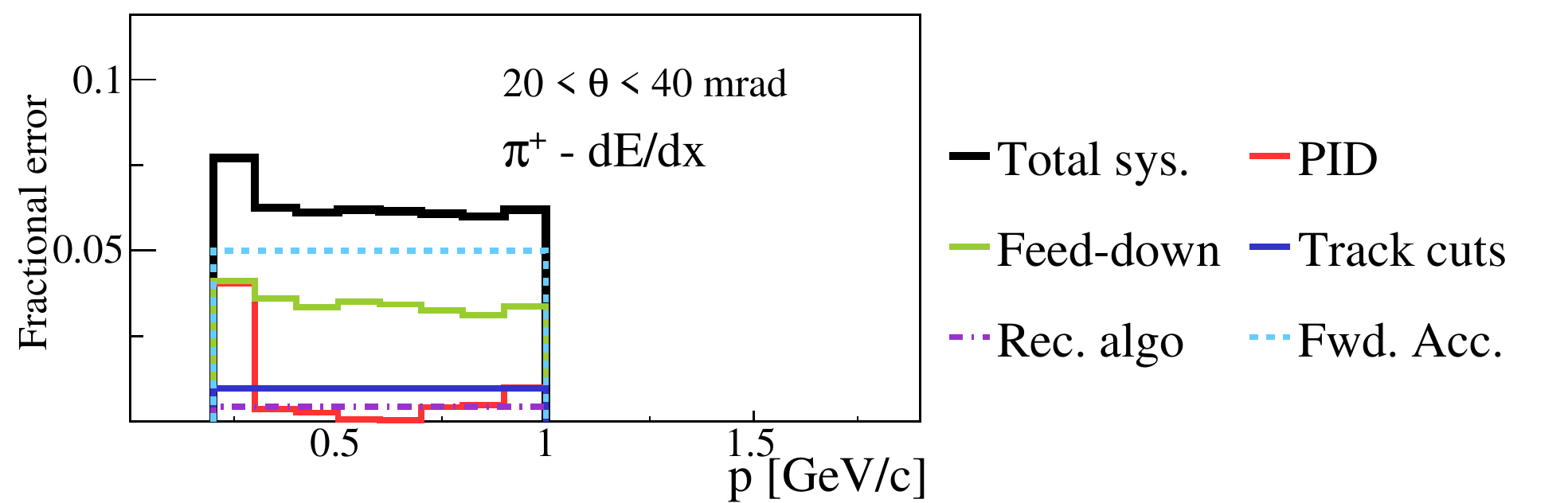}
\caption{
\label{fig:pip_err_breakdown_dedx} 
(Colour online)
Breakdown of systematic uncertainties of $\pi^+$ spectra from the $\dd E/\dd x$ analysis, presented as a function 
of momentum for the [20,40]~mrad angular interval.
}
\includegraphics[width = 0.9\textwidth]{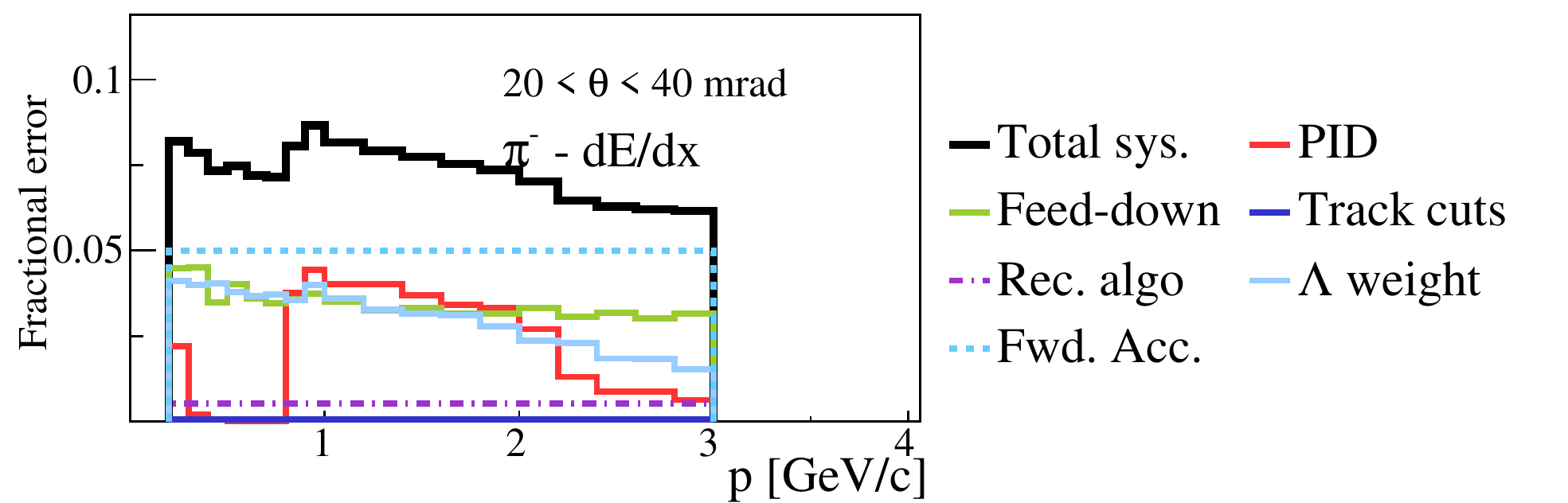}
\caption{
\label{fig:pim_err_breakdown_dedx} 
(Colour online)
Breakdown of systematic uncertainties of $\pi^-$ spectra from the $\dd E/\dd x$ analysis, presented as a function 
of momentum for the [20,40]~mrad angular interval.
}
\end{figure*}

In the $\dd E/\dd x$ analysis a number of sources of systematic uncertainties were considered. 
Some are specific to the $\dd E/\dd x$ analysis, others were already described in previous sections. 
\begin{enumerate}[(i)]
\item {\it PID ($\dd E/\dd x$)}. 
The BB parametrization used in the analysis was fitted to the
experimental data.
In order to estimate the uncertainty related to the precision
of the parametrization the results were calculated using
the BB curves shifted by $\pm 1\%$. 
This value was chosen based on small discrepancies observed between the fitted BB parametrization 
and the measured energy loss for pions.
The resulting relative systematic uncertainty  ($p_\text{BB}$)
was calculted as:
\begin{equation}
p_\text{BB} \equiv \frac{n_{\text{iden,BB},\pi^\pm}}{n_{\text{iden},\pi^\pm}}~,
\end{equation}
where $n_{\text{iden,BB},\pi^\pm}$ represents the number of identified $\pi^\pm$ 
when the BB curves are shifted by $\pm 1\%$, 
while $n_{\text{iden},\pi^\pm}$ gives the original number of identified charged pions. 

Clearly the calibration uncertainty of the BB function is important only for the momentum  
bins where the BB curves for two particle hypotheses are close to each other. 
A particularly difficult momentum region is [0.8-2.2]\,GeV/$c$ where 
the BB curve for kaons approaches and crosses that for pions. 
The kaon relative abundance is low causing additional instability in the fit. 
Therefore a conservative estimate of the resulting uncertainty of the  pion 
abundance in this momentum range was obtained by allowing the kaon abundance $m_K$ to vary between 0 and 0.5\%. 
The limit was adjusted while studying neighboring bins with $p=[0.7,0.9]$\,GeV/$c$, 
where the fitted abundace $m_K$ was not larger than 0.1\%. 
The corresponding relative systematic uncertainty ($p_{m_K}$) was calculated as:
\begin{equation}
p_{m_K} \equiv \frac{n_{\text{iden},m_K,\pi^\pm}}{n_{\text{iden},\pi^\pm}}~, 
\end{equation}
where $n_{\text{iden},m_K,\pi^\pm}$ represents the fitted number of identified $\pi^\pm$ 
with the 0.5\% limit set on the fitted relative kaon fraction. \\
One has to keep in mind that $p_\text{BB}$ and  $p_{m_K}$ are correlated. 
Therefore the larger value ($p_f\equiv \max[p_\text{BB},p_{m_K}]$) was taken as the estimate 
of the systematic uncertainty $p_f$ coming from the $\dd E/\dd x$ identification procedure.

\item {\it Forward acceptance.}
The uncertainties of the acceptance correction in the forward region  were determined as 
described in Section~4.4.2 item (iv) also for low momentum tracks.
A systematic uncertainty  
of 5\% was assigned for low angle intervals ($\theta<$ 60 mrad), 
and 3\% for other intervals.

\item {\it Feed-down corrections.} 
An uncertainty of 30\% was assigned to the corrections for feed-down
from non-$\Lambda$ (purely simulation based) and $\Lambda$ (simulation and data based) decays.  
This uncertainty is particularly significant in the low momentum region 
studied in the $\dd E/\dd x$ analysis. 
For further information see Section~\ref{subsec:lambda_reweighting}.

\item {\it Reconstruction efficiency.} 
For estimating the uncertainty of the reconstruction efficiency corrected results for $\pi$ spectra from the 
$\dd E/\dd x$ analysis using different reconstruction algorithms were compared, see Section~4.4.2 item (iii).  
For most $\theta$ angles this uncertainty is below 2\%.

\item {\it Track cuts.}  
The impact of the dominant track cut in the $\dd E/\dd x$ analysis 
was studied by changing the selection cut on the measured number of points 
by 10\% from the starting value of 30. The change of results is below 1\%
and thus the associated systematic uncertainty is mostly negligible.

\end{enumerate}

Figures~\ref{fig:pip_err_breakdown_dedx} 
and~\ref{fig:pim_err_breakdown_dedx} show a breakdown of 
the total systematic uncertainty in the $\dd E/\dd x$ analysis
for the example of the angular interval [20,40]~mrad
for $\pi^+$ and $\pi^-$, respectively.

\subsection{The $h^-$ analysis method} \label{Sec:h-}

The $h^-$ method utilizes the observation that  negatively charged pions  account 
for more than 90\% of primary negatively charged particles
produced in p+C interactions at 31\,GeV/$c$. Therefore $\pi^-$ spectra
can be obtained without the  use of particle identification  by correcting 
negatively charged particle spectra for the small non-pion contribution calculated 
by simulations. 
The method is not applicable to positively charged  hadrons due to the larger 
contribution of $K^+$ and protons.   
The $\pi^-$ spectra obtained from the $h^-$ analysis method cover a wide momentum range 
from about 0.2\,GeV/$c$ to 20\,GeV/$c$, providing 
an independent cross-check in the region of overlap with the other two 
analysis methods discussed above.

The $h^-$ analysis follows the previously developed procedure described in Refs.~\cite{pion_paper,Tomek}. 
In the first step, the raw $h^-$ yields were obtained in \pth bins.
Standard cuts, described in Section~\ref{sec:track_sel}, were applied  to provide good quality events and tracks.
In case the track had less than 12 hits in the VTPCs, at least 4 hits 
were required in the GTPC  
to ensure a long enough track segment in the first part of the spectrometer. 
This is  important for a good determination of the track parameters,
particularly for tracks that start in the GTPC and continue to the MTPCs.     
A selection based on the $\dd E/\dd x$ of the track was introduced (not used in the analysis of the 2007 data)
to improve the purity of the track sample by rejecting the large  electron contamination
at low $\theta$ angles. This was done to remove, in particular, electrons 
emitted by beam protons ($\delta$-rays) passing through the \NASixtyOne detector that are not accounted for
in the simulations.
The effect was not prominent in the 2007 data  
because the intensity of the  beam was significantly lower than in 2009.  
The separation between electrons and mostly pions was found to be sufficient for all $\theta$ bins when a 
momentum-dependent cut on $\dd E/\dd x$
was applied for each $\theta$ bin separately.
The contribution of electrons  was  the largest for  $\theta$ below  20\,mrad  and  momentum below 4\,GeV/$c$ 
reaching about 10\%. It decreased to 2\% in bins of larger $\theta$.  

In the next step, the raw yields of selected negatively charged  hadrons were corrected 
using the standard \NASixtyOne simulation chain with the \VenusLong model 
as primary event generator and GEANT3 for particle propagation and generation of secondary interactions.    
Like in the previous $h^-$ analysis one global correction factor was calculated for each \pth bin. 
The correction includes the contribution of primary $K^-$ and $\bar{p}$,  
the contribution from   weak decays of strange particles 
and  secondary interactions in the target as well as  in the detector material.
It also accounts for track reconstruction efficiency and losses due to the limited geometrical acceptance.

Given the large available statistics, 
the $h^-$ analysis technique provides the best precision 
for the $\pi^-$ spectra in the full \pth range covered by \NASixtyOne.

\subsubsection{Systematic uncertainties of the $h^-$  analysis}

\begin{figure*}
\centering
\includegraphics[width = 0.9\textwidth]{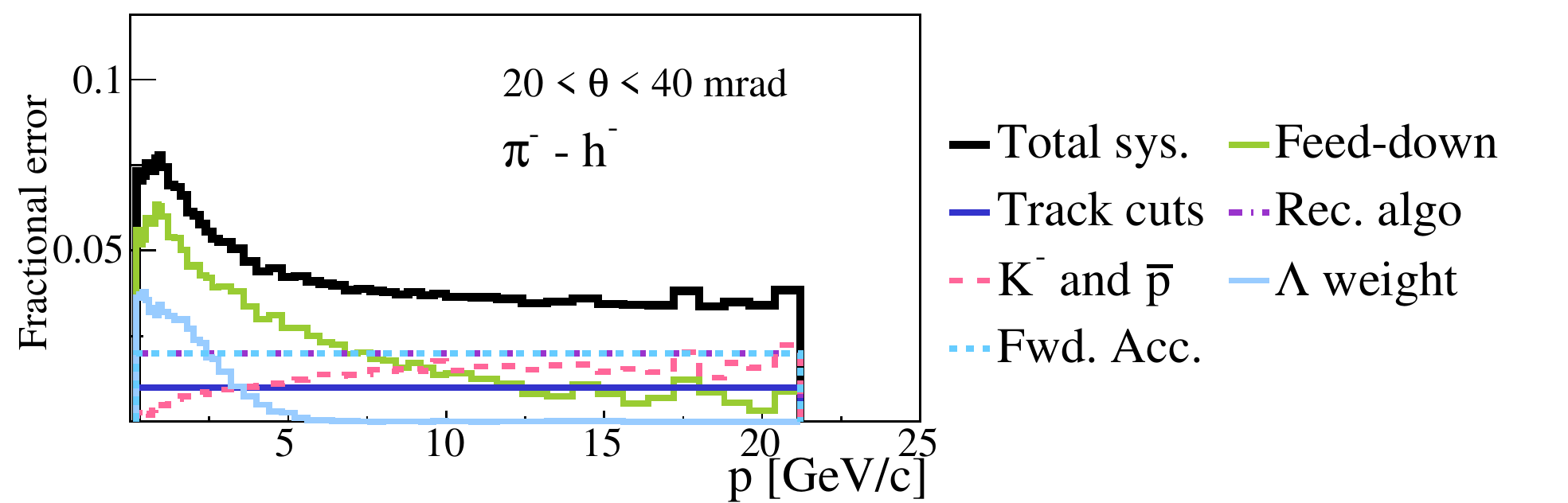}
\caption{
\label{fig:pim_err_breakdown_hminus} 
(Colour online)
Breakdown of $\pi^-$ systematic uncertainties for the $h^-$ analysis, presented as a function 
of momentum for the [20,40]~mrad angular interval as an example.
}
\end{figure*}

Most of the systematic uncertainties were evaluated in the same way as for the other charged hadron spectra analyses.
Error sources which are specific to the  $h^-$  analysis include uncertainties of the contamination 
by $K^-$, $\bar{p}$ and of the electron rejection procedure. 
The following effects were taken into account for the estimate of systematic uncertainties: 
\begin{enumerate}[(i)]

\item {\it Feed-down.}
The uncertainty related to the correction for secondary interactions and for weak decays of strange
particles remains the dominant contribution to the systematic uncertainty at low momentum. For both of these sources
a 30\% error was assigned as described in Ref.~\cite{pion_paper}.  
The significant reduction of the uncertainty, compared to the 2007 analysis, comes from  
the use of new \NASixtyOne measurements of $\Lambda$ production
(see Subsection~\ref{subsec:lambda_reweighting} for more details)
and the removal of electrons at low $\theta$ and low momentum.  
 
\item {\it Forward acceptance.}
The impact of the GTPC detector was evaluated  
following the procedure described in Section~\ref{sec:FA}.
The systematic uncertainty deduced from the difference of
the spectra obtained with the reconstruction algorithms 
with and without use of the GTPC   
was found to be 4\% for $0<\theta<10$\,mrad and 2\% for $10<\theta<40$\,mrad
in the overlapping acceptance regions.  
The variation of the number of hits required for tracks starting in the GTPC  
resulted in a systematic uncertainty of up to 
4-5\% for the most forward region $\theta<20$\,mrad and 
at momenta $p>5$\,GeV/$c$.

 \item {\it $K^-$ and $\bar{p}$ contamination.}
The admixture of $K^-$ and $\bar{p}$ was estimated from
simulations with the \VenusLong model as the primary event generator
and was assigned a systematic uncertainty of 20\%.
The contribution of $K^-$ and $\bar{p}$  varies 
with momentum from  5\% to 10\% at low $\theta$ angles. 
For $\theta>$~100\,mrad it increases with momentum  
and varies from 5\% to  20\%. 
 
 \item {\it Reconstruction efficiency.}
The  reconstruction efficiency was evaluated by comparing results 
for different reconstruction algorithms used in the experiment.
Differences of about 2\% were found. Since no  dependence was 
observed on momentum and angle $\theta$ this value was assigned 
to the systematic uncertainty.

  \item {\it Track selection method.}
The  uncertainty was estimated by varying the following selections:
required minimum total number of points on the track, $\dd E/\dd x$ cuts for electron removal, and 
maximally allowed impact parameter. 
The requirement on the number of points on the track was changed by 10\% 
from the nominal value of 30 and the corresponding uncertainty was found to be up to 1\%. 
The uncertainty associated with the cut to reject electrons was estimated  by varying  
the  cut value in  $\dd E/\dd x$ for each of 
the $\theta$ bins by $\pm$3\%.  The separation of electrons and mostly pions is 
good, so the impact on the spectra is negligible compared to that of varying the 
requirement on the number of points and no additional uncertainty was assigned.
To test the sensitivity to the impact parameter cut, the analysis was repeated  
without the cut and  the effect was found to be negligible. 
\end{enumerate}

Figure~\ref{fig:pim_err_breakdown_hminus} shows a breakdown of 
the total systematic uncertainty for  $\pi^-$ as a function of momentum 
in the angular interval [20,40]~mrad
as an example.

\section{Results on spectra} \label{sec:Results}

This section presents results on spectra of $\pi^\pm$,  $K^\pm$, and protons, 
as well as $K^0_S$ and $\Lambda$ in p+C interactions at 31\,GeV/$c$.   
The large statistics of the 2009 dataset resulted 
in improved precision 
and larger coverage compared to the published 2007 data
and allowed measurements of $K^-$ spectra in \NASixtyOne for the first time.  
The comparison between the 2009 and 2007 results is
discussed. 

The new measurements  are presented together with selected predictions 
of hadron production models, which are described in Section~\ref{Sec:models}. 
Complete comparisons are shown in Ref.~\cite{CERNpreprint-2015-278}.
In order to avoid uncertainties related to the different treatment
of quasi-elastic
interactions and to the absence of predictions for inclusive cross sections,
spectra are normalized to the mean particle multiplicity
in all production interactions.
For the data, the normalization relies on
the p+C inclusive production cross section $\sigma_\text{prod}$ which
was found to be 230.7\,mb (see Section~\ref{Sec:inelprodxsec}).
The production cross section is calculated
from the inelastic cross section by subtracting the quasi-elastic
contribution. Therefore production processes are defined as those in
which only new hadrons are present in the final state. Details of the
cross section analysis procedure can be found in Section~\ref{Sec:norm}
and in Refs.~\cite{Davide,pion_paper,claudia}.

The experimental measurements are shown with total uncertainties
which correspond to the statistical and systematic
uncertainties added in quadrature. 
The overall uncertainty due to the normalization procedure 
(discussed in Section~\ref{Sec:norma}) is not shown.

\subsection{$\pi^\pm$ results}

\begin{figure*}
\centering
\includegraphics[width=0.9\textwidth]{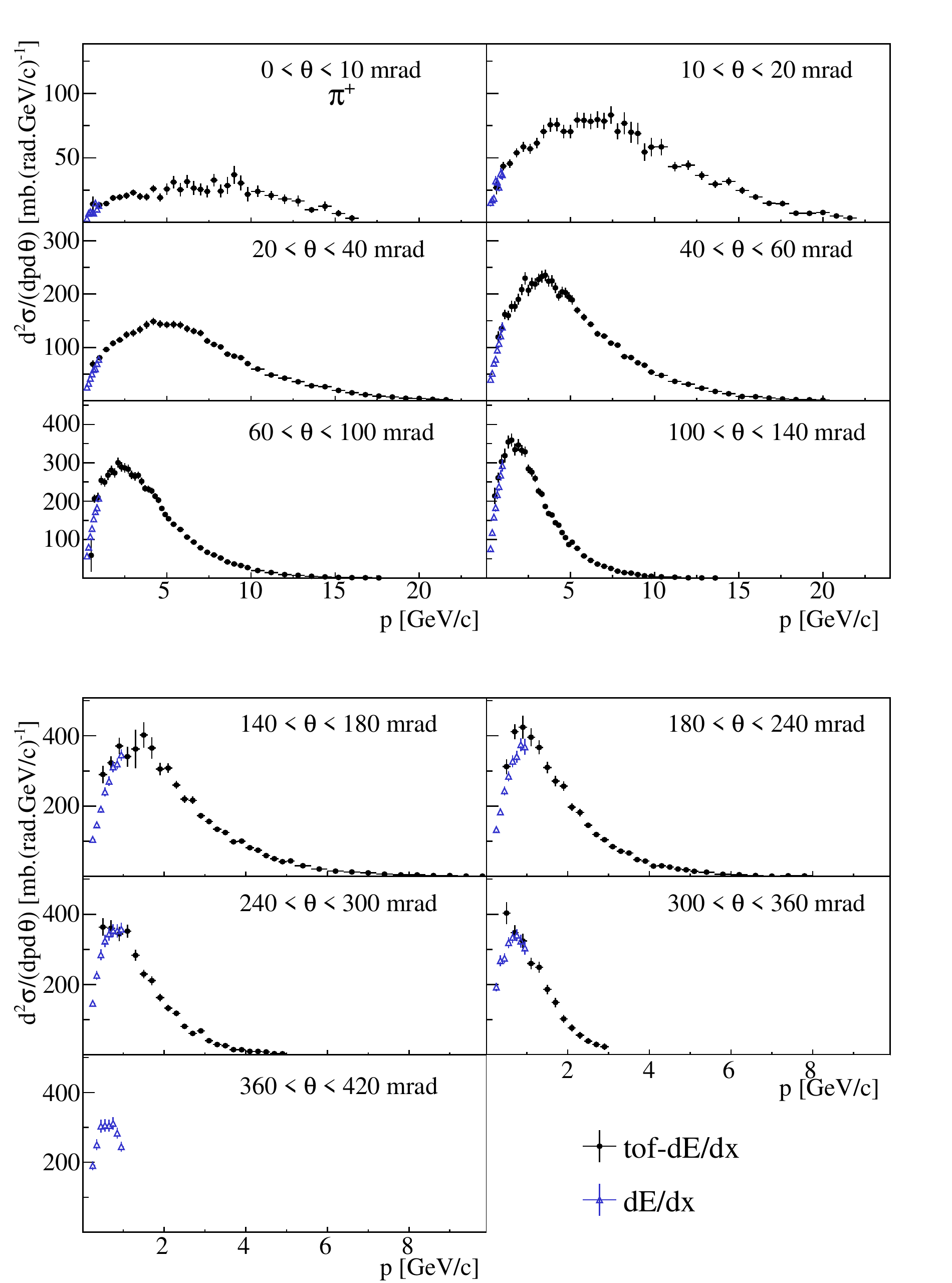}
\caption{\label{fig:pip_multiplicity_ana}
(Colour online)
Laboratory momentum distributions of $\pi^+$ produced in p+C interactions at 31\,GeV/$c$ production processes in different polar angle intervals ($\theta$). Error bars indicate the statistical and systematic uncertainties added in quadrature. The overall uncertainty due to the normalization procedure is not shown. Results obtained with two different analysis techniques are presented: open blue triangles are the $dE/dx$ analysis and full black circles are the $tof$-$dE/dx$ analysis.}
\end{figure*}

\begin{figure*}
\centering
\includegraphics[width=0.9\textwidth]{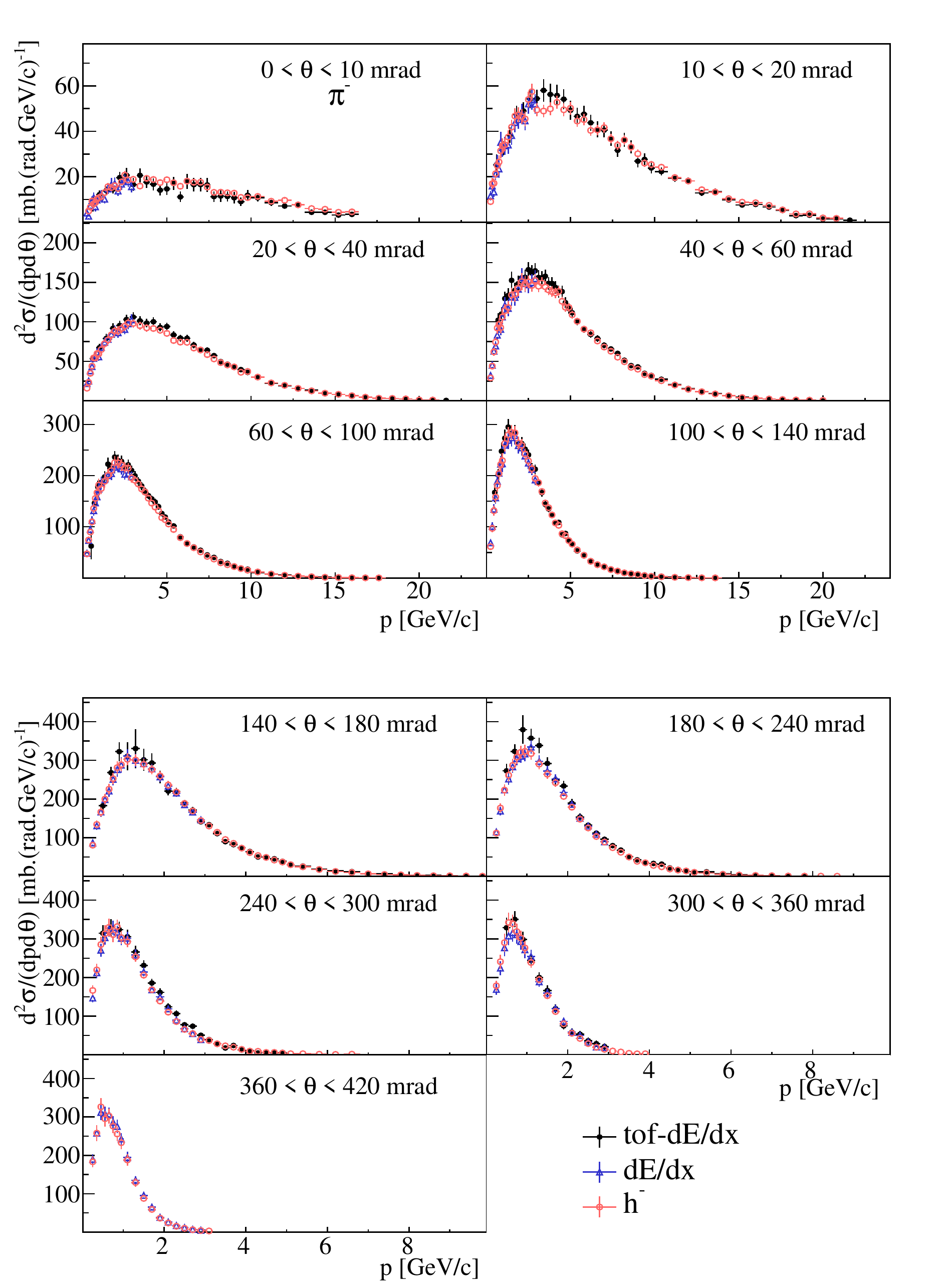}
\caption{\label{fig:pim_multiplicity_ana}
(Colour online)
Laboratory momentum distributions of $\pi^-$ produced in p+C interactions at 31\,GeV/$c$ production processes in different polar angle intervals ($\theta$). Error bars indicate the the statistical and systematic uncertainties added in quadrature. The overall uncertainty due to the normalization procedure is not shown. Results obtained with three different analysis techniques are presented: open blue triangles are the $dE/dx$ analysis, open red circles are the $h^-$ analysis and full black circles are the $tof$-$dE/dx$ analysis.}
\end{figure*}

\begin{figure*}
\centering
\includegraphics[width = 0.9\textwidth]{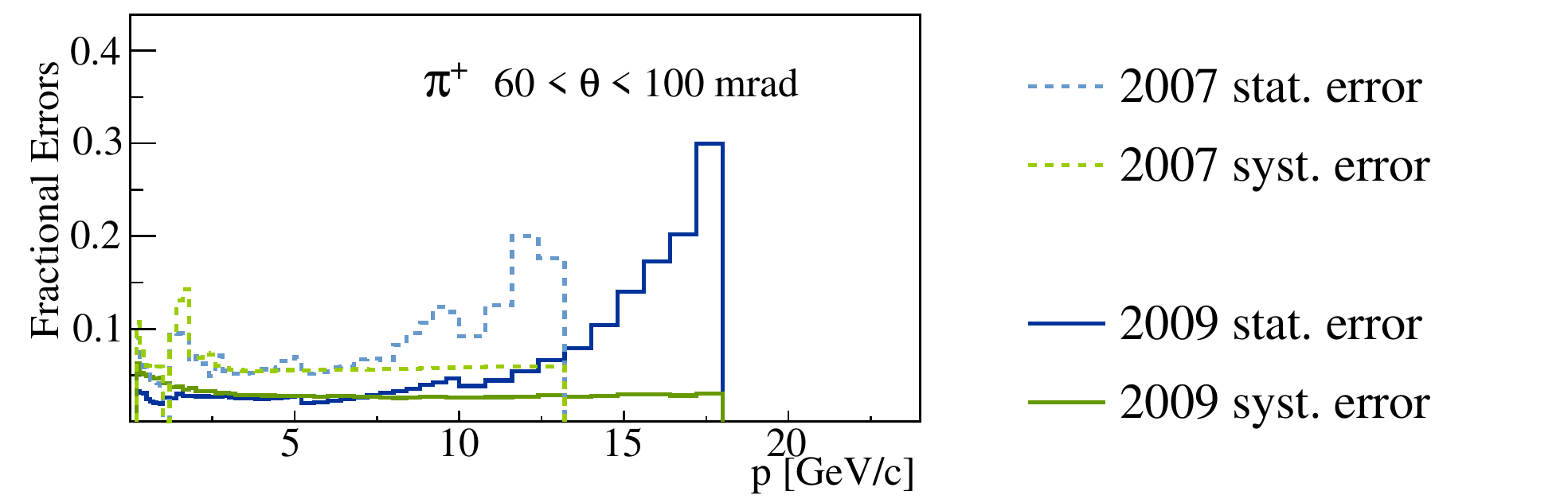}
\caption{\label{fig:07_09_err_pip} 
(Colour online)
Comparison between \NASixtyOne statistical and systematic uncertainties obtained
using the 2007~\cite{pion_paper} and the 2009 datasets 
in a selected angular interval [60,100]~mrad
for $\pi^+$. 
}
\end{figure*}

\begin{figure*}
\centering
\includegraphics[width = 0.9\textwidth]{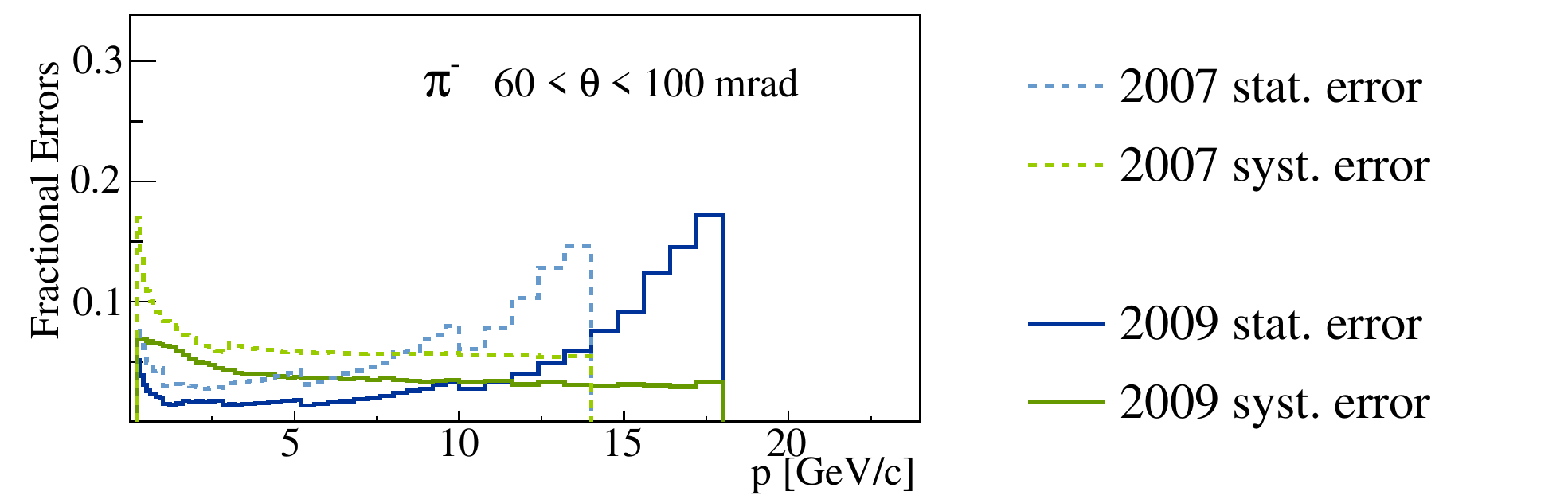}
\caption{\label{fig:07_09_err_pim} 
(Colour online)
Comparison between \NASixtyOne statistical and systematic uncertainties obtained
using the 2007~\cite{pion_paper} and the 2009 datasets
in a selected angular interval [60,100]~mrad 
for $\pi^-$. 
}
\end{figure*}

The spectra of $\pi^\pm$ mesons were obtained using three different
analysis techniques \emph{tof}-$\dd E/\dd x$, $\dd E/\dd x$ and $h^-$ described in 
Sections~\ref{Sec:dedxtof}, \ref{Sec:dedx} and~\ref{Sec:h-}.  
Within the corresponding systematic uncertainties, the results of various  
methods were found to be in a good agreement as shown in 
Fig.~\ref{fig:pip_multiplicity_ana} and Fig.~\ref{fig:pim_multiplicity_ana} 
for $\pi^+$ and $\pi^-$, respectively. In order to present 
a single spectrum for positively and negatively changed pions, the
results were combined. 
For $\pi^+$ mesons, where there is complementarity of acceptance for different analysis techniques 
results from the $\dd E/\dd x$ analysis are used for all angular intervals up to 420~mrad 
in the momentum range below 1~GeV/$c$.
The momentum region above 1~GeV/$c$ is covered by the \emph{tof}-$\dd E/\dd x$ analysis for angular intervals up to 360~mrad. 
In case of $\pi^-$ mesons, results of the $h^-$ analysis are used in the full angular range up to 420~mrad since they provide the smallest total uncertainty 
in the region of the overlap between methods. 
The final spectra are shown for  $\pi^+$ in Fig.~\ref{fig:pip_data_vs_models} 
and for $\pi^-$ in Fig.~\ref{fig:pim_data_vs_models}. 
Numerical results are given in Tables~\ref{tab:pi+_results} 
and~\ref{tab:pi-_results} for $\pi^+$ and $\pi^-$, respectively. 
Thus,
the final results span a broad kinematic range. 
When comparing the new $\pi^\pm$ results to the previously published 
measurements based on the 2007 data~\cite{pion_paper} 
one should note that the present results are shown in the form 
of $d^2 \sigma /(d p d \theta)$,
while the form $d \sigma /(d p)$ was used in previous 
publications~Refs.~\cite{pion_paper,kaon_paper}. 
The deviations between the two data sets are consistent within the quoted errors 
and are distributed uniformly over the phase space.
Thus, the new measurements confirm the published 2007 results
which were questioned in Ref.~\cite{Chvala_Fischer}.
Both the statistical and the total systematic uncertainties
are considerably smaller for the new 2009 results as shown 
for the angular interval [60,100]~mrad
for $\pi^+$ in Fig.~\ref{fig:07_09_err_pip}  
and for $\pi^-$ in Fig.~\ref{fig:07_09_err_pim}. 
At low momentum systematic uncertainties dominate whereas at higher momentum 
the statistical uncertainty 
gives the largest contribution to the total uncertainty.

\subsection{$K^\pm$ results}

\begin{figure*}
\centering
\includegraphics[width = 0.9\textwidth]{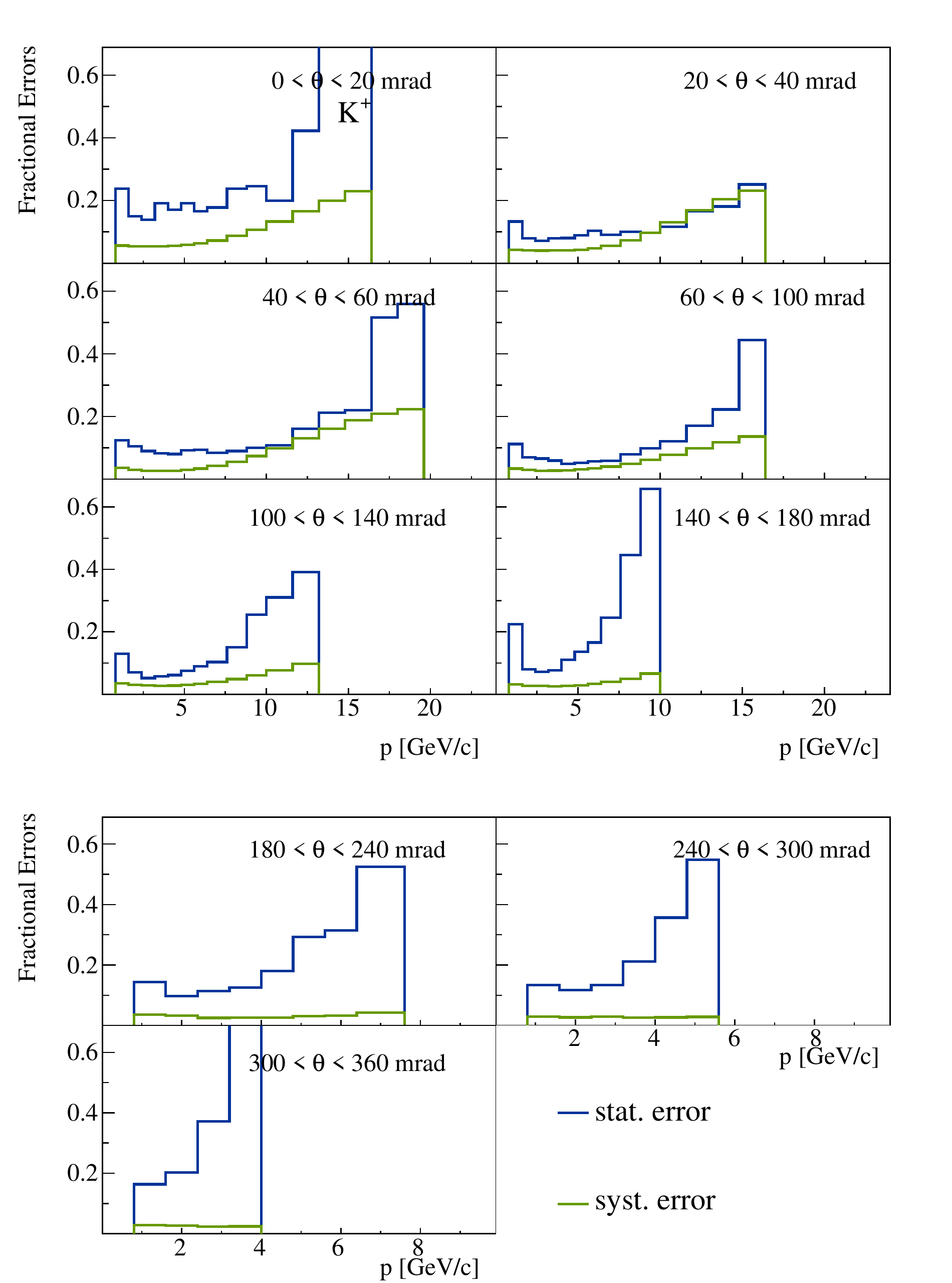}
\caption{\label{fig:09_err_kp} 
(Colour online)
Statistical and systematic uncertainties obtained
using the 2009 dataset for $K^+$. 
}
\end{figure*}

The \emph{tof}-$\dd E/\dd x$ analysis technique was used to  obtain 
the $K^+$ spectra shown in Fig.~\ref{fig:kp_data_vs_models} 
and the $K^-$ spectra plotted in Fig.~\ref{fig:km_data_vs_models}. 
The large 2009 dataset allowed for the first measurements of $K^-$ yields 
in p+C interactions by \NASixtyOne.  
Large statistics for $K^+$ made it possible to use narrow \pth bins 
in the analysis which is 
important for tuning the neutrino flux predictions in T2K. 
Numerical results are given in Tables~\ref{tab:k+_results} 
and~\ref{tab:k-_results} for $K^+$ and $K^-$, respectively. 
The analysis of $K^+$ was repeated, as a cross-check, 
with a coarser \pth binning that corresponds to the previously published
measurements based on the 2007 data~\cite{kaon_paper}. Good agreement 
was found.
The total uncertainty for the 2009 data remains dominated 
by the statistical uncertainty as shown in Fig.~\ref{fig:09_err_kp} 
for $K^+$ as an example.

\subsection{Proton results}

The \emph{tof}-$\dd E/\dd x$ analysis technique was  
used to extract proton spectra. 
These measurements, shown in Fig.~\ref{fig:p_data_vs_models}, 
cover a wide kinematic range above 1~GeV/$c$ and are important  
for T2K since they provide the possibility to constrain the contributions from re-interactions in 
the long carbon target and in the elements of the beamline.
Numerical results are given in Table~\ref{tab:proton_results}. 
The new proton spectra are in a good agreement with preliminary results 
obtained using the 2007 data~\cite{Status_Report_2012}.

\subsection{$V^0$ results}

The $K_S^0$ and $\Lambda$ spectra are shown 
in Fig.~\ref{fig:k0_09DataMixVenusEpos}  
and Fig.~\ref{fig:lambda_data_vs_models} 
and
numerical results are given in Tables~\ref{tab:k0_results} 
and~\ref{tab:l_results}, respectively. 
As explained in Section~\ref{subsec:lambda_reweighting},
those measurements are used
to reduce the dominant systematic
uncertainties due to the feed-down correction in the charged hadron analyses.
Results are consistent within the quoted systematic uncertainties 
with previously published measurements~\cite{V0_2007}
which were obtained with a much coarser \pth binning due to the lower 
statistics of the 2007 data.

The spectra of $K^0_S$ can be cross-checked by measurements of $K^\pm$ yields, 
thanks to the unique capability 
to measure these three types of kaons simultaneously 
in the \NASixtyOne experiment. 
So far, only a few experiments have been able to perform such measurements  
and to test 
two different theoretical hypotheses that predict $K^0_S$ yields 
from $K^\pm$ production rates giving inconclusive results.   
The first approach assumes isospin symmetry in kaon production 
and predicts (see e.g. Ref.~\cite{Gazdzicki:1991ih}):

\begin{equation}
N(K^0_S) = \frac 1 2 \, (N(K^+) + N(K^-))~.
\end{equation}
The second method uses a quark-counting argument~\cite{BMPT}, 
with a simplified quark parton model. 
The following assumptions are made on the number of sea and valence quarks:

\begin{align}
&u_\text{s} =\bar{u}_\text{s} = d_\text{s} 
= \bar{d}_\text{s}, \, s_\text{s} = \bar{s}_\text{s}~,
\\
&n \equiv u_\text{v} / d_\text{v}~.
\end{align}
Taking into account that interactions are with a carbon nucleus 
($n = 2$ for p+p collisions, $n = 1$ for p+n), 
the relation between mean multiplicity  of $K^0_S$, $K^+$ and $K^-$ is:

\begin{equation}
N(K^0_S) = \frac 1 8 \, (3\,N(K^+)+5\,N(K^-))~.
\end{equation}
The second method is currently used to tune $K^0_L$ production in 
the T2K flux simulation chain~\cite{T2Kflux}. 
Figure~\ref{fig:k0_09DataMixVenusEpos} shows a comparison of the measured spectra 
of $K^0_S$ 
to the predictions from  measured charged kaon yields 
and reasonable agreement is observed.  
Unfortunately, the large uncertainties associated with the $K^0_S$ measurements do not allow 
us to discriminate between the two hypotheses.

\section{Comparisons with hadron production models } \label{Sec:models}

\begin{figure*}
\centering
\includegraphics[width=0.9\textwidth]{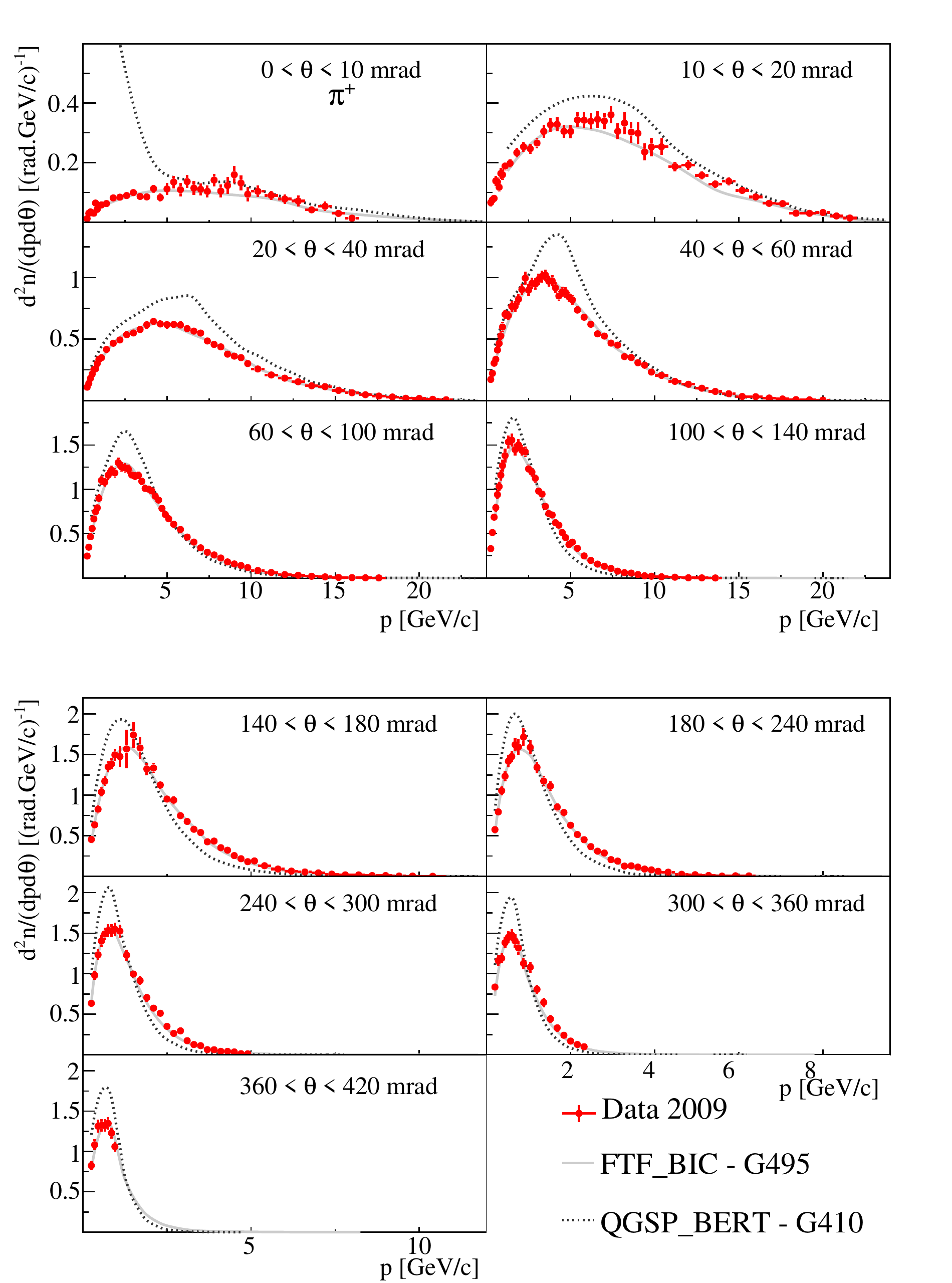}
\caption{
  (Colour online)
  Laboratory momentum distributions of $\pi^+$ mesons
  produced in p+C interactions at 31\,GeV/$c$ in different
  polar angle intervals. Distributions are normalized to
  the mean  $\pi^+$ multiplicity in all production p+C interactions.
  Vertical bars show the statistical and systematic uncertainties
  added in quadrature, horizontal bars indicate the
  size of the momentum bin. The overall uncertainty due to the normalization 
  procedure is not shown. The spectra are compared to
  predictions of the FTF\_BIC-G495 and QGSP\_BERT-G410 models.
  Ref.~\cite{CERNpreprint-2015-278} shows predictions for all models considered in Sec.~\ref{Sec:models}.
}
\label{fig:pip_data_vs_models}
\end{figure*}

\begin{figure*}
\centering
\includegraphics[width=0.9\textwidth]{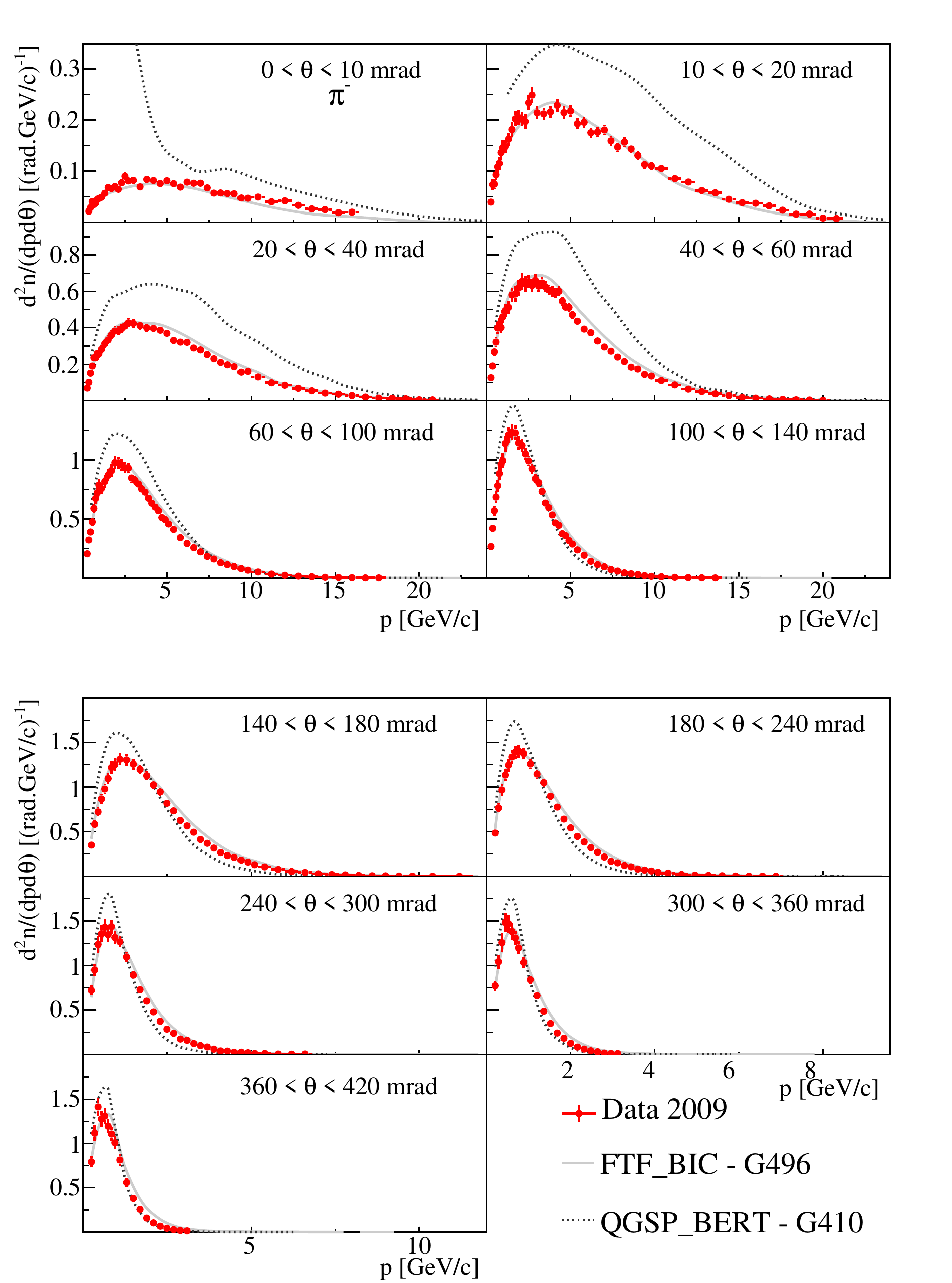}
\caption{
  (Colour online)
  Laboratory momentum distributions of $\pi^-$ mesons
  produced in p+C interactions at 31\,GeV/$c$ in different
  polar angle intervals. Distributions are normalized to
  the mean  $\pi^-$ multiplicity in all production p+C interactions.
  Vertical bars show the statistical and systematic uncertainties
  added in quadrature, horizontal bars indicate the
  size of the momentum bin. The overall uncertainty due to the normalization 
  procedure is not shown. The spectra are compared to
  predictions of the FTF\_BIC-G496 and QGSP\_BERT-G410 models.
  Ref.~\cite{CERNpreprint-2015-278} shows predictions for all models considered in Sec.~\ref{Sec:models}.
}
\label{fig:pim_data_vs_models}
\end{figure*}

\begin{figure*}
\centering
\includegraphics[width=0.9\textwidth]{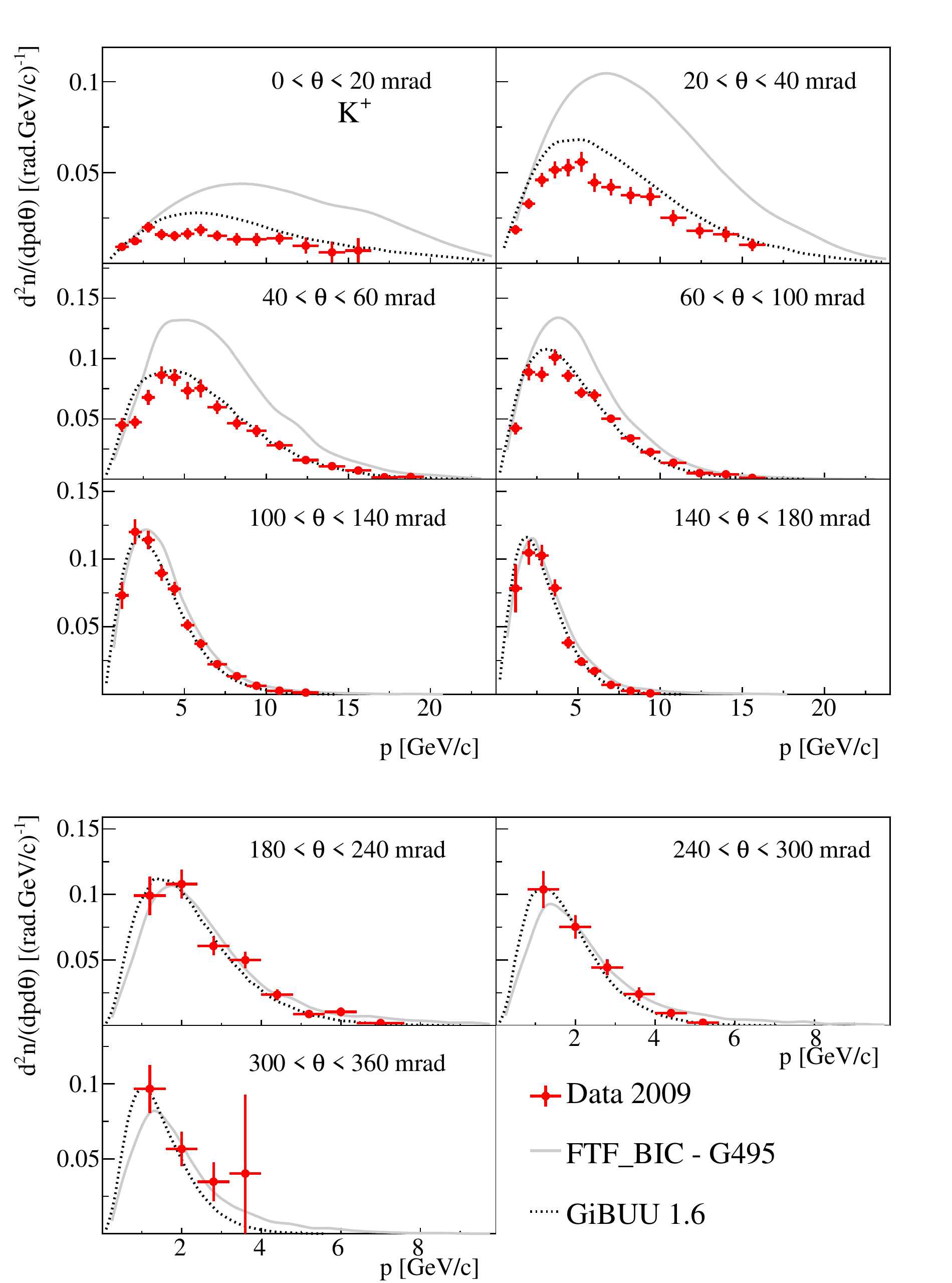}
\caption{
  (Colour online)
  Laboratory momentum distributions of $K^+$ mesons
  produced in p+C interactions at 31\,GeV/$c$ in different
  polar angle intervals. Distributions are normalized to
  the mean  $K^+$  multiplicity in all production p+C interactions.
  Vertical bars show the statistical and systematic uncertainties
  added in quadrature, horizontal bars indicate the
  size of the momentum bin. The overall uncertainty due to the normalization 
  procedure is not shown. The spectra are compared to
  predictions of the FTF\_BIC-G495 and \GiBUULong  models.
  Ref.~\cite{CERNpreprint-2015-278} shows predictions for all models considered in Sec.~\ref{Sec:models}.
}
\label{fig:kp_data_vs_models}
\end{figure*}

\begin{figure*}
\centering
\includegraphics[width=0.9\textwidth]{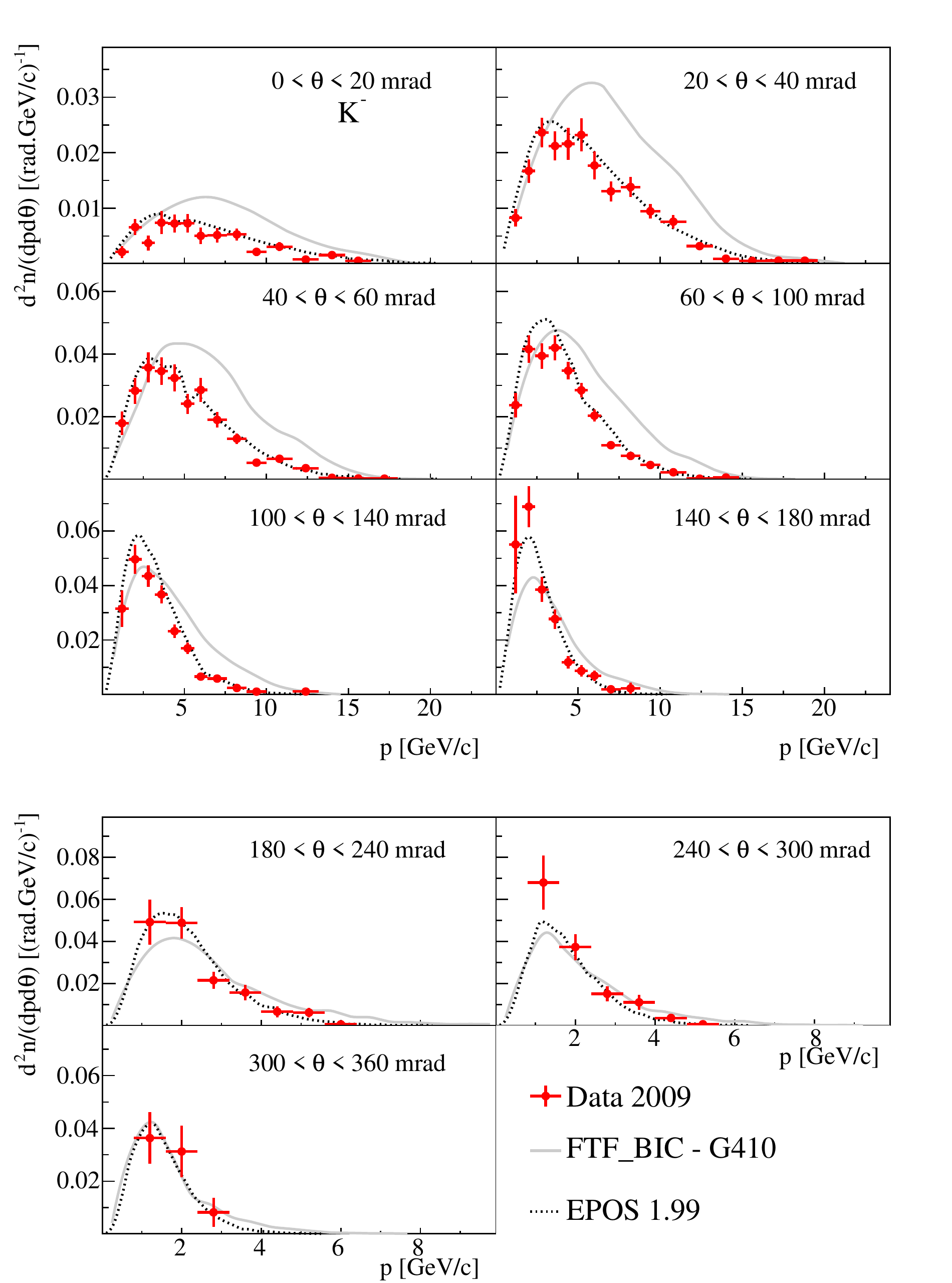}
\caption{
  (Colour online)
  Laboratory momentum distributions of $K^-$ mesons
  produced in p+C interactions at 31\,GeV/$c$ in different
  polar angle intervals. Distributions are normalized to
  the mean  $K^-$  multiplicity in all production p+C interactions.
  Vertical bars show the statistical and systematic uncertainties
  added in quadrature, horizontal bars indicate the
  size of the momentum bin. The overall uncertainty due to the normalization 
  procedure is not shown. The spectra are compared to
  predictions of the FTF\_BIC-G410 and \EposLong models.
  Ref.~\cite{CERNpreprint-2015-278} shows predictions for all models considered in Sec.~\ref{Sec:models}.
}
\label{fig:km_data_vs_models}
\end{figure*}

\begin{figure*}
\centering
\includegraphics[width=0.9\textwidth]{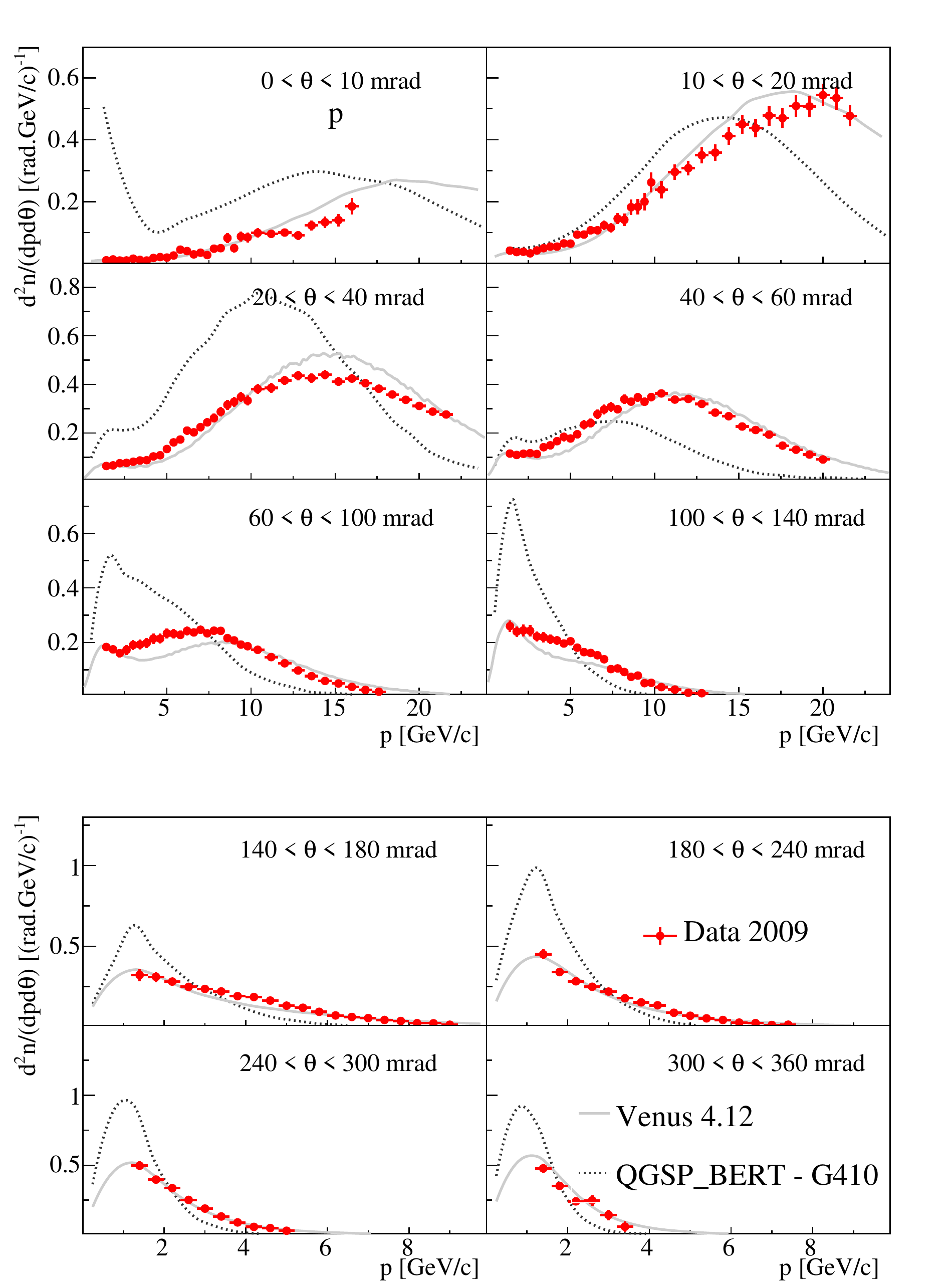}
\caption{
  (Colour online)
  Laboratory momentum distributions of protons
  produced in p+C interactions at 31\,GeV/$c$ in different
  polar angle intervals. Distributions are normalized to
  the mean  proton  multiplicity in all production p+C interactions.
  Vertical bars show the statistical and systematic uncertainties
  added in quadrature, horizontal bars indicate the
  size of the momentum bin. The overall uncertainty due to the normalization 
  procedure is not shown. The spectra are compared to
  predictions of the QGSP\_BERT-G410 and \VenusLong models.
  Ref.~\cite{CERNpreprint-2015-278} shows predictions for all models considered in Sec.~\ref{Sec:models}.
}
\label{fig:p_data_vs_models}
\end{figure*}

\begin{figure*}
\centering
\includegraphics[width=0.9\textwidth]{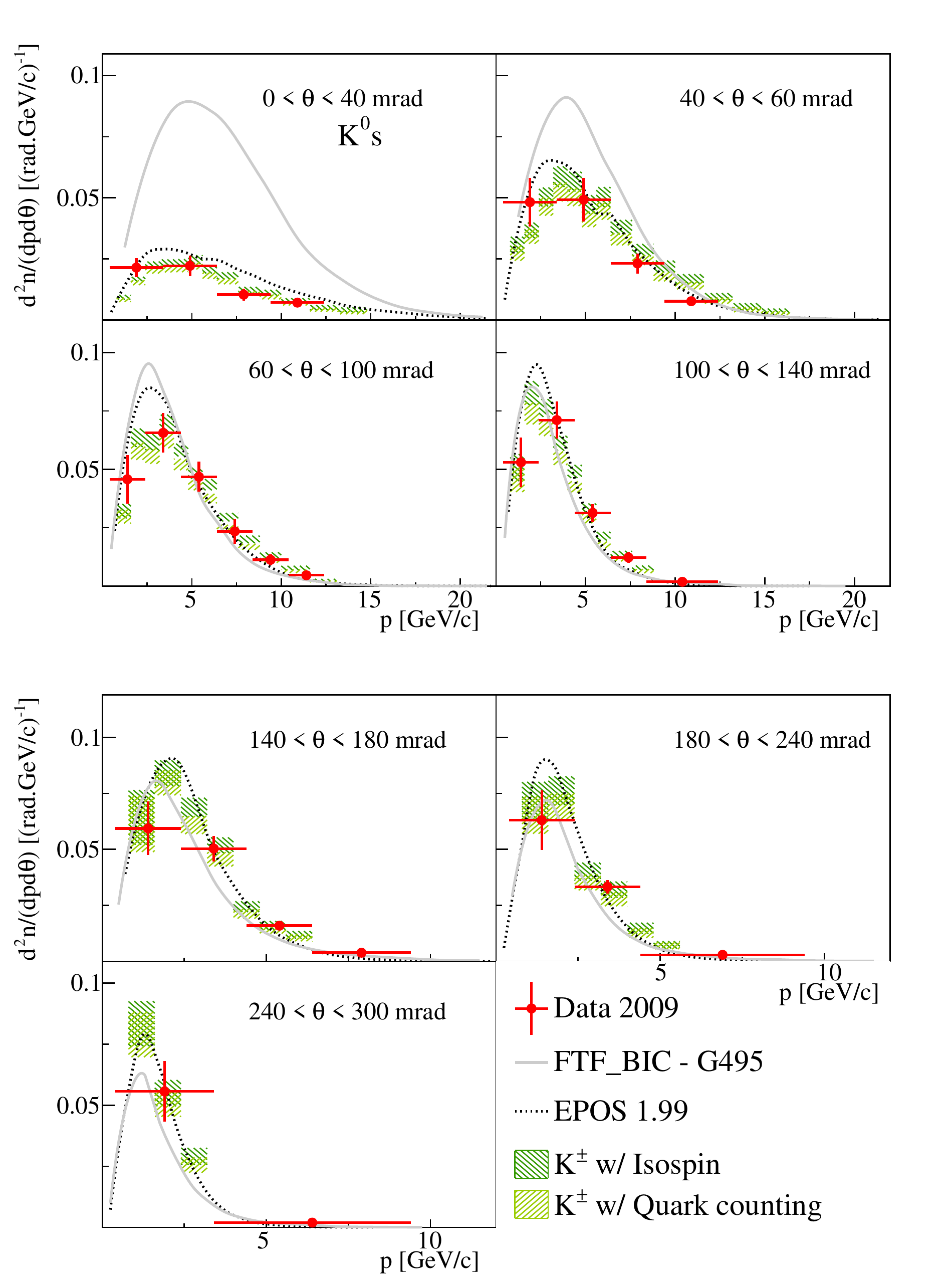}
\caption{
  (Colour online)
  Laboratory momentum distributions of $K^0_S$ mesons
  produced in p+C interactions at 31\,GeV/$c$ in different
  polar angle intervals. Distributions are normalized to
  the mean  $K^0_S$  multiplicity in all production p+C interactions. 
  Vertical bars show the statistical and systematic uncertainties
  added in quadrature, horizontal bars indicate the
  size of the momentum bin. The overall uncertainty due to the normalization 
  procedure is not shown. 
  Shaded boxes show predictions obtained from $K^\pm$ yields
  using isospin (dark green) and quark counting (light green) hypotheses.
  The spectra are compared to predictions of the FTF\_BIC-G495 and \EposLong models.
  Ref.~\cite{CERNpreprint-2015-278} shows predictions for all models considered in Sec.~\ref{Sec:models}.
}
\label{fig:k0_09DataMixVenusEpos}
\end{figure*}

\begin{figure*}
\centering
\includegraphics[width=0.9\textwidth]{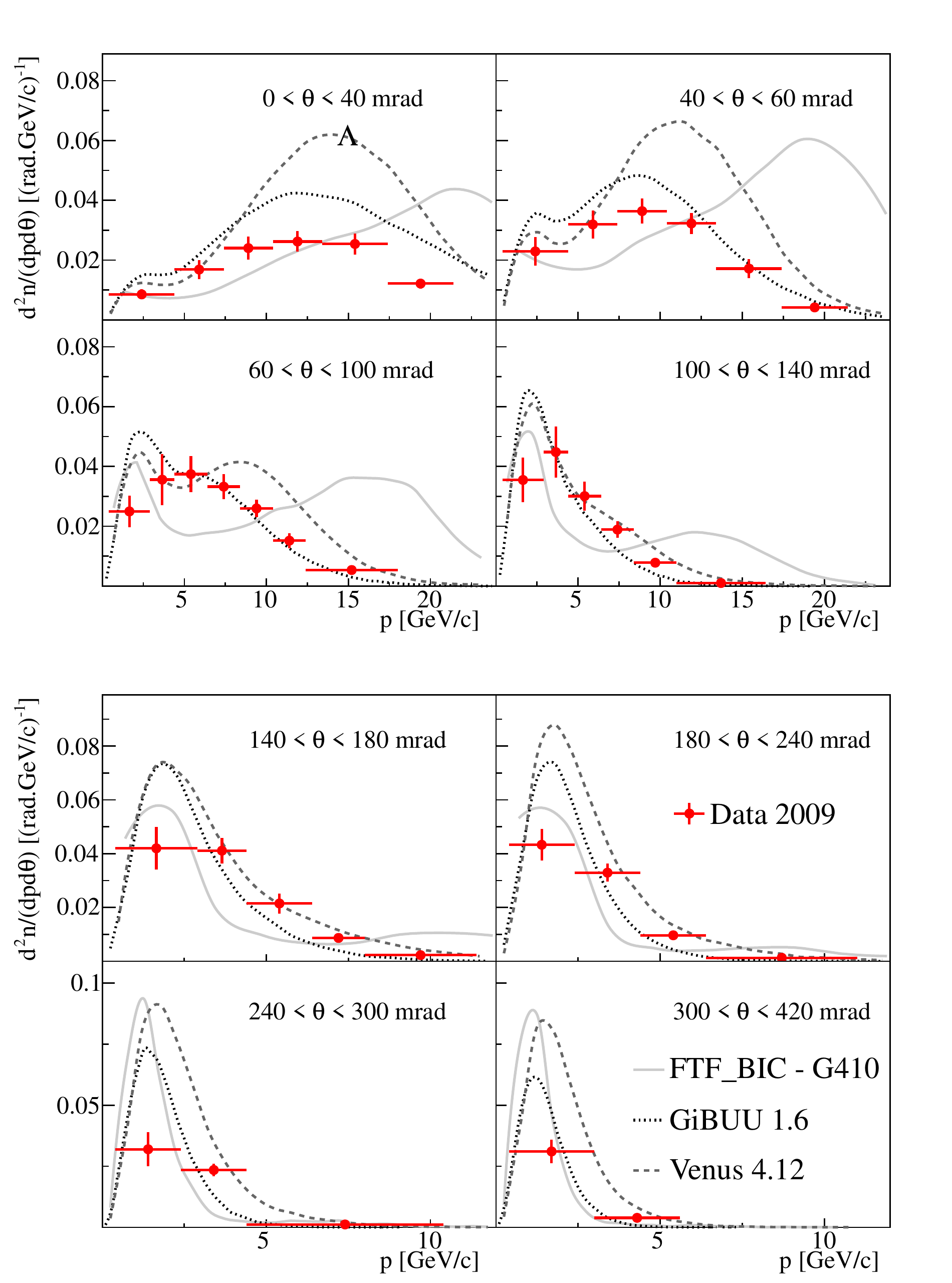}
\caption{
  (Colour online)
  Laboratory momentum distributions of $\Lambda$ hyperons
  produced in p+C interactions at 31\,GeV/$c$ in different
  polar angle intervals. Distributions are normalized to
  the mean  $\Lambda$  multiplicity in all production p+C interactions.
  Vertical bars show the statistical and systematic uncertainties
  added in quadrature, horizontal bars indicate the
  size of the momentum bin. The overall uncertainty due to the normalization 
  procedure is not shown. The spectra are compared to
  predictions of the FTF\_BIC-G410, \GiBUULong and \VenusLong models.
  Ref.~\cite{CERNpreprint-2015-278} shows predictions for all models considered in Sec.~\ref{Sec:models}.
}
\label{fig:lambda_data_vs_models}
\end{figure*}

The spectra shown in Figs.~\ref{fig:pip_data_vs_models}--\ref{fig:lambda_data_vs_models}
are compared to predictions of seven hadronic event generators. These are \\
\VenusLong~\cite{Venus1,Venus2}, \EposLong~\cite{EPOS}~\footnote{Note that the \Epos model is used here
in an energy domain for which it was originally not designed (below 100 GeV in the laboratory).},
\GiBUULong~\cite{GiBUU,Gallmeister}, as well as FTF\_BIC-G495, FTF\_BIC-G496, FTF\_BIC-G410 and QGSP\_BERT physics lists
defined in the GEANT4 toolkit~\footnote{The predictions of the different versions of FTF-based physics lists 
from the consecutive GEANT4 releases 4.9.5, 4.9.6 and 4.10 are significantly different.}.
For clarity of presentation predictions of only two models
are plotted with the spectra of each hadron type. They were selected to show examples of predictions with
small and large deviations from the measurements. The comparison of the measured spectra 
with all model predictions is available in Ref.~\cite{CERNpreprint-2015-278}.

None of the models provides a satisfactory description of all the measured spectra.
The FTF-based physics lists of GEANT4 provide a reasonable 
description of $\pi^\pm$ and $K^+$ spectra, but do not  
reproduce $K^-$ and proton spectra. 
The \EposLong and \GiBUULong models show good agreement with the measured 
$K^\pm$ spectra.   
The best description of the proton spectra is achieved by 
the \VenusLong model which was used to calculate the corrections 
applied to obtain the results presented in this paper.

\clearpage

\clearpage

\clearpage

\clearpage

\clearpage

\clearpage

\clearpage

\section{Summary} \label{Sec:summary}

This paper presents precise measurements of interaction and production cross sections 
as well as of spectra of $\pi^\pm$, $K^\pm$, protons, $K^0_S$ and $\Lambda$
in p+C interactions at 31\,GeV/$c$.
These data are crucial for  predictions of the
initial neutrino fluxes in the T2K long baseline neutrino oscillation
experiment in Japan. 
Furthermore, they provide important input to
improve hadron production models needed for the interpretation of air
showers initiated by ultra high energy cosmic particles.  
The measurements were performed with the large acceptance \NASixtyOne
spectrometer at the CERN SPS.  
A set of data collected with a 4\%~$\lambda_{\mathrm{I}}$ isotropic 
graphite target during the high-statistics \NASixtyOne run in 2009 
was used for the analysis.  
The measured spectra are compared with 
predictions of several hadron production models. None of the
models provides a satisfactory description of all the spectra.

\begin{acknowledgements}

We would like to thank the CERN PH, BE and EN Departments for the
strong support of \NASixtyOne.

This work was supported by
the Hungarian Scientific Research Fund (grants OTKA 68506 and 71989),
the J\'anos Bolyai Research Scholarship of
the Hungarian Academy of Sciences,
the Polish Ministry of Science and Higher Education (grants 667\slash N-CERN\slash2010\slash0, NN\,202\,48\,4339 and NN\,202\,23\,1837),
the Polish National Center for Science (grants~2011\slash03\slash N\slash ST2\slash03691, 2012\slash04\slash M\slash ST2\slash00816 and 
2013\slash11\slash N\slash ST2\slash03879),
the Foundation for Polish Science --- MPD program, co-financed by the European Union within the European Regional Development Fund,
the Federal Agency of Education of the Ministry of Education and Science of the
Russian Federation (SPbSU research grant 11.38.193.2014),
the Russian Academy of Science and the Russian Foundation for Basic Research (grants 08-02-00018, 09-02-00664 and 12-02-91503-CERN),
the Ministry of Education, Culture, Sports, Science and Tech\-no\-lo\-gy, Japan, Grant-in-Aid for Sci\-en\-ti\-fic Research (grants 18071005, 19034011, 19740162, 20740160 and 20039012),
the German Research Foundation (grant GA\,1480/2-2),
the U. S. Department of Energy,
the EU-funded Marie Curie Outgoing Fellowship,
Grant PIOF-GA-2013-624803,
the Bulgarian Nuclear Regulatory Agency and the Joint Institute for
Nuclear Research, Dubna (bilateral contract No. 4418-1-15\slash 17),
Ministry of Education and Science of the Republic of Serbia (grant OI171002),
Swiss Nationalfonds Foundation (grant 200020\-117913/1)
and ETH Research Grant TH-01\,07-3.

\end{acknowledgements}

\bibliographystyle{spphys}

\bibliography{submit}

\clearpage
\appendix

\begin{table}[h]
\begin{tabular}{rcrrcrrrr}
\toprule
\multicolumn{3}{c}{$\theta$} & \multicolumn{3}{c}{$p$} &  $\frac{\dd^{2}\sigma}{\dd p \dd\theta}(\pi^{+})$ & $\Delta_\text{stat}$  & $\Delta_\text{syst}$  \\
\midrule
\\[2mm]
0 & - & 10 & 0.2 & - & 0.3 & 2.8 & 1.4 & 0.3 \\
 & & & 0.3 & - & 0.4 & 7.0 & 1.3 & 0.6 \\
 & & & 0.4 & - & 0.5 & 8.1 & 1.5 & 0.6 \\
 & & & 0.5 & - & 0.6 & 7.3 & 1.2 & 0.6 \\
 & & & 0.6 & - & 0.7 & 7.1 & 1.4 & 0.6 \\
 & & & 0.7 & - & 0.8 & 14.9 & 1.9 & 1.2 \\
\bottomrule
\end{tabular}
\end{table}

\begin{table}[h]
\begin{tabular}{rcrrcrrrr}
\toprule
\multicolumn{3}{c}{$\theta$} & \multicolumn{3}{c}{$p$} &  $\frac{\dd^{2}\sigma}{\dd p \dd\theta}(\pi^{+})$ & $\Delta_\text{stat}$ & $\Delta_\text{syst}$ \\
\midrule
\\[2mm]
10 & - & 20 & 0.2 & - & 0.3 & 15.3 & 2.4 & 1.6 \\
 & & & 0.3 & - & 0.4 & 17.4 & 1.9 & 1.4 \\
 & & & 0.4 & - & 0.5 & 18.5 & 2.1 & 1.5 \\
 & & & 0.5 & - & 0.6 & 32.1 & 2.9 & 2.4 \\
 & & & 0.6 & - & 0.7 & 31.2 & 2.7 & 2.4 \\
 & & & 0.7 & - & 0.8 & 27.1 & 2.6 & 2.1 \\
\bottomrule
\end{tabular}
\end{table}

\begin{table}[h]
\begin{tabular}{rcrrcrrrr}
\toprule
\multicolumn{3}{c}{$\theta$} & \multicolumn{3}{c}{$p$} &  $\frac{\dd^{2}\sigma}{\dd p \dd\theta}(\pi^{+})$ & $\Delta_\text{stat}$ & $\Delta_\text{syst}$  \\
\midrule
\\[2mm]
20 & - & 40 & 0.2 & - & 0.3 & 25.5 & 2.0 & 2.0 \\
 & & & 0.3 & - & 0.4 & 32.6 & 2.0 & 2.0 \\
 & & & 0.4 & - & 0.5 & 41.9 & 2.3 & 2.6 \\
 & & & 0.5 & - & 0.6 & 50.1 & 2.4 & 3.1 \\
 & & & 0.6 & - & 0.7 & 58.5 & 2.6 & 3.6 \\
 & & & 0.7 & - & 0.8 & 60.0 & 2.6 & 3.6 \\
\bottomrule
\end{tabular}
\end{table}

\begin{table}[h]
\begin{tabular}{rcrrcrrrr}
\toprule
\multicolumn{3}{c}{$\theta$} & \multicolumn{3}{c}{$p$} &  $\frac{\dd^{2}\sigma}{\dd p \dd\theta}(\pi^{+})$ & $\Delta_\text{stat}$ & $\Delta_\text{syst}$ \\
\midrule
\\[2mm]
40 & - & 60 & 0.2 & - & 0.3 & 39.9 & 2.2 & 3.2 \\
 & & & 0.3 & - & 0.4 & 51.3 & 2.6 & 3.4 \\
 & & & 0.4 & - & 0.5 & 70.3 & 3.2 & 4.6 \\
 & & & 0.5 & - & 0.6 & 77.6 & 2.9 & 5.1 \\
 & & & 0.6 & - & 0.7 & 95.0 & 3.2 & 6.2 \\
 & & & 0.7 & - & 0.8 & 107.3 & 3.5 & 6.8 \\
\bottomrule
\end{tabular}
\end{table}

\begin{table}[h]
\begin{tabular}{rcrrcrrrr}
\toprule
\multicolumn{3}{c}{$\theta$} & \multicolumn{3}{c}{$p$} &  $\frac{\dd^{2}\sigma}{\dd p \dd\theta}(\pi^{+})$ & $\Delta_\text{stat}$ & $\Delta_\text{syst}$ \\
\midrule
\\[2mm]
60 & - & 100 & 0.2 & - & 0.3 & 57.2 & 1.9 & 3.6 \\
 & & & 0.3 & - & 0.4 & 80.3 & 2.5 & 4.2 \\
 & & & 0.4 & - & 0.5 & 107.6 & 3.3 & 5.5 \\
 & & & 0.5 & - & 0.6 & 128.8 & 3.2 & 6.3 \\
 & & & 0.6 & - & 0.7 & 153.8 & 3.3 & 7.4 \\
 & & & 0.7 & - & 0.8 & 172.8 & 3.5 & 8.1 \\
\bottomrule
\end{tabular}
\end{table}

\begin{table}[h]
\begin{tabular}{rcrrcrrrr}
\toprule
\multicolumn{3}{c}{$\theta$} & \multicolumn{3}{c}{$p$} &  $\frac{\dd^{2}\sigma}{\dd p \dd\theta}(\pi^{+})$ & $\Delta_\text{stat}$ & $\Delta_\text{syst}$ \\
\midrule
\\[2mm]
100 & - & 140 & 0.2 & - & 0.3 & 76.5 & 2.9 & 4.9 \\
 & & & 0.3 & - & 0.4 & 118.4 & 3.9 & 6.5 \\
 & & & 0.4 & - & 0.5 & 158.5 & 4.5 & 8.5 \\
 & & & 0.5 & - & 0.6 & 183.3 & 4.0 & 9.4 \\
 & & & 0.6 & - & 0.7 & 217.2 & 4.4 & 10.7 \\
 & & & 0.7 & - & 0.8 & 238.1 & 4.4 & 11.5 \\
\bottomrule
\end{tabular}
\end{table}

\begin{table}[h]
\begin{tabular}{rcrrcrrrr}
\toprule
\multicolumn{3}{c}{$\theta$} & \multicolumn{3}{c}{$p$} &  $\frac{\dd^{2}\sigma}{\dd p \dd\theta}(\pi^{+})$ & $\Delta_\text{stat}$ & $\Delta_\text{syst}$ \\
\midrule
\\[2mm]
140 & - & 180 & 0.2 & - & 0.3 & 105.0 & 4.2 & 6.3 \\
 & & & 0.3 & - & 0.4 & 146.5 & 4.7 & 7.5 \\
 & & & 0.4 & - & 0.5 & 190.9 & 4.2 & 9.4 \\
 & & & 0.5 & - & 0.6 & 240.2 & 4.5 & 12.0 \\
 & & & 0.6 & - & 0.7 & 270.5 & 4.8 & 12.7 \\
 & & & 0.7 & - & 0.8 & 311.2 & 5.1 & 14.9 \\
\bottomrule
\end{tabular}
\end{table}

\begin{table}[h]
\begin{tabular}{rcrrcrrrr}
\toprule
\multicolumn{3}{c}{$\theta$} & \multicolumn{3}{c}{$p$} &  $\frac{\dd^{2}\sigma}{\dd p \dd\theta}(\pi^{+})$ & $\Delta_\text{stat}$ & $\Delta_\text{syst}$ \\
\midrule
\\[2mm]
180 & - & 240 & 0.2 & - & 0.3 & 133.0 & 3.2 & 7.6 \\
 & & & 0.3 & - & 0.4 & 183.3 & 3.4 & 9.5 \\
 & & & 0.4 & - & 0.5 & 243.5 & 3.7 & 12.5 \\
 & & & 0.5 & - & 0.6 & 284.7 & 4.0 & 14.2 \\
 & & & 0.6 & - & 0.7 & 327.3 & 4.3 & 16.3 \\
 & & & 0.7 & - & 0.8 & 340.7 & 4.3 & 16.3 \\
\bottomrule
\end{tabular}
\end{table}

\begin{table}[h]
\begin{tabular}{rcrrcrrrr}
\toprule
\multicolumn{3}{c}{$\theta$} & \multicolumn{3}{c}{$p$} &  $\frac{\dd^{2}\sigma}{\dd p \dd\theta}(\pi^{+})$ & $\Delta_\text{stat}$ & $\Delta_\text{syst}$ \\
\midrule
\\[2mm]
300 & - & 360 & 0.2 & - & 0.3 & 192.7 & 6.2 & 10.5 \\
 & & & 0.3 & - & 0.4 & 267.9 & 7.1 & 13.5 \\
 & & & 0.4 & - & 0.5 & 274.7 & 5.0 & 13.8 \\
 & & & 0.5 & - & 0.6 & 319.1 & 5.4 & 15.4 \\
 & & & 0.6 & - & 0.7 & 334.3 & 5.5 & 16.0 \\
 & & & 0.7 & - & 0.8 & 340.2 & 5.6 & 16.3 \\
 & & & 0.8 & - & 0.9 & 323.5 & 5.5 & 18.1 \\
 & & & 0.9 & - & 1.0 & 303.9 & 5.4 & 18.6 \\
 & & & 1.0 & - & 1.2 & 259.8 & 12.8 & 9.8 \\
 & & & 1.2 & - & 1.4 & 249.0 & 13.5 & 8.6 \\
 & & & 1.4 & - & 1.6 & 186.2 & 12.3 & 5.9 \\
 & & & 1.6 & - & 1.8 & 149.2 & 12.5 & 5.0 \\
 & & & 1.8 & - & 2.0 & 102.3 & 10.4 & 3.8 \\
 & & & 2.0 & - & 2.2 & 76.7 & 9.8 & 3.2 \\
 & & & 2.2 & - & 2.4 & 55.6 & 9.7 & 2.0 \\
 & & & 2.4 & - & 2.6 & 39.2 & 8.2 & 1.6 \\
 & & & 2.6 & - & 2.8 & 29.4 & 8.2 & 1.4 \\
 & & & 2.8 & - & 3.0 & 23.0 & 8.9 & 2.0 \\
\\[2mm]
360 & - & 420 & 0.2 & - & 0.3 & 190.5 & 6.5 & 11.1 \\
 & & & 0.3 & - & 0.4 & 250.0 & 7.2 & 14.0 \\
 & & & 0.4 & - & 0.5 & 303.2 & 8.0 & 16.6 \\
 & & & 0.5 & - & 0.6 & 305.5 & 8.0 & 16.3 \\
 & & & 0.6 & - & 0.7 & 305.0 & 8.1 & 15.9 \\
 & & & 0.7 & - & 0.8 & 311.3 & 8.4 & 15.7 \\
 & & & 0.8 & - & 0.9 & 283.4 & 8.2 & 14.2 \\
 & & & 0.9 & - & 1.0 & 244.7 & 7.5 & 11.9 \\
\bottomrule
\end{tabular}
\caption{\label{tab:pi+_results} The \NASixtyOne results on the $\pi^+$ double differential cross section, $d^2 \sigma_{\pi^+}/(d p d \theta)$ (in [mb/rad/(GeV/$c$)]), in the laboratory system for p+C interactions at 31\,GeV$/c$. The results are presented as a function of momentum, $p$ (in [GeV/$c$]), in different angular intervals, $\theta$ (in [mrad]). The statistical $\Delta_\text{stat}$ (in [mb/rad/(GeV/$c$)]) and systematic $\Delta_\text{syst}$ (in [mb/rad/(GeV/$c$)]) errors are quoted. The overall uncertainty due to the normalization procedure is not included.}
\end{table}

\clearpage
\begin{table}[h]
\begin{tabular}{rcrrcrrrr}
\toprule
\multicolumn{3}{c}{$\theta$} & \multicolumn{3}{c}{$p$} &  $\frac{\dd^{2}\sigma}{\dd p \dd\theta}(\pi^{-})$ & $\Delta_\text{stat}$ & $\Delta_\text{syst}$  \\
\midrule
\\[2mm]
0 & - & 10 & 0.3 & - & 0.4 & 5.0 & 0.8 & 0.4 \\
 & & & 0.4 & - & 0.5 & 6.6 & 1.0 & 0.6 \\
 & & & 0.5 & - & 0.6 & 9.3 & 1.3 & 0.7 \\
 & & & 0.6 & - & 0.7 & 8.0 & 1.1 & 0.6 \\
 & & & 0.7 & - & 0.8 & 8.7 & 1.0 & 0.7 \\
 & & & 0.8 & - & 0.9 & 10.2 & 1.2 & 0.9 \\
\bottomrule
\end{tabular}
\end{table}

\begin{table}[h]
\begin{tabular}{rcrrcrrrr}
\toprule
\multicolumn{3}{c}{$\theta$} & \multicolumn{3}{c}{$p$} &  $\frac{\dd^{2}\sigma}{\dd p \dd\theta}(\pi^{-})$ & $\Delta_\text{stat}$ & $\Delta_\text{syst}$ \\
\midrule
\\[2mm]
10 & - & 20 & 0.2 & - & 0.3 & 9.1 & 1.2 & 0.7 \\
 & & & 0.3 & - & 0.4 & 17.0 & 2.2 & 1.2 \\
 & & & 0.4 & - & 0.5 & 17.2 & 1.7 & 1.2 \\
 & & & 0.5 & - & 0.6 & 21.4 & 1.7 & 1.4 \\
 & & & 0.6 & - & 0.7 & 25.0 & 1.9 & 2.0 \\
 & & & 0.7 & - & 0.8 & 26.5 & 1.9 & 2.1 \\
\bottomrule
\end{tabular}
\end{table}

\begin{table}[h]
\begin{tabular}{rcrrcrrrr}
\toprule
\multicolumn{3}{c}{$\theta$} & \multicolumn{3}{c}{$p$} &  $\frac{\dd^{2}\sigma}{\dd p \dd\theta}(\pi^{-})$ & $\Delta_\text{stat}$ & $\Delta_\text{syst}$ \\
\midrule
\\[2mm]
20 & - & 40 & 0.2 & - & 0.3 & 15.9 & 1.1 & 1.2 \\
 & & & 0.3 & - & 0.4 & 23.2 & 1.7 & 1.6 \\
 & & & 0.4 & - & 0.5 & 34.8 & 1.8 & 2.5 \\
 & & & 0.5 & - & 0.6 & 43.8 & 1.8 & 3.3 \\
 & & & 0.6 & - & 0.7 & 54.3 & 2.0 & 4.0 \\
 & & & 0.7 & - & 0.8 & 54.2 & 1.9 & 4.0 \\
\bottomrule
\end{tabular}
\end{table}

\begin{table}[h]
\begin{tabular}{rcrrcrrrr}
\toprule
\multicolumn{3}{c}{$\theta$} & \multicolumn{3}{c}{$p$} &  $\frac{\dd^{2}\sigma}{\dd p \dd\theta}(\pi^{-})$ & $\Delta_\text{stat}$ & $\Delta_\text{syst}$ \\
\midrule
\\[2mm]
40 & - & 60 & 0.2 & - & 0.3 & 29.0 & 2.3 & 2.4 \\
 & & & 0.3 & - & 0.4 & 43.9 & 2.8 & 3.0 \\
 & & & 0.4 & - & 0.5 & 61.9 & 2.8 & 4.1 \\
 & & & 0.5 & - & 0.6 & 74.3 & 2.8 & 5.3 \\
 & & & 0.6 & - & 0.7 & 92.1 & 3.0 & 6.6 \\
 & & & 0.7 & - & 0.8 & 96.8 & 3.0 & 6.9 \\
\bottomrule
\end{tabular}
\end{table}

\begin{table}[h]
\begin{tabular}{rcrrcrrrr}
\toprule
\multicolumn{3}{c}{$\theta$} & \multicolumn{3}{c}{$p$} &  $\frac{\dd^{2}\sigma}{\dd p \dd\theta}(\pi^{-})$ & $\Delta_\text{stat}$ & $\Delta_\text{syst}$ \\
\midrule
\\[2mm]
60 & - & 100 & 0.2 & - & 0.3 & 47.2 & 2.4 & 3.4 \\
 & & & 0.3 & - & 0.4 & 74.2 & 2.8 & 5.1 \\
 & & & 0.4 & - & 0.5 & 89.8 & 2.8 & 6.1 \\
 & & & 0.5 & - & 0.6 & 109.5 & 2.8 & 7.1 \\
 & & & 0.6 & - & 0.7 & 135.9 & 3.1 & 9.2 \\
 & & & 0.7 & - & 0.8 & 155.5 & 3.6 & 10.2 \\
\bottomrule
\end{tabular}
\end{table}

\begin{table}[h]
\begin{tabular}{rcrrcrrrr}
\toprule
\multicolumn{3}{c}{$\theta$} & \multicolumn{3}{c}{$p$} &  $\frac{\dd^{2}\sigma}{\dd p \dd\theta}(\pi^{-})$ & $\Delta_\text{stat}$ & $\Delta_\text{syst}$ \\
\midrule
\\[2mm]
100 & - & 140 & 0.2 & - & 0.3 & 61.3 & 3.0 & 4.2 \\
 & & & 0.3 & - & 0.4 & 96.9 & 3.5 & 6.5 \\
 & & & 0.4 & - & 0.5 & 131.4 & 3.4 & 9.0 \\
 & & & 0.5 & - & 0.6 & 158.2 & 3.5 & 10.6 \\
 & & & 0.6 & - & 0.7 & 180.5 & 3.6 & 11.1 \\
 & & & 0.7 & - & 0.8 & 204.1 & 3.9 & 12.8 \\
\bottomrule
\end{tabular}
\end{table}

\begin{table}[h]
\begin{tabular}{rcrrcrrrr}
\toprule
\multicolumn{3}{c}{$\theta$} & \multicolumn{3}{c}{$p$} &  $\frac{\dd^{2}\sigma}{\dd p \dd\theta}(\pi^{-})$ & $\Delta_\text{stat}$ & $\Delta_\text{syst}$ \\
\midrule
\\[2mm]
140 & - & 180 & 0.2 & - & 0.3 & 80.6 & 3.4 & 5.7 \\
 & & & 0.3 & - & 0.4 & 134.3 & 3.6 & 9.4 \\
 & & & 0.4 & - & 0.5 & 166.3 & 3.6 & 11.1 \\
 & & & 0.5 & - & 0.6 & 199.9 & 3.8 & 12.9 \\
 & & & 0.6 & - & 0.7 & 225.5 & 3.9 & 13.4 \\
 & & & 0.7 & - & 0.8 & 252.6 & 4.2 & 14.5 \\
\bottomrule
\end{tabular}
\end{table}

\begin{table}[h]
\begin{tabular}{rcrrcrrrr}
\toprule
\multicolumn{3}{c}{$\theta$} & \multicolumn{3}{c}{$p$} &  $\frac{\dd^{2}\sigma}{\dd p \dd\theta}(\pi^{-})$ & $\Delta_\text{stat}$ & $\Delta_\text{syst}$ \\
\midrule
\\[2mm]
180 & - & 240 & 0.2 & - & 0.3 & 111.8 & 2.6 & 7.8 \\
 & & & 0.3 & - & 0.4 & 176.8 & 3.0 & 11.5 \\
 & & & 0.4 & - & 0.5 & 223.2 & 3.3 & 14.3 \\
 & & & 0.5 & - & 0.6 & 261.7 & 3.5 & 15.9 \\
 & & & 0.6 & - & 0.7 & 287.1 & 3.7 & 16.9 \\
 & & & 0.7 & - & 0.8 & 308.5 & 3.9 & 17.0 \\
\bottomrule
\end{tabular}
\end{table}

\begin{table}[h]
\begin{tabular}{rcrrcrrrr}
\toprule
\multicolumn{3}{c}{$\theta$} & \multicolumn{3}{c}{$p$} &  $\frac{\dd^{2}\sigma}{\dd p \dd\theta}(\pi^{-})$ & $\Delta_\text{stat}$ & $\Delta_\text{syst}$ \\
\midrule
\\[2mm]
240 & - & 300 & 0.2 & - & 0.3 & 166.6 & 4.9 & 12.0 \\
 & & & 0.3 & - & 0.4 & 219.6 & 5.1 & 15.1 \\
 & & & 0.4 & - & 0.5 & 285.3 & 5.8 & 20.0 \\
 & & & 0.5 & - & 0.6 & 313.2 & 6.0 & 21.1 \\
 & & & 0.6 & - & 0.7 & 329.9 & 6.2 & 21.6 \\
 & & & 0.7 & - & 0.8 & 310.9 & 5.9 & 18.4 \\
 & & & 0.8 & - & 0.9 & 331.6 & 4.7 & 17.6 \\
 & & & 0.9 & - & 1.0 & 303.4 & 4.4 & 15.9 \\
 & & & 1.0 & - & 1.2 & 292.4 & 3.1 & 14.3 \\
 & & & 1.2 & - & 1.4 & 253.2 & 2.9 & 11.7 \\
 & & & 1.4 & - & 1.6 & 206.1 & 2.6 & 9.3 \\
 & & & 1.6 & - & 1.8 & 168.4 & 2.4 & 7.6 \\
 & & & 1.8 & - & 2.0 & 139.0 & 2.1 & 6.1 \\
 & & & 2.0 & - & 2.2 & 110.4 & 1.9 & 5.0 \\
 & & & 2.2 & - & 2.4 & 86.1 & 1.7 & 4.0 \\
 & & & 2.4 & - & 2.6 & 65.8 & 1.5 & 3.0 \\
 & & & 2.6 & - & 2.8 & 55.0 & 1.4 & 2.7 \\
 & & & 2.8 & - & 3.0 & 40.2 & 1.1 & 1.8 \\
 & & & 3.0 & - & 3.2 & 37.1 & 1.6 & 1.7 \\
 & & & 3.2 & - & 3.4 & 28.6 & 1.5 & 1.4 \\
 & & & 3.4 & - & 3.6 & 23.2 & 1.3 & 1.1 \\
 & & & 3.6 & - & 3.8 & 19.7 & 1.3 & 0.9 \\
 & & & 3.8 & - & 4.0 & 14.1 & 1.0 & 0.6 \\
 & & & 4.0 & - & 4.2 & 9.6 & 0.8 & 0.4 \\
 & & & 4.2 & - & 4.4 & 9.1 & 0.8 & 0.4 \\
 & & & 4.4 & - & 4.6 & 6.2 & 0.6 & 0.4 \\
 & & & 4.6 & - & 4.8 & 6.3 & 0.7 & 0.4 \\
 & & & 4.8 & - & 5.0 & 4.8 & 0.6 & 0.2 \\
 & & & 5.0 & - & 5.2 & 2.9 & 0.4 & 0.1 \\
 & & & 5.2 & - & 5.6 & 2.9 & 0.3 & 0.1 \\
 & & & 5.6 & - & 6.0 & 1.5 & 0.2 & 0.1 \\
 & & & 6.0 & - & 6.4 & 1.0 & 0.2 & 0.1 \\
 & & & 6.4 & - & 6.8 & 1.7 & 0.3 & 0.1 \\
\bottomrule
\end{tabular}
\end{table}

\begin{table}[h]
\begin{tabular}{rcrrcrrrr}
\toprule
\multicolumn{3}{c}{$\theta$} & \multicolumn{3}{c}{$p$} &  $\frac{\dd^{2}\sigma}{\dd p \dd\theta}(\pi^{-})$ & $\Delta_\text{stat}$ & $\Delta_\text{syst}$ \\
\midrule
\\[2mm]
300 & - & 360 & 0.2 & - & 0.3 & 178.6 & 5.5 & 12.2 \\
 & & & 0.3 & - & 0.4 & 240.9 & 6.1 & 16.8 \\
 & & & 0.4 & - & 0.5 & 290.1 & 6.7 & 22.8 \\
 & & & 0.5 & - & 0.6 & 342.3 & 7.7 & 24.3 \\
 & & & 0.6 & - & 0.7 & 339.2 & 7.4 & 22.0 \\
 & & & 0.7 & - & 0.8 & 318.9 & 7.2 & 19.7 \\
\bottomrule
\end{tabular}
\caption{\label{tab:pi-_results} 
The \NASixtyOne results on the $\pi^-$ double differential cross section, $d^2 \sigma_{\pi^-}/(d p d \theta)$ (in [mb/rad/(GeV/$c$)]), in the laboratory system for p+C interactions at 31\,GeV$/c$. The results are presented as a function of momentum, $p$ (in [GeV/$c$]), in different angular intervals, $\theta$ (in [mrad]). The statistical $\Delta_\text{stat}$ (in [mb/rad/(GeV/$c$)]) and systematic $\Delta_\text{syst}$ (in [mb/rad/(GeV/$c$)]) errors are quoted. The overall uncertainty due to the normalization procedure is not included.
}
\end{table}

\clearpage
\begin{table}[h]
\begin{tabular}{rcrrcrrrr}
\toprule
\multicolumn{3}{c}{$\theta$} & \multicolumn{3}{c}{$p$} &  $\frac{\dd^{2}\sigma}{\dd p \dd\theta}(K^{+})$ & $\Delta_\text{stat}$ & $\Delta_\text{syst}$ \\
\midrule
\\[2mm]
0 & - & 20 & 0.8 & - & 1.6 & 2.1 & 0.5 & 0.1 \\
 & & & 1.6 & - & 2.4 & 2.8 & 0.4 & 0.2 \\
 & & & 2.4 & - & 3.2 & 4.6 & 0.6 & 0.2 \\
 & & & 3.2 & - & 4.0 & 3.7 & 0.7 & 0.2 \\
 & & & 4.0 & - & 4.8 & 3.5 & 0.6 & 0.2 \\
 & & & 4.8 & - & 5.6 & 3.8 & 0.7 & 0.2 \\
 & & & 5.6 & - & 6.4 & 4.3 & 0.7 & 0.3 \\
 & & & 6.4 & - & 7.6 & 3.5 & 0.6 & 0.3 \\
 & & & 7.6 & - & 8.8 & 3.1 & 0.7 & 0.3 \\
 & & & 8.8 & - & 10.0 & 3.0 & 0.7 & 0.3 \\
 & & & 10.0 & - & 11.6 & 3.2 & 0.6 & 0.4 \\
 & & & 11.6 & - & 13.2 & 2.3 & 1.0 & 0.4 \\
 & & & 13.2 & - & 14.8 & 1.4 & 1.3 & 0.3 \\
 & & & 14.8 & - & 16.4 & 1.6 & 1.6 & 0.4 \\
\\[2mm]
20 & - & 40 & 0.8 & - & 1.6 & 4.3 & 0.6 & 0.2 \\
 & & & 1.6 & - & 2.4 & 7.6 & 0.6 & 0.3 \\
 & & & 2.4 & - & 3.2 & 10.6 & 0.8 & 0.4 \\
 & & & 3.2 & - & 4.0 & 11.9 & 0.9 & 0.5 \\
 & & & 4.0 & - & 4.8 & 12.2 & 1.0 & 0.5 \\
 & & & 4.8 & - & 5.6 & 12.9 & 1.1 & 0.6 \\
 & & & 5.6 & - & 6.4 & 10.3 & 1.1 & 0.5 \\
 & & & 6.4 & - & 7.6 & 9.7 & 0.9 & 0.5 \\
 & & & 7.6 & - & 8.8 & 8.6 & 0.9 & 0.6 \\
 & & & 8.8 & - & 10.0 & 8.5 & 0.8 & 0.8 \\
 & & & 10.0 & - & 11.6 & 5.8 & 0.7 & 0.8 \\
 & & & 11.6 & - & 13.2 & 4.1 & 0.7 & 0.7 \\
 & & & 13.2 & - & 14.8 & 3.7 & 0.7 & 0.8 \\
 & & & 14.8 & - & 16.4 & 2.4 & 0.6 & 0.5 \\
\\[2mm]
40 & - & 60 & 0.8 & - & 1.6 & 10.4 & 1.3 & 0.4 \\
 & & & 1.6 & - & 2.4 & 10.9 & 1.1 & 0.3 \\
 & & & 2.4 & - & 3.2 & 15.7 & 1.4 & 0.4 \\
 & & & 3.2 & - & 4.0 & 19.9 & 1.6 & 0.5 \\
 & & & 4.0 & - & 4.8 & 19.5 & 1.6 & 0.5 \\
 & & & 4.8 & - & 5.6 & 17.0 & 1.6 & 0.5 \\
 & & & 5.6 & - & 6.4 & 17.4 & 1.6 & 0.6 \\
 & & & 6.4 & - & 7.6 & 13.8 & 1.2 & 0.6 \\
 & & & 7.6 & - & 8.8 & 10.7 & 1.0 & 0.6 \\
 & & & 8.8 & - & 10.0 & 9.2 & 0.9 & 0.7 \\
 & & & 10.0 & - & 11.6 & 6.5 & 0.7 & 0.6 \\
 & & & 11.6 & - & 13.2 & 3.7 & 0.6 & 0.5 \\
 & & & 13.2 & - & 14.8 & 2.5 & 0.5 & 0.4 \\
 & & & 14.8 & - & 16.4 & 1.7 & 0.4 & 0.3 \\
 & & & 16.4 & - & 18.0 & 0.4 & 0.2 & 0.1 \\
 & & & 18.0 & - & 19.6 & 0.4 & 0.2 & 0.1 \\
\\[2mm]
60 & - & 100 & 0.8 & - & 1.6 & 9.7 & 1.1 & 0.3 \\
 & & & 1.6 & - & 2.4 & 20.5 & 1.4 & 0.6 \\
 & & & 2.4 & - & 3.2 & 20.0 & 1.3 & 0.5 \\
 & & & 3.2 & - & 4.0 & 23.3 & 1.4 & 0.6 \\
 & & & 4.0 & - & 4.8 & 19.8 & 1.0 & 0.6 \\
 & & & 4.8 & - & 5.6 & 16.5 & 0.9 & 0.5 \\
 & & & 5.6 & - & 6.4 & 16.1 & 0.9 & 0.5 \\
 & & & 6.4 & - & 7.6 & 11.6 & 0.7 & 0.5 \\
 & & & 7.6 & - & 8.8 & 7.8 & 0.6 & 0.4 \\
 & & & 8.8 & - & 10.0 & 5.2 & 0.5 & 0.3 \\
 & & & 10.0 & - & 11.6 & 3.2 & 0.4 & 0.2 \\
 & & & 11.6 & - & 13.2 & 1.2 & 0.2 & 0.1 \\
 & & & 13.2 & - & 14.8 & 0.9 & 0.2 & 0.1 \\
 & & & 14.8 & - & 16.4 & 0.3 & 0.1 & 0.04 \\
\bottomrule
\end{tabular}
\end{table}

\begin{table}[h]
\begin{tabular}{rcrrcrrrr}
\toprule
\multicolumn{3}{c}{$\theta$} & \multicolumn{3}{c}{$p$} &  $\frac{\dd^{2}\sigma}{\dd p \dd\theta}(K^{+})$ & $\Delta_\text{stat}$ & $\Delta_\text{syst}$ \\
\midrule
\\[2mm]
100 & - & 140 & 0.8 & - & 1.6 & 16.9 & 2.2 & 0.6 \\
 & & & 1.6 & - & 2.4 & 27.7 & 2.0 & 0.8 \\
 & & & 2.4 & - & 3.2 & 26.3 & 1.4 & 0.7 \\
 & & & 3.2 & - & 4.0 & 20.7 & 1.2 & 0.6 \\
 & & & 4.0 & - & 4.8 & 18.0 & 1.1 & 0.5 \\
 & & & 4.8 & - & 5.6 & 11.8 & 0.9 & 0.4 \\
 & & & 5.6 & - & 6.4 & 8.6 & 0.8 & 0.3 \\
 & & & 6.4 & - & 7.6 & 5.2 & 0.5 & 0.2 \\
 & & & 7.6 & - & 8.8 & 3.1 & 0.5 & 0.2 \\
 & & & 8.8 & - & 10.0 & 1.5 & 0.4 & 0.1 \\
 & & & 10.0 & - & 11.6 & 0.6 & 0.2 & 0.05 \\
 & & & 11.6 & - & 13.2 & 0.3 & 0.1 & 0.03 \\
\\[2mm]
140 & - & 180 & 0.8 & - & 1.6 & 18.1 & 4.1 & 0.6 \\
 & & & 1.6 & - & 2.4 & 24.2 & 1.9 & 0.7 \\
 & & & 2.4 & - & 3.2 & 23.7 & 1.7 & 0.6 \\
 & & & 3.2 & - & 4.0 & 18.2 & 1.4 & 0.5 \\
 & & & 4.0 & - & 4.8 & 8.8 & 1.0 & 0.2 \\
 & & & 4.8 & - & 5.6 & 5.6 & 0.8 & 0.2 \\
 & & & 5.6 & - & 6.4 & 4.0 & 0.7 & 0.1 \\
 & & & 6.4 & - & 7.6 & 1.6 & 0.4 & 0.1 \\
 & & & 7.6 & - & 8.8 & 0.6 & 0.3 & 0.03 \\
 & & & 8.8 & - & 10.0 & 0.2 & 0.1 & 0.01 \\
\\[2mm]
180 & - & 240 & 0.8 & - & 1.6 & 22.9 & 3.3 & 0.8 \\
 & & & 1.6 & - & 2.4 & 24.9 & 2.4 & 0.8 \\
 & & & 2.4 & - & 3.2 & 14.0 & 1.6 & 0.4 \\
 & & & 3.2 & - & 4.0 & 11.5 & 1.4 & 0.3 \\
 & & & 4.0 & - & 4.8 & 5.4 & 1.0 & 0.1 \\
 & & & 4.8 & - & 5.6 & 2.0 & 0.6 & 0.1 \\
 & & & 5.6 & - & 6.4 & 2.4 & 0.8 & 0.1 \\
 & & & 6.4 & - & 7.6 & 0.4 & 0.2 & 0.02 \\
\\[2mm]
240 & - & 300 & 0.8 & - & 1.6 & 24.0 & 3.2 & 0.7 \\
 & & & 1.6 & - & 2.4 & 17.4 & 2.0 & 0.5 \\
 & & & 2.4 & - & 3.2 & 10.2 & 1.4 & 0.3 \\
 & & & 3.2 & - & 4.0 & 5.5 & 1.2 & 0.1 \\
 & & & 4.0 & - & 4.8 & 2.2 & 0.8 & 0.1 \\
 & & & 4.8 & - & 5.6 & 0.5 & 0.3 & 0.01 \\
\bottomrule
\end{tabular}
\caption{\label{tab:k+_results} 
The \NASixtyOne results on the $K^+$ double differential cross section, $d^2 \sigma_{K^+}/(d p d \theta)$ (in [mb/rad/(GeV/$c$)]), in the laboratory system for p+C interactions at 31\,GeV$/c$. The results are presented as a function of momentum, $p$ (in [GeV/$c$]), in different angular intervals, $\theta$ (in [mrad]). The statistical $\Delta_\text{stat}$ (in [mb/rad/(GeV/$c$)]) and systematic $\Delta_\text{syst}$ (in [mb/rad/(GeV/$c$)]) errors are quoted.The overall uncertainty due to the normalization procedure is not included. 
}
\end{table}

\clearpage
\begin{table}[h]
\begin{tabular}{rcrrcrrrr}
\toprule
\multicolumn{3}{c}{$\theta$} & \multicolumn{3}{c}{$p$} &  $\frac{\dd^{2}\sigma}{\dd p \dd\theta}(K^{-})$ & $\Delta_\text{stat}$ & $\Delta_\text{syst}$ \\
\midrule
\\[2mm]
0 & - & 20 & 0.8 & - & 1.6 & 0.5 & 0.2 & 0.03 \\
 & & & 1.6 & - & 2.4 & 1.5 & 0.3 & 0.1 \\
 & & & 2.4 & - & 3.2 & 0.9 & 0.3 & 0.1 \\
 & & & 3.2 & - & 4.0 & 1.7 & 0.4 & 0.2 \\
 & & & 4.0 & - & 4.8 & 1.7 & 0.3 & 0.2 \\
 & & & 4.8 & - & 5.6 & 1.7 & 0.3 & 0.2 \\
 & & & 5.6 & - & 6.4 & 1.2 & 0.3 & 0.1 \\
 & & & 6.4 & - & 7.6 & 1.2 & 0.3 & 0.1 \\
 & & & 7.6 & - & 8.8 & 1.2 & 0.2 & 0.1 \\
 & & & 8.8 & - & 10.0 & 0.5 & 0.2 & 0.04 \\
 & & & 10.0 & - & 11.6 & 0.7 & 0.2 & 0.1 \\
 & & & 11.6 & - & 13.2 & 0.2 & 0.1 & 0.02 \\
 & & & 13.2 & - & 14.8 & 0.4 & 0.1 & 0.1 \\
 & & & 14.8 & - & 16.4 & 0.1 & 0.2 & 0.02 \\
\\[2mm]
20 & - & 40 & 0.8 & - & 1.6 & 1.9 & 0.4 & 0.1 \\
 & & & 1.6 & - & 2.4 & 3.9 & 0.5 & 0.2 \\
 & & & 2.4 & - & 3.2 & 5.5 & 0.6 & 0.2 \\
 & & & 3.2 & - & 4.0 & 4.9 & 0.6 & 0.2 \\
 & & & 4.0 & - & 4.8 & 5.0 & 0.6 & 0.2 \\
 & & & 4.8 & - & 5.6 & 5.4 & 0.6 & 0.2 \\
 & & & 5.6 & - & 6.4 & 4.1 & 0.6 & 0.2 \\
 & & & 6.4 & - & 7.6 & 3.0 & 0.4 & 0.1 \\
 & & & 7.6 & - & 8.8 & 3.2 & 0.4 & 0.2 \\
 & & & 8.8 & - & 10.0 & 2.2 & 0.3 & 0.1 \\
 & & & 10.0 & - & 11.6 & 1.7 & 0.2 & 0.1 \\
 & & & 11.6 & - & 13.2 & 0.7 & 0.2 & 0.1 \\
 & & & 13.2 & - & 14.8 & 0.2 & 0.1 & 0.02 \\
 & & & 14.8 & - & 16.4 & 0.1 & 0.06 & 0.01 \\
 & & & 16.4 & - & 18.0 & 0.1 & 0.06 & 0.02 \\
 & & & 18.0 & - & 19.6 & 0.1 & 0.06 & 0.02 \\
\\[2mm]
40 & - & 60 & 0.8 & - & 1.6 & 4.1 & 0.9 & 0.1 \\
 & & & 1.6 & - & 2.4 & 6.5 & 0.9 & 0.2 \\
 & & & 2.4 & - & 3.2 & 8.2 & 1.1 & 0.2 \\
 & & & 3.2 & - & 4.0 & 8.0 & 1.0 & 0.2 \\
 & & & 4.0 & - & 4.8 & 7.5 & 1.0 & 0.2 \\
 & & & 4.8 & - & 5.6 & 5.6 & 0.7 & 0.2 \\
 & & & 5.6 & - & 6.4 & 6.6 & 0.9 & 0.2 \\
 & & & 6.4 & - & 7.6 & 4.4 & 0.6 & 0.2 \\
 & & & 7.6 & - & 8.8 & 3.0 & 0.4 & 0.1 \\
 & & & 8.8 & - & 10.0 & 1.2 & 0.2 & 0.1 \\
 & & & 10.0 & - & 11.6 & 1.5 & 0.2 & 0.1 \\
 & & & 11.6 & - & 13.2 & 0.8 & 0.1 & 0.04 \\
 & & & 13.2 & - & 14.8 & 0.1 & 0.1 & 0.01 \\
 & & & 14.8 & - & 16.4 & 0.1 & 0.04 & 0.00 \\
 & & & 16.4 & - & 18.0 & 0.1 & 0.04 & 0.00 \\
\\[2mm]
60 & - & 100 & 0.8 & - & 1.6 & 5.5 & 0.9 & 0.2 \\
 & & & 1.6 & - & 2.4 & 9.6 & 1.0 & 0.3 \\
 & & & 2.4 & - & 3.2 & 9.1 & 0.9 & 0.3 \\
 & & & 3.2 & - & 4.0 & 9.7 & 0.9 & 0.3 \\
 & & & 4.0 & - & 4.8 & 8.0 & 0.6 & 0.2 \\
 & & & 4.8 & - & 5.6 & 6.6 & 0.5 & 0.2 \\
 & & & 5.6 & - & 6.4 & 4.7 & 0.4 & 0.1 \\
 & & & 6.4 & - & 7.6 & 2.5 & 0.2 & 0.1 \\
 & & & 7.6 & - & 8.8 & 1.7 & 0.2 & 0.1 \\
 & & & 8.8 & - & 10.0 & 1.1 & 0.2 & 0.05 \\
 & & & 10.0 & - & 11.6 & 0.5 & 0.1 & 0.03 \\
 & & & 11.6 & - & 13.2 & 0.1 & 0.03 & 0.00 \\
 & & & 13.2 & - & 14.8 & 0.1 & 0.1 & 0.01 \\
 & & & 14.8 & - & 16.4 & 0.01 & 0.01 & 0.00 \\
 & & & 16.4 & - & 18.0 & 0.01 & 0.01 & 0.00 \\
\bottomrule
\end{tabular}
\end{table}

\begin{table}[h]
\begin{tabular}{rcrrcrrrr}
\toprule
\multicolumn{3}{c}{$\theta$} & \multicolumn{3}{c}{$p$} &  $\frac{\dd^{2}\sigma}{\dd p \dd\theta}(K^{-})$ & $\Delta_\text{stat}$ & $\Delta_\text{syst}$ \\
\midrule
\\[2mm]
100 & - & 140 & 0.8 & - & 1.6 & 7.3 & 1.5 & 0.2 \\
 & & & 1.6 & - & 2.4 & 11.4 & 1.2 & 0.4 \\
 & & & 2.4 & - & 3.2 & 10.0 & 0.9 & 0.3 \\
 & & & 3.2 & - & 4.0 & 8.5 & 0.7 & 0.2 \\
 & & & 4.0 & - & 4.8 & 5.4 & 0.6 & 0.1 \\
 & & & 4.8 & - & 5.6 & 3.9 & 0.5 & 0.1 \\
 & & & 5.6 & - & 6.4 & 1.5 & 0.3 & 0.05 \\
 & & & 6.4 & - & 7.6 & 1.4 & 0.2 & 0.05 \\
 & & & 7.6 & - & 8.8 & 0.6 & 0.1 & 0.02 \\
 & & & 8.8 & - & 10.0 & 0.2 & 0.1 & 0.01 \\
 & & & 10.0 & - & 11.6 & 0.02 & 0.02 & 0.00 \\
 & & & 11.6 & - & 13.2 & 0.3 & 0.3 & 0.01 \\
\\[2mm]
140 & - & 180 & 0.8 & - & 1.6 & 12.7 & 4.1 & 0.4 \\
 & & & 1.6 & - & 2.4 & 15.9 & 1.7 & 0.5 \\
 & & & 2.4 & - & 3.2 & 8.9 & 1.0 & 0.2 \\
 & & & 3.2 & - & 4.0 & 6.4 & 0.8 & 0.2 \\
 & & & 4.0 & - & 4.8 & 2.7 & 0.5 & 0.1 \\
 & & & 4.8 & - & 5.6 & 2.0 & 0.5 & 0.1 \\
 & & & 5.6 & - & 6.4 & 1.6 & 0.4 & 0.05 \\
 & & & 6.4 & - & 7.6 & 0.4 & 0.2 & 0.02 \\
 & & & 7.6 & - & 8.8 & 0.5 & 0.5 & 0.02 \\
\\[2mm]
180 & - & 240 & 0.8 & - & 1.6 & 11.3 & 2.5 & 0.4 \\
 & & & 1.6 & - & 2.4 & 11.3 & 1.7 & 0.4 \\
 & & & 2.4 & - & 3.2 & 5.0 & 0.9 & 0.1 \\
 & & & 3.2 & - & 4.0 & 3.6 & 0.9 & 0.1 \\
 & & & 4.0 & - & 4.8 & 1.5 & 0.6 & 0.04 \\
 & & & 4.8 & - & 5.6 & 1.4 & 0.5 & 0.04 \\
 & & & 5.6 & - & 6.4 & 0.1 & 0.1 & 0.01 \\
\bottomrule
\end{tabular}
\caption{\label{tab:k-_results} 
The \NASixtyOne results on the $K^-$ double differential cross section, $d^2 \sigma_{K^-}/(d p d \theta)$ (in [mb/rad/(GeV/$c$)]), in the laboratory system for p+C interactions at 31\,GeV$/c$. The results are presented as a function of momentum, $p$ (in [GeV/$c$]), in different angular intervals, $\theta$ (in [mrad]). The statistical $\Delta_\text{stat}$ (in [mb/rad/(GeV/$c$)]) and systematic $\Delta_\text{syst}$ (in [mb/rad/(GeV/$c$)]) errors are quoted. The overall uncertainty due to the normalization procedure is not included.
}
\end{table}

\clearpage
\begin{table}[h]
\begin{tabular}{rcrrcrrrr}
\toprule
\multicolumn{3}{c}{$\theta$} & \multicolumn{3}{c}{$p$} &  $\frac{\dd^{2}\sigma}{\dd p \dd\theta}(p)$ & $\Delta_\text{stat}$ & $\Delta_\text{syst}$ \\
\midrule
\\[2mm]
0 & - & 10 & 1.2 & - & 1.6 & 2.4 & 0.5 & 0.2 \\
 & & & 1.6 & - & 2.0 & 3.1 & 0.5 & 0.3 \\
 & & & 2.0 & - & 2.4 & 2.2 & 0.4 & 0.2 \\
 & & & 2.4 & - & 2.8 & 2.3 & 0.4 & 0.3 \\
 & & & 2.8 & - & 3.2 & 3.6 & 0.5 & 0.5 \\
 & & & 3.2 & - & 3.6 & 2.7 & 0.6 & 0.3 \\
 & & & 3.6 & - & 4.0 & 2.3 & 0.5 & 0.2 \\
 & & & 4.0 & - & 4.4 & 4.1 & 0.7 & 0.5 \\
 & & & 4.4 & - & 4.8 & 4.8 & 0.8 & 0.5 \\
 & & & 4.8 & - & 5.2 & 4.5 & 1.2 & 0.4 \\
 & & & 5.2 & - & 5.6 & 6.0 & 1.7 & 0.4 \\
 & & & 5.6 & - & 6.0 & 10.3 & 2.2 & 0.9 \\
 & & & 6.0 & - & 6.4 & 9.3 & 1.9 & 0.8 \\
 & & & 6.4 & - & 6.8 & 6.8 & 1.5 & 0.6 \\
 & & & 6.8 & - & 7.2 & 8.1 & 1.7 & 0.6 \\
 & & & 7.2 & - & 7.6 & 6.4 & 1.5 & 0.5 \\
 & & & 7.6 & - & 8.0 & 11.1 & 2.0 & 0.7 \\
 & & & 8.0 & - & 8.4 & 11.4 & 2.6 & 1.3 \\
 & & & 8.4 & - & 8.8 & 19.0 & 3.0 & 2.3 \\
 & & & 8.8 & - & 9.2 & 11.5 & 2.8 & 1.4 \\
 & & & 9.2 & - & 9.6 & 20.2 & 3.0 & 2.4 \\
 & & & 9.6 & - & 10.0 & 19.5 & 2.9 & 2.3 \\
 & & & 10.0 & - & 10.8 & 22.9 & 2.4 & 2.7 \\
 & & & 10.8 & - & 11.6 & 22.2 & 2.3 & 1.4 \\
 & & & 11.6 & - & 12.4 & 23.1 & 2.7 & 1.5 \\
 & & & 12.4 & - & 13.2 & 21.0 & 3.1 & 1.4 \\
 & & & 13.2 & - & 14.0 & 28.4 & 3.4 & 1.9 \\
 & & & 14.0 & - & 14.8 & 30.8 & 3.7 & 2.1 \\
 & & & 14.8 & - & 15.6 & 32.4 & 4.0 & 2.3 \\
 & & & 15.6 & - & 16.4 & 42.6 & 5.3 & 3.0 \\
\bottomrule
\end{tabular}
\end{table}

\begin{table}[h]
\begin{tabular}{rcrrcrrrr}
\toprule
\multicolumn{3}{c}{$\theta$} & \multicolumn{3}{c}{$p$} &  $\frac{\dd^{2}\sigma}{\dd p \dd\theta}(p)$ & $\Delta_\text{stat}$ & $\Delta_\text{syst}$ \\
\midrule
\\[2mm]
10 & - & 20 & 1.2 & - & 1.6 & 9.6 & 0.9 & 0.8 \\
 & & & 1.6 & - & 2.0 & 8.6 & 0.8 & 0.8 \\
 & & & 2.0 & - & 2.4 & 8.8 & 0.8 & 1.0 \\
 & & & 2.4 & - & 2.8 & 7.6 & 0.8 & 0.8 \\
 & & & 2.8 & - & 3.2 & 9.8 & 1.4 & 1.0 \\
 & & & 3.2 & - & 3.6 & 11.4 & 1.3 & 1.3 \\
 & & & 3.6 & - & 4.0 & 12.6 & 1.6 & 1.5 \\
 & & & 4.0 & - & 4.4 & 12.6 & 1.4 & 1.3 \\
 & & & 4.4 & - & 4.8 & 15.0 & 1.5 & 1.4 \\
 & & & 4.8 & - & 5.2 & 14.9 & 1.6 & 1.5 \\
 & & & 5.2 & - & 5.6 & 21.6 & 2.2 & 1.8 \\
 & & & 5.6 & - & 6.0 & 21.6 & 2.4 & 1.8 \\
 & & & 6.0 & - & 6.4 & 24.8 & 2.7 & 1.9 \\
 & & & 6.4 & - & 6.8 & 24.8 & 2.6 & 1.8 \\
 & & & 6.8 & - & 7.2 & 28.5 & 3.0 & 2.2 \\
 & & & 7.2 & - & 7.6 & 26.7 & 3.3 & 1.7 \\
 & & & 7.6 & - & 8.0 & 33.4 & 3.5 & 2.1 \\
 & & & 8.0 & - & 8.4 & 32.6 & 3.5 & 3.2 \\
 & & & 8.4 & - & 8.8 & 42.0 & 4.3 & 4.1 \\
 & & & 8.8 & - & 9.2 & 42.3 & 3.9 & 4.2 \\
 & & & 9.2 & - & 9.6 & 46.2 & 4.6 & 4.5 \\
 & & & 9.6 & - & 10.0 & 60.5 & 4.7 & 5.8 \\
 & & & 10.0 & - & 10.8 & 55.1 & 3.5 & 5.3 \\
 & & & 10.8 & - & 11.6 & 68.3 & 4.3 & 3.9 \\
 & & & 11.6 & - & 12.4 & 71.2 & 3.9 & 3.9 \\
 & & & 12.4 & - & 13.2 & 80.8 & 4.4 & 4.4 \\
 & & & 13.2 & - & 14.0 & 82.7 & 4.3 & 4.5 \\
 & & & 14.0 & - & 14.8 & 95.1 & 4.3 & 5.0 \\
 & & & 14.8 & - & 15.6 & 103.8 & 4.4 & 5.5 \\
 & & & 15.6 & - & 16.4 & 101.0 & 4.5 & 5.3 \\
 & & & 16.4 & - & 17.2 & 110.3 & 4.8 & 5.8 \\
 & & & 17.2 & - & 18.0 & 108.5 & 4.6 & 5.7 \\
 & & & 18.0 & - & 18.8 & 117.6 & 5.2 & 6.2 \\
 & & & 18.8 & - & 19.6 & 117.2 & 5.0 & 6.2 \\
 & & & 19.6 & - & 20.4 & 125.6 & 5.1 & 6.7 \\
 & & & 20.4 & - & 21.2 & 123.4 & 5.4 & 6.6 \\
 & & & 21.2 & - & 22.0 & 110.2 & 5.3 & 6.0 \\
\bottomrule
\end{tabular}
\end{table}

\begin{table}[h]
\begin{tabular}{rcrrcrrrr}
\toprule
\multicolumn{3}{c}{$\theta$} & \multicolumn{3}{c}{$p$} &  $\frac{\dd^{2}\sigma}{\dd p \dd\theta}(p)$ & $\Delta_\text{stat}$ & $\Delta_\text{syst}$ \\
\midrule
\\[2mm]
20 & - & 40 & 1.2 & - & 1.6 & 14.8 & 0.7 & 1.3 \\
 & & & 1.6 & - & 2.0 & 15.4 & 0.7 & 1.6 \\
 & & & 2.0 & - & 2.4 & 17.5 & 0.8 & 2.0 \\
 & & & 2.4 & - & 2.8 & 17.7 & 1.0 & 2.0 \\
 & & & 2.8 & - & 3.2 & 18.9 & 1.1 & 2.1 \\
 & & & 3.2 & - & 3.6 & 20.1 & 1.1 & 2.0 \\
 & & & 3.6 & - & 4.0 & 20.3 & 1.2 & 1.9 \\
 & & & 4.0 & - & 4.4 & 23.9 & 1.2 & 1.9 \\
 & & & 4.4 & - & 4.8 & 25.0 & 1.3 & 2.4 \\
 & & & 4.8 & - & 5.2 & 31.0 & 1.6 & 2.4 \\
 & & & 5.2 & - & 5.6 & 37.3 & 1.8 & 3.1 \\
 & & & 5.6 & - & 6.0 & 40.0 & 1.9 & 3.0 \\
 & & & 6.0 & - & 6.4 & 48.4 & 1.9 & 3.5 \\
 & & & 6.4 & - & 6.8 & 46.8 & 2.0 & 3.1 \\
 & & & 6.8 & - & 7.2 & 51.8 & 2.3 & 3.3 \\
 & & & 7.2 & - & 7.6 & 56.3 & 2.2 & 3.3 \\
 & & & 7.6 & - & 8.0 & 60.3 & 2.4 & 3.6 \\
 & & & 8.0 & - & 8.4 & 66.3 & 2.4 & 3.8 \\
 & & & 8.4 & - & 8.8 & 73.0 & 2.7 & 3.8 \\
 & & & 8.8 & - & 9.2 & 75.8 & 2.8 & 4.0 \\
 & & & 9.2 & - & 9.6 & 80.4 & 2.8 & 4.1 \\
 & & & 9.6 & - & 10.0 & 76.9 & 2.9 & 3.8 \\
 & & & 10.0 & - & 10.8 & 88.0 & 2.2 & 4.2 \\
 & & & 10.8 & - & 11.6 & 89.0 & 2.2 & 4.2 \\
 & & & 11.6 & - & 12.4 & 96.1 & 2.3 & 3.8 \\
 & & & 12.4 & - & 13.2 & 100.7 & 2.5 & 3.9 \\
 & & & 13.2 & - & 14.0 & 98.4 & 2.5 & 3.7 \\
 & & & 14.0 & - & 14.8 & 101.5 & 2.5 & 3.8 \\
 & & & 14.8 & - & 15.6 & 95.0 & 2.5 & 3.5 \\
 & & & 15.6 & - & 16.4 & 97.9 & 2.5 & 3.5 \\
 & & & 16.4 & - & 17.2 & 93.5 & 2.4 & 3.3 \\
 & & & 17.2 & - & 18.0 & 88.2 & 2.4 & 3.1 \\
 & & & 18.0 & - & 18.8 & 82.8 & 2.3 & 2.9 \\
 & & & 18.8 & - & 19.6 & 77.7 & 2.3 & 2.7 \\
 & & & 19.6 & - & 20.4 & 71.9 & 2.3 & 2.5 \\
 & & & 20.4 & - & 21.2 & 66.5 & 2.1 & 2.3 \\
 & & & 21.2 & - & 22.0 & 63.8 & 2.2 & 2.2 \\
\bottomrule
\end{tabular}
\end{table}

\begin{table}[h]
\begin{tabular}{rcrrcrrrr}
\toprule
\multicolumn{3}{c}{$\theta$} & \multicolumn{3}{c}{$p$} &  $\frac{\dd^{2}\sigma}{\dd p \dd\theta}(p)$ & $\Delta_\text{stat}$ & $\Delta_\text{syst}$ \\
\midrule
\\[2mm]
40 & - & 60 & 1.2 & - & 1.6 & 26.6 & 1.3 & 1.9 \\
 & & & 1.6 & - & 2.0 & 25.3 & 1.3 & 2.3 \\
 & & & 2.0 & - & 2.4 & 26.5 & 1.4 & 2.5 \\
 & & & 2.4 & - & 2.8 & 27.0 & 1.6 & 2.5 \\
 & & & 2.8 & - & 3.2 & 26.2 & 1.6 & 2.5 \\
 & & & 3.2 & - & 3.6 & 32.7 & 2.0 & 2.8 \\
 & & & 3.6 & - & 4.0 & 34.5 & 2.1 & 3.2 \\
 & & & 4.0 & - & 4.4 & 38.4 & 2.2 & 2.8 \\
 & & & 4.4 & - & 4.8 & 42.5 & 2.4 & 3.5 \\
 & & & 4.8 & - & 5.2 & 40.9 & 2.2 & 2.8 \\
 & & & 5.2 & - & 5.6 & 45.0 & 2.5 & 3.3 \\
 & & & 5.6 & - & 6.0 & 54.0 & 2.8 & 3.8 \\
 & & & 6.0 & - & 6.4 & 55.6 & 2.8 & 3.5 \\
 & & & 6.4 & - & 6.8 & 64.0 & 3.1 & 3.6 \\
 & & & 6.8 & - & 7.2 & 68.5 & 3.1 & 3.5 \\
 & & & 7.2 & - & 7.6 & 71.0 & 3.2 & 3.6 \\
 & & & 7.6 & - & 8.0 & 68.7 & 2.8 & 3.1 \\
 & & & 8.0 & - & 8.4 & 78.2 & 2.9 & 3.4 \\
 & & & 8.4 & - & 8.8 & 76.0 & 2.7 & 3.1 \\
 & & & 8.8 & - & 9.2 & 80.1 & 2.9 & 3.3 \\
 & & & 9.2 & - & 9.6 & 75.8 & 2.6 & 3.0 \\
 & & & 9.6 & - & 10.0 & 80.3 & 2.7 & 3.1 \\
 & & & 10.0 & - & 10.8 & 83.8 & 1.9 & 3.1 \\
 & & & 10.8 & - & 11.6 & 77.9 & 1.9 & 2.7 \\
 & & & 11.6 & - & 12.4 & 78.5 & 1.9 & 2.6 \\
 & & & 12.4 & - & 13.2 & 73.8 & 1.8 & 2.4 \\
 & & & 13.2 & - & 14.0 & 65.5 & 1.6 & 2.1 \\
 & & & 14.0 & - & 14.8 & 62.1 & 1.6 & 1.9 \\
 & & & 14.8 & - & 15.6 & 52.3 & 1.4 & 1.6 \\
 & & & 15.6 & - & 16.4 & 49.0 & 1.4 & 1.4 \\
 & & & 16.4 & - & 17.2 & 44.6 & 1.3 & 1.3 \\
 & & & 17.2 & - & 18.0 & 34.2 & 1.1 & 0.9 \\
 & & & 18.0 & - & 18.8 & 30.3 & 1.1 & 0.8 \\
 & & & 18.8 & - & 19.6 & 25.7 & 1.0 & 0.7 \\
 & & & 19.6 & - & 20.4 & 21.0 & 0.9 & 0.6 \\
\\[2mm]
60 & - & 100 & 1.2 & - & 1.6 & 42.3 & 1.6 & 3.1 \\
 & & & 1.6 & - & 2.0 & 40.3 & 1.6 & 3.2 \\
 & & & 2.0 & - & 2.4 & 37.1 & 1.7 & 3.1 \\
 & & & 2.4 & - & 2.8 & 39.9 & 1.7 & 3.9 \\
 & & & 2.8 & - & 3.2 & 44.1 & 1.9 & 3.9 \\
 & & & 3.2 & - & 3.6 & 44.4 & 1.6 & 3.9 \\
 & & & 3.6 & - & 4.0 & 45.8 & 1.6 & 3.9 \\
 & & & 4.0 & - & 4.4 & 49.5 & 1.6 & 4.1 \\
 & & & 4.4 & - & 4.8 & 49.4 & 1.6 & 3.9 \\
 & & & 4.8 & - & 5.2 & 53.8 & 1.7 & 3.7 \\
 & & & 5.2 & - & 5.6 & 53.6 & 1.6 & 3.6 \\
 & & & 5.6 & - & 6.0 & 52.6 & 1.6 & 3.2 \\
 & & & 6.0 & - & 6.4 & 56.0 & 1.7 & 3.3 \\
 & & & 6.4 & - & 6.8 & 54.7 & 1.7 & 2.9 \\
 & & & 6.8 & - & 7.2 & 56.9 & 1.7 & 2.8 \\
 & & & 7.2 & - & 7.6 & 53.9 & 1.7 & 2.4 \\
 & & & 7.6 & - & 8.0 & 56.2 & 1.7 & 2.5 \\
 & & & 8.0 & - & 8.4 & 56.0 & 1.7 & 2.3 \\
 & & & 8.4 & - & 8.8 & 49.9 & 1.6 & 1.9 \\
 & & & 8.8 & - & 9.2 & 47.9 & 1.5 & 1.9 \\
 & & & 9.2 & - & 9.6 & 44.4 & 1.5 & 1.6 \\
 & & & 9.6 & - & 10.0 & 43.0 & 1.5 & 1.5 \\
 & & & 10.0 & - & 10.8 & 40.0 & 1.0 & 1.4 \\
 & & & 10.8 & - & 11.6 & 33.8 & 0.9 & 1.1 \\
 & & & 11.6 & - & 12.4 & 28.6 & 0.9 & 0.9 \\
 & & & 12.4 & - & 13.2 & 22.5 & 0.8 & 0.7 \\
 & & & 13.2 & - & 14.0 & 17.7 & 0.7 & 0.5 \\
 & & & 14.0 & - & 14.8 & 13.6 & 0.6 & 0.4 \\
 & & & 14.8 & - & 15.6 & 11.4 & 0.5 & 0.3 \\
 & & & 15.6 & - & 16.4 & 8.6 & 0.5 & 0.2 \\
 & & & 16.4 & - & 17.2 & 6.0 & 0.4 & 0.2 \\
 & & & 17.2 & - & 18.0 & 4.5 & 0.3 & 0.1 \\
\bottomrule
\end{tabular}
\end{table}

\begin{table}[h]
\begin{tabular}{rcrrcrrrr}
\toprule
\multicolumn{3}{c}{$\theta$} & \multicolumn{3}{c}{$p$} &  $\frac{\dd^{2}\sigma}{\dd p \dd\theta}(p)$ & $\Delta_\text{stat}$ & $\Delta_\text{syst}$ \\
\midrule
\\[2mm]
100 & - & 140 & 1.2 & - & 1.6 & 60.0 & 2.7 & 4.0 \\
 & & & 1.6 & - & 2.0 & 55.4 & 2.5 & 3.7 \\
 & & & 2.0 & - & 2.4 & 56.4 & 2.1 & 4.7 \\
 & & & 2.4 & - & 2.8 & 56.0 & 1.9 & 4.5 \\
 & & & 2.8 & - & 3.2 & 51.2 & 1.8 & 4.0 \\
 & & & 3.2 & - & 3.6 & 50.7 & 1.8 & 3.9 \\
 & & & 3.6 & - & 4.0 & 48.9 & 1.8 & 3.7 \\
 & & & 4.0 & - & 4.4 & 47.9 & 1.8 & 3.4 \\
 & & & 4.4 & - & 4.8 & 45.3 & 1.8 & 2.8 \\
 & & & 4.8 & - & 5.2 & 47.2 & 1.9 & 2.7 \\
 & & & 5.2 & - & 5.6 & 41.8 & 1.7 & 2.2 \\
 & & & 5.6 & - & 6.0 & 38.0 & 1.7 & 1.9 \\
 & & & 6.0 & - & 6.4 & 37.3 & 1.7 & 1.6 \\
 & & & 6.4 & - & 6.8 & 35.4 & 1.6 & 1.5 \\
 & & & 6.8 & - & 7.2 & 32.0 & 1.6 & 1.3 \\
 & & & 7.2 & - & 7.6 & 23.6 & 1.3 & 1.0 \\
 & & & 7.6 & - & 8.0 & 24.2 & 1.4 & 0.9 \\
 & & & 8.0 & - & 8.4 & 21.2 & 1.3 & 0.7 \\
 & & & 8.4 & - & 8.8 & 17.1 & 1.1 & 0.6 \\
 & & & 8.8 & - & 9.2 & 18.2 & 1.2 & 0.6 \\
 & & & 9.2 & - & 9.6 & 11.8 & 1.0 & 0.4 \\
 & & & 9.6 & - & 10.0 & 12.0 & 0.9 & 0.4 \\
 & & & 10.0 & - & 10.8 & 8.4 & 0.6 & 0.3 \\
 & & & 10.8 & - & 11.6 & 6.3 & 0.5 & 0.2 \\
 & & & 11.6 & - & 12.4 & 3.7 & 0.4 & 0.1 \\
 & & & 12.4 & - & 13.2 & 3.2 & 0.3 & 0.1 \\
\\[2mm]
140 & - & 180 & 1.2 & - & 1.6 & 74.5 & 6.2 & 6.5 \\
 & & & 1.6 & - & 2.0 & 71.7 & 3.0 & 6.8 \\
 & & & 2.0 & - & 2.4 & 65.0 & 2.6 & 4.6 \\
 & & & 2.4 & - & 2.8 & 57.4 & 2.4 & 3.9 \\
 & & & 2.8 & - & 3.2 & 54.4 & 2.4 & 3.6 \\
 & & & 3.2 & - & 3.6 & 50.8 & 2.3 & 3.4 \\
 & & & 3.6 & - & 4.0 & 44.2 & 2.2 & 2.7 \\
 & & & 4.0 & - & 4.4 & 42.9 & 2.2 & 2.9 \\
 & & & 4.4 & - & 4.8 & 37.9 & 2.1 & 1.9 \\
 & & & 4.8 & - & 5.2 & 30.6 & 1.8 & 1.4 \\
 & & & 5.2 & - & 5.6 & 27.6 & 1.8 & 1.5 \\
 & & & 5.6 & - & 6.0 & 21.8 & 1.6 & 0.9 \\
 & & & 6.0 & - & 6.4 & 16.6 & 1.4 & 0.7 \\
 & & & 6.4 & - & 6.8 & 14.4 & 1.3 & 0.5 \\
 & & & 6.8 & - & 7.2 & 12.8 & 1.3 & 0.5 \\
 & & & 7.2 & - & 7.6 & 10.1 & 1.0 & 0.3 \\
 & & & 7.6 & - & 8.0 & 8.7 & 1.1 & 0.3 \\
 & & & 8.0 & - & 8.4 & 5.6 & 0.8 & 0.2 \\
 & & & 8.4 & - & 8.8 & 5.4 & 0.7 & 0.2 \\
 & & & 8.8 & - & 9.2 & 3.3 & 0.6 & 0.1 \\
 & & & 9.2 & - & 9.6 & 1.3 & 0.4 & 0.04 \\
 & & & 9.6 & - & 10.0 & 1.3 & 0.4 & 0.05 \\
\bottomrule
\end{tabular}
\end{table}

\begin{table}[h]
\begin{tabular}{rcrrcrrrr}
\toprule
\multicolumn{3}{c}{$\theta$} & \multicolumn{3}{c}{$p$} &  $\frac{\dd^{2}\sigma}{\dd p \dd\theta}(p)$ & $\Delta_\text{stat}$ & $\Delta_\text{syst}$ \\
\midrule
\\[2mm]
180 & - & 240 & 1.2 & - & 1.6 & 104.0 & 4.4 & 5.9 \\
 & & & 1.6 & - & 2.0 & 78.5 & 3.6 & 4.2 \\
 & & & 2.0 & - & 2.4 & 65.4 & 3.1 & 3.6 \\
 & & & 2.4 & - & 2.8 & 57.5 & 2.7 & 3.4 \\
 & & & 2.8 & - & 3.2 & 50.7 & 2.5 & 2.7 \\
 & & & 3.2 & - & 3.6 & 41.1 & 2.1 & 2.0 \\
 & & & 3.6 & - & 4.0 & 35.3 & 2.0 & 1.7 \\
 & & & 4.0 & - & 4.4 & 31.1 & 1.9 & 1.4 \\
 & & & 4.4 & - & 4.8 & 20.4 & 1.4 & 0.9 \\
 & & & 4.8 & - & 5.2 & 16.1 & 1.2 & 0.6 \\
 & & & 5.2 & - & 5.6 & 12.4 & 1.2 & 0.5 \\
 & & & 5.6 & - & 6.0 & 10.1 & 1.0 & 0.4 \\
 & & & 6.0 & - & 6.4 & 6.4 & 0.8 & 0.2 \\
 & & & 6.4 & - & 6.8 & 5.6 & 0.7 & 0.2 \\
 & & & 6.8 & - & 7.2 & 3.0 & 0.5 & 0.1 \\
 & & & 7.2 & - & 7.6 & 3.4 & 0.6 & 0.1 \\
\\[2mm]
240 & - & 300 & 1.2 & - & 1.6 & 114.5 & 4.5 & 5.8 \\
 & & & 1.6 & - & 2.0 & 91.6 & 4.0 & 4.4 \\
 & & & 2.0 & - & 2.4 & 77.5 & 3.8 & 3.5 \\
 & & & 2.4 & - & 2.8 & 58.1 & 3.3 & 2.6 \\
 & & & 2.8 & - & 3.2 & 43.8 & 2.9 & 1.9 \\
 & & & 3.2 & - & 3.6 & 30.7 & 2.4 & 1.3 \\
 & & & 3.6 & - & 4.0 & 20.9 & 2.0 & 0.9 \\
 & & & 4.0 & - & 4.4 & 13.6 & 1.6 & 0.6 \\
 & & & 4.4 & - & 4.8 & 11.6 & 1.7 & 0.4 \\
 & & & 4.8 & - & 5.2 & 7.2 & 1.2 & 0.2 \\
\\[2mm]
300 & - & 360 & 1.2 & - & 1.6 & 110.2 & 5.4 & 4.9 \\
 & & & 1.6 & - & 2.0 & 81.2 & 5.5 & 3.2 \\
 & & & 2.0 & - & 2.4 & 55.7 & 5.5 & 2.2 \\
 & & & 2.4 & - & 2.8 & 57.3 & 8.1 & 2.6 \\
 & & & 2.8 & - & 3.2 & 33.3 & 8.2 & 3.0 \\
 & & & 3.2 & - & 3.6 & 13.9 & 8.8 & 0.3 \\
\bottomrule
\end{tabular}
\caption{\label{tab:proton_results} 
The \NASixtyOne results on the proton double differential cross section, $d^2 \sigma_\text{p}/(d p d \theta)$ (in [mb/rad/(GeV/$c$)]), in the laboratory system for p+C interactions at 31\,GeV$/c$. The results are presented as a function of momentum, $p$ (in [GeV/$c$]), in different angular intervals, $\theta$ (in [mrad]). The statistical $\Delta_\text{stat}$ (in [mb/rad/(GeV/$c$)]) and systematic $\Delta_\text{syst}$ (in [mb/rad/(GeV/$c$)]) errors are quoted. The overall uncertainty due to the normalization procedure is not included.
}
\end{table}

\begin{table}[h]
\begin{tabular}{rcrrcrrrr}
\toprule
\multicolumn{3}{c}{$\theta$} & \multicolumn{3}{c}{$p$} &  $\frac{\dd^2\sigma}{\dd p \dd\theta}(K^0_S)$ & $\Delta_\text{stat}$ & $\Delta_\text{syst}$ 
\\
\midrule
\\[2mm]
0 & - & 40 & 0.4 & - & 3.4 & 4.9 & 0.6 & 0.7 \\
 & & & 3.4 & - & 6.4 & 5.1 & 0.5 & 0.9 \\
 & & & 6.4 & - & 9.4 & 2.4 & 0.3 & 0.5 \\
 & & & 9.4 & - & 12.4 & 1.6 & 0.2 & 0.3 \\ 
\\[2mm] 
40 & - & 60 & 0.4 & - & 3.4 & 11.1 & 1.3 & 1.9 \\ 
& & & 3.4 & - & 6.4 & 11.4 & 0.9 & 1.8 \\
 & & & 6.4 & - & 9.4 & 5.3 & 0.6 & 0.7 \\
 & & & 9.4 & - & 12.4 & 1.8 & 0.3 & 0.3 \\
\\[2mm] 
60 & - & 100 & 0.4 & - & 2.4 & 10.6 & 1.1 & 2.1 \\ 
& & & 2.4 & - & 4.4 & 15.1 & 1.0 & 1.7 \\
 & & & 4.4 & - & 6.4 & 10.8 & 0.7 & 1.3 \\
 & & & 6.4 & - & 8.4 & 5.4 & 0.4 & 1.1 \\
 & & & 8.4 & - & 10.4 & 2.6 & 0.3 & 0.5 \\
 & & & 10.4 & - & 12.4 & 1.1 & 0.2 & 0.2 \\
\\[2mm] 
100 & - & 140 & 0.4 & - & 2.4 & 12.2 & 1.3 & 2.1 \\ 
 & & & 2.4 & - & 4.4 & 16.4 & 1.0 & 1.6 \\
 & & & 4.4 & - & 6.4 & 7.2 & 0.5 & 0.7 \\
 & & & 6.4 & - & 8.4 & 2.8 & 0.3 & 0.5 \\
 & & & 8.4 & - & 12.4 & 0.4 & 0.1 & 0.1 \\
\\[2mm] 
140 & - & 180 & 0.4 & - & 2.4 & 13.7 & 1.8 & 2.1 \\ 
 & & & 2.4 & - & 4.4 & 11.6 & 0.8 & 1.1 \\
 & & & 4.4 & - & 6.4 & 3.7 & 0.3 & 0.3 \\
 & & & 6.4 & - & 9.4 & 0.9 & 0.1 & 0.1 \\
\\[2mm] 
180 & - & 240 & 0.4 & - & 2.4 & 14.5 & 1.8 & 2.5 \\ 
 & & & 2.4 & - & 4.4 & 7.7 & 0.5 & 0.5 \\
 & & & 4.4 & - & 9.4 & 0.7 & 0.1 & 0.1 \\
\\[2mm] 
240 & - & 300 & 0.4 & - & 3.4 & 12.8 & 1.3 & 2.6 \\ 
 & & & 3.4 & - & 9.4 & 0.4 & 0.1 & 0.1 \\ 
\bottomrule
\end{tabular}
\caption{\label{tab:k0_results} 
The \NASixtyOne results on the $K^0_S$ double differential cross section, $d^2 \sigma_{K^0_S}/(d p d \theta)$ (in [mb/rad/(GeV/$c$)]), in the laboratory system for p+C interactions at 31\,GeV$/c$. The results are presented as a function of momentum, $p$ (in [GeV/$c$]), in different angular intervals, $\theta$ (in [mrad]). The statistical $\Delta_\text{stat}$ (in [mb/rad/(GeV/$c$)]) and systematic $\Delta_\text{syst}$ (in [mb/rad/(GeV/$c$)]) errors are quoted. The overall uncertainty due to the normalization procedure is not included.
}
\end{table}

\begin{table}[h]
\centering
\begin{tabular}{rcrrcrrrr}
\toprule
\multicolumn{3}{c}{$\theta$} & \multicolumn{3}{c}{$p$} &  $\frac{\dd^2\sigma}{\dd p\dd\theta}(\Lambda)$ & $\Delta_\text{stat}$ & $\Delta_\text{syst}$ \\
\midrule
\\[2mm]
0 & - & 40 & 0.4 & - & 4.4 & 2.0 & 0.3 & 0.2 \\ 
& & & 4.4 & - & 7.4 & 3.9 & 0.4 & 0.6 \\
 & & & 7.4 & - & 10.4 & 5.5 & 0.4 & 0.8 \\
 & & & 10.4 & - & 13.4 & 6.0 & 0.4 & 0.7 \\
 & & & 13.4 & - & 17.4 & 5.9 & 0.4 & 0.7 \\
 & & & 17.4 & - & 21.4 & 2.8 & 0.3 & 0.2 \\
\\[2mm] 
 40 & - & 60 & 0.4 & - & 4.4 & 5.3 & 0.8 & 0.7 \\ 
  & & & 4.4 & - & 7.4 & 7.4 & 0.7 & 0.8 \\
 & & & 7.4 & - & 10.4 & 8.4 & 0.6 & 0.8 \\
 & & & 10.4 & - & 13.4 & 7.4 & 0.5 & 0.6 \\
 & & & 13.4 & - & 17.4 & 4.0 & 0.3 & 0.7 \\
 & & & 17.4 & - & 21.4 & 1.0 & 0.1 & 0.1 \\
\\[2mm] 
60 & - & 100 & 0.4 & - & 2.9 & 5.8 & 0.9 & 0.8 \\ 
 & & & 2.9 & - & 4.4 & 8.2 & 0.8 & 1.8 \\
 & & & 4.4 & - & 6.4 & 8.6 & 0.5 & 1.3 \\
 & & & 6.4 & - & 8.4 & 7.7 & 0.4 & 0.8 \\
 & & & 8.4 & - & 10.4 & 6.0 & 0.3 & 0.6 \\
 & & & 10.4 & - & 12.4 & 3.5 & 0.3 & 0.5 \\
 & & & 12.4 & - & 18.0 & 1.2 & 0.1 & 0.1 \\
 \\[2mm] 
100 & - & 140 & 0.4 & - & 2.9 & 8.2 & 0.9 & 1.5 \\ 
 & & & 2.9 & - & 4.4 & 10.3 & 0.7 & 1.9 \\
 & & & 4.4 & - & 6.4 & 6.9 & 0.4 & 1.0 \\
 & & & 6.4 & - & 8.4 & 4.4 & 0.3 & 0.6 \\
 & & & 8.4 & - & 11.0 & 1.8 & 0.2 & 0.3 \\
 & & & 11.0 & - & 16.4 & 0.2 & 0.04 & 0.03 \\
\\[2mm] 
140 & - & 180 & 0.4 & - & 2.9 & 9.7 & 1.0 & 1.5 \\ 
& & & 2.9 & - & 4.4 & 9.5 & 0.6 & 0.9 \\
 & & & 4.4 & - & 6.4 & 5.0 & 0.3 & 0.8 \\
 & & & 6.4 & - & 8.0 & 2.0 & 0.2 & 0.3 \\
 & & & 8.0 & - & 11.4 & 0.5 & 0.1 & 0.1 \\
\\[2mm] 
180 & - & 240 & 0.4 & - & 2.4 & 10.0 & 1.0 & 0.9 \\ 
& & & 2.4 & - & 4.4 & 7.6 & 0.4 & 0.7 \\
 & & & 4.4 & - & 6.4 & 2.2 & 0.2 & 0.3 \\
 & & & 6.4 & - & 11.0 & 0.3 & 0.1 & 0.1 \\
\\[2mm] 
240 & - & 300 & 0.4 & - & 2.4 & 7.4 & 1.0 & 1.3 \\ 
 & & & 2.4 & - & 4.4 & 5.4 & 0.4 & 0.5 \\
 & & & 4.4 & - & 10.4 & 0.3 & 0.04 & 0.1 \\
\\[2mm] 
300 & - & 420 & 0.4 & - & 3.0 & 7.2 & 0.7 & 0.9 \\ 
 & & & 3.0 & - & 5.6 & 0.9 & 0.1 & 0.2 \\ 
\bottomrule
\end{tabular}
\caption{\label{tab:l_results} 
The \NASixtyOne results on the $\Lambda$ double differential cross section, $d^2 \sigma_{\Lambda}/(d p d \theta)$ (in [mb/rad/(GeV/$c$)]), in the laboratory system for p+C interactions at 31\,GeV$/c$. The results are presented as a function of momentum, $p$ (in [GeV/$c$]), in different angular intervals, $\theta$ (in [mrad]). The statistical $\Delta_\text{stat}$ (in [mb/rad/(GeV/$c$)]) and systematic $\Delta_\text{syst}$ (in [mb/rad/(GeV/$c$)]) errors are quoted. The overall uncertainty due to the normalization procedure is not included.
}
\end{table}

\end{document}